\documentclass[11pt,a4paper]{article}
\usepackage{fullpage}
\usepackage{latexsym}

\usepackage[centertags]{amsmath}
\allowdisplaybreaks[4]
\usepackage{amsfonts}
\usepackage{amssymb}
\usepackage{bm}
\usepackage[dvips]{graphicx}

\newcommand{\psla}{\mbox{$\not{\! p}$}}
\newcommand{\qsla}{\mbox{$\not{\! q}$}}
\newcommand{\ksla}{\mbox{$\not{\! k}$}}
\newcommand{\Qsla}{\mbox{$\not{\! Q}$}}
\newcommand{\Msla}{\mbox{$\not{\! \! M}$}}
\newcommand{\varepssla}{\mbox{$\not{\! \varepsilon}$}}

\newcommand{\dsp}{\displaystyle}

\def\dem {\noindent {\bf Proof : }}

\begin{document}

\begin{flushright}
\begin{tabular}{r}
LAPTH-1241/08\\
Avril 2008
\end{tabular}
\end{flushright}
\vspace{1cm}
\begin{center}
\large\bfseries Light-by-light scattering amplitudes from
generalized unitarity in massive QED \\ [0.5cm]
\normalsize\normalfont C.~Bernicot
\\
\small\itshape LAPTH, Universit\'e de Savoie, CNRS,
\\
\small\itshape
 B.P. 110, F-74941
Annecy-le-Vieux Cedex, France
\\
\end{center}

\vspace{3cm}

\begin{abstract}
We calculate all the four-photon helicity amplitudes at the
one-loop level in a massive theory using multiple-cut methods. The
amplitudes are derived in scalar QED, QED and $\textrm{QED}^{{\cal
N} =1}$ theories. We will see the origin of rational terms. We
extend the calculation to the simplest six-photon helicity
amplitude where all photons have the same helicity.
\end{abstract}

\newpage

\section{Introduction}

\hfil

The light-by-light scattering is a good laboratory to find
efficient methods to compute a massive loop multi-leg amplitude,
because gauge invariance and IR/UV finiteness lead to enormous
cancellations when we sum all the Feynman diagram. The first
calculation of the four-photon amplitude in massive QED was done
by Karplus and al. \cite{karplus}. They straightforwardly
calculated each Feynman diagram. Then B. de Tollis \cite{tollis}
computed the four-photon amplitude with Cutkosky rules. Ten years
ago, Bern and Morgan calculated the four-gluon helicity amplitudes
with a massive loop \cite{Bern:massiveloop}. They used the two-cut
unitarity methods in $n$ dimensions. Recently, Bern et al.
calculated the two loop corrections QCD and QED to light-by-light
scattering by fermion loops in the ultrarelativistic limit
\cite{4gammabern:2001}. Then Binoth and al.~\cite{gamsusy}
calculated the four-photon helicity amplitudes in QED, scalar QED,
supersymmetric $\textrm{QED}^{{\cal N}=1}$ and
$\textrm{QED}^{{\cal N}=2}$ in massless theory. They used a
n-dimensional projection method. Two years ago, Brandhuber and al.
calculated the four-gluon one loop helicity amplitudes with
generalized unitarity cuts~\cite{method8}.

\hfil

\noindent Here we want to compute the four-photon amplitude, in
massive QED, with a recent method: the generalized unitarity cuts.
Even at the energy of LHC, 14 TeV, the mass of heavy quark $t$ can
have a significant effect. This article aims at explaining how to
use new unitarity-cut methods in the case of massive theories.
Here we apply this new technology for a $2\rightarrow2$ process.
However, the future project is to calculate easily some $2
\rightarrow 4$ QCD processes, like $ gg \rightarrow
\overline{t}t\overline{t}t$ with heavy quarks, present in the
background of LHC. The knowledge of the background is very
important to detect new particles like Higgs.

\hfil

\noindent In this paper, we calculate the four-photon amplitude at
one loop order in three massive QED theories: scalar QED, QED and
supersymmetric $\textrm{QED}^{{\cal N}=1}$. Our result agree with
\cite{4gammabern:2001,gamsusy}. We call $A^{scalar}_{4}$
(respectively $A^{spinor}_{4}$ and $A^{{\cal N}=1}_{4}$ ) the
four-photon amplitude in scalar QED (respectively QED and
supersymmetric $\textrm{QED}^{{\cal N}=1}$). We obtain very
compact expressions contrary to Karplus and we can deduce easily
the origin of the rational terms. Actually, we can link easily
these three amplitudes with a supersymmetric decomposition. All
diagrams of the four-photon amplitude in QED have the same
pattern: four external photons coupled to a fermion loop. However,
using the fact that degrees of freedom for internal lines can be
added and subtracted~\cite{Bern:super,Dixon:TASI}, we write the
internal fermion as a supersymmetric contribution and a scalar:
\begin{equation}
    f \ = \ -2 \ s + (f+2s) \ \ \Rightarrow \ \ A^{fermion}_{4} \ = \ -2 \ A^{scalar}_{4} + A^{{\cal N}
    =1}_{4}.
\label{supersymmetricdecomposition}
\end{equation}
\noindent This formula is true for massless and massive theories.
Moreover we point out that this formula $\left(
\ref{supersymmetricdecomposition} \right)$ is true for any number
of photons.

\hfil

\noindent We calculate all $A^{scalar}_{4}$ helicity amplitudes
with generalized unitarity cuts in sections
$\ref{amplitudescalar++++},\ref{amplitudescalar-+++},\ref{amplitudescalar--++}
$. Then, we use extensively the supersymmetric decomposition $
\left( \ref{supersymmetricdecomposition} \right)$ to calculate the
QED amplitude $A^{fermion}_{4}$ in section
$\ref{amplitudefermions}$ and the amplitude $A^{{\cal N}=1}_{4}$
in section $\ref{amplitudesuper}$. Indeed, if we write the
four-photon amplitude in QED, the supersymmetric decomposition
imposes the pattern of the supersymmetric amplitude $A^{{\cal
N}=1}_{4}$ function of magnetic moments. We don't need to
calculate all the supersymmetric diagrams. After, we discuss about
the origin of the rational terms and the different cut-techniques
in section $\ref{discussions}$. Finally we derive the most simple
helicity amplitude of the six-photon amplitudes in Section
$\ref{amplitude6photons}$.

\hfil

\noindent But first, some notations and explanations on
generalized unitary cuts will be introduced.

\hfil

\section{Notations and explanations}

\hfil

\subsection{The structure of the amplitude $\gamma_{1}+\gamma_{2}+\gamma_{3}+ \gamma_{4}
\rightarrow 0$}

\hfil

We study the process $ \gamma_{1}+\gamma_{2}+\gamma_{3}+
\gamma_{4} \rightarrow 0$. The momenta of the ingoing photon
called $i$ is $p_{i}^{\mu}$. In this paper we suppose that all the
photons are on-shell, so we have the first relation $ \forall i
\in [1..4], \ p_{i}^{2} =0$. Diagrams at tree order are impossible
and the first non vanishing order is one-loop order.

\hfil

The QED Lagrangian contains one vertex, whereas the scalar QED
Lagrangian gives us two vertices. We recall them in Appendix
$\ref{vertex}$. Therefore in QED we have six one-loop diagrams
whereas in scalar QED we have twenty one one-loop diagrams. As we
have only four external particles entering in the loop, the power
counting tells us that individual diagrams are UV divergent. So we
regularized the divergence in calculating loops in $n=4-2
\epsilon$ dimensions. Nevertheless, thanks to gauge invariance,
the sum over all diagrams has no UV divergences. So we have to
observe compensations. Moreover, as we suppose that the four
photons are on-shell, therefore we decide to decompose the
amplitudes $A^{spinor}_{4}$, $A^{scalar}_{4}$ and $A^{{\cal N} =
1}_{4}$ on a basis of master integrals in $n$ and $n+2$
dimensions:
\begin{align}
    A_{4} = \sum_{i \in {1,2,3,4} } \left( a_{i} {I_{4}^{n+2}}_{i} + b_{i}
    {I_{3}^{n}(\textrm{1mass})}_{i} + c_{i} { I_{2}^{n}(\textrm{1mass}) }_{i} \right) + \textrm{rational
    terms}, \label{decomposition}
\end{align}
\noindent where $I_{4}^{n+2}$ is the no external mass box in $n+2$
dimensions, $I_{3}^{n}(\textrm{1mass})$ is the three-point
function with one external mass and
$I_{2}^{n}(\textrm{\textrm{1mass}})$ is the two-point function
with one external mass. The exact definition of $I_{4}^{n+2}$,
$I_{3}^{n}(\textrm{1mass})$ and $I_{2}^{n}$ can be found, for
example in \cite{Bern:massiveloop,method8,Binoth:2001vm,
Bern:1993kr}; nevertheless to be self consistent we recall them in
Appendix $\ref{Masterintegrals}$. The interest of this basis of
master integrals, is to separate IR/UV, rational and analytic
terms. But this basis is not unique. $I_{4}^{n+2}$ has an analytic
structure with polylogarithms whereas the IR divergences, in
massless theories are carried by the function $I_{3}^{n}(1m)$ and
the UV one, in massless and massive theories by the function
$I_{2}^{n}$. Each diagram is UV divergent so each diagram has
scalar bubbles $I_{2}^{n}$, but the amplitude (sum of diagrams) is
UV finite so we expect to have compensations between the different
scalar bubbles to eliminate the divergences.

\hfil

The amplitude is totally defined by the coefficients
$a_{i},b_{i},c_{i}$ and the $\textrm{rational terms}$. To
calculate all these coefficients, we use the spinor formalism and
the method of the helicity amplitudes.

\hfil

\subsection{Spinor formalism and helicity amplitudes}

\hfil

We use the spinor helicity formalism developed in
~\cite{spinor:chinois}. For the spinorial product, we introduce
the following notation:
\begin{align}
    \langle p_{a}- | p_{b}+ \rangle & = \langle ab \rangle &\\
    \langle p_{a}+ | p_{b}-\rangle &= [ab] &\\
    \langle p_{a}- | \mbox{$\not{\! p_{b}}$} | p_{c} - \rangle &= \langle abc
    \rangle = [cba] = \langle p_{c}+ | \mbox{$\not{\! p_{b}}$} | p_{a} + \rangle = [ p_{c}p_{b}]  \langle p_{b}p_{a}\rangle &\\
    \langle p_{a}+ | \mbox{$\not{\! p_{b}}$}\mbox{$\not{\! p_{c}}$} | p_{d} - \rangle & = [ abcd ] = - [ dcba ] =
    -\langle p_{d}+ | \mbox{$\not{\! p_{c}}$} \mbox{$\not{\! p_{b}}$} | p_{a} - \rangle . &
\end{align}
\noindent Moreover we introduce the classical Mandelstam variables
:
\begin{equation}
    s \ = \ s_{12} \ = \ \langle 12 \rangle [21] \quad t \ = \ s_{14} \ = \ \langle 14 \rangle [41] \quad u \ =
    \ s_{13} \ = \ \langle 13 \rangle [31].
\end{equation}
\noindent All coefficients $a_{i},b_{i},c_{i}$ and rational terms
are described as products of spinor and Mandelstam variables.

\hfil

We calculate all helicity amplitudes. A photon has two helicity
states $\sigma = \pm$. The amplitude is the sum of all helicity
states.
\begin{equation}
    A_{4} \ = \ \dsp \sum_{ \sigma_{1} = \pm,\sigma_{2} = \pm,\sigma_{3} = \pm,\sigma_{4}
    =\pm } A_{4}\left(\sigma_{1},\sigma_{2},\sigma_{3},\sigma_{4}
    \right).
\end{equation}
\noindent As we have two helicity states per external photons and
four external photons, so the amplitude $A_{4}$ is the sum of
$2^{4}=16$ helicity states. However they are not all independent.
In fact we have only three independent helicity states
$A_{4}(++++),A_{4}(-+++)$ and $A_{4}(--++)$. Others are obtained
by permutations and parity.

\hfil

\noindent It is necessary to introduce the polarisation vectors of
the external photons. The advantage of the helicity amplitudes is
that we can express easily all the photon-polarisation vectors
with spinors. Those expressions come from \cite{spinor:chinois}:
\begin{equation}
    \dsp {\varepsilon_{1}^{+}}^{\mu} \ = \ \frac{\langle R \gamma^{\mu} 1\rangle}{\sqrt{2} \langle
    R1\rangle}\quad \quad {\varepsilon_{1}^{-}}^{\mu} \ = \ \frac{[ r \gamma^{\mu}
    1]}{\sqrt{2} [1r]}, \label{vectorpolarisation}
\end{equation}
\noindent where $|R\rangle $ and $|r\rangle $ are two arbitrary
light-like vectors. Before giving results on $A_{4}^{scalar}$, we
give explanations and notations on regularization, propagators and
integrals.

\hfil

\subsection{The regularization scheme and notation of
integrals}\label{schemeregularization}

\hfil

All diagrams have the same pattern and all are UV divergent. To
regularize, we put the loop momentum $Q^{\mu}$ in $n=4-2\epsilon$
dimensions. Capital letters describe vectors in $n$ dimensions
whereas small letters describe vectors in four dimensions. So we
decompose the loop momentum $Q^{\mu} = q^{\mu}+\mu^{\mu} $ with
$q^{\mu}$ the four-dimensional part and $\mu^{\mu} $ the
$-2\epsilon$ part. The four-dimensional space and the
$-2\epsilon$-dimensional space are orthogonal so one has: $Q^{2} =
q^{2}-\mu^{2} $. And we operate in the ``four-dimensional helicity
scheme'', in which all external momenta are in four dimensions. In
$n$ dimensions, we take the prescription where the propagators are
defined by:
\begin{align}
    \textrm{fermion}& \quad \quad i\frac{\Qsla +m}{Q^{2}-m^{2}
    +i\lambda} \ = \ i \frac{\Qsla +m}{q^{2}-\mu^{2}-m^{2}
    +i\lambda} \ = \ i\frac{\Qsla +m}{D^{2}_{q}}, & \label{propagatorfermions} \\
    \textrm{scalar} & \quad \quad i\frac{1}{Q^{2}-m^{2}
    +i\lambda} \ = \ i \frac{1}{q^{2}-\mu^{2}-m^{2}
    +i\lambda} \ = \ i \frac{1}{D^{2}_{q}}. & \label{propagatorscalar}
\end{align}
\noindent The formulas $\left(
\ref{propagatorfermions},\ref{propagatorscalar} \right)$ show that
the running particle seems to have a mass: $m^{2}+\mu^{2}$. We
will observe this phenomena in the analytic expression of the
amplitude $A_{4}$. The generalized unitarity-cut imposes the
calculation of some trees in $n$ dimensions, where the running
particle enters with a mass $ m^{2} + \mu^{2}$. The
$n$-dimensional calculation introduces some integrals with powers
of $\mu$ in the numerator; those integrals are called
extra-dimension integrals. We express, in Appendix
$\ref{integraldimsup}$, those integrals in terms of higher
dimension loop scalar integrals. But here we give the definitions
of some scalar integrals and extra-dimension scalar integrals,
used in this paper \cite{Bern:massiveloop}:
\begin{align}
    I_{N}^{n} \ & = \ \frac{1}{i \pi ^{n/2}} \int d^{n}Q \frac{1}{D_{1}^{2} ...
    D_{N}^{2}}, & \\
    J_{N}^{n} \ & = \ \frac{1}{i \pi ^{n/2}} \int d^{n}Q \frac{m^{2}+\mu^{2}}{D_{1}^{2} ...
    D_{N}^{2}}, & \\
    K_{N}^{n} \ & = \ \frac{1}{i \pi ^{n/2}} \int d^{n}Q \frac{\left( m^{2}+\mu^{2}\right) ^{2}}{D_{1}^{2} ...
    D_{N}^{2}}.
\end{align}
\noindent We explain in Section $\ref{rationnalterms}$ the origin
of those extra scalar integrals. In our convention, we ignore a
factor $K = i (4\pi)^{-n/2}$ in front of each scalar integral of
the amplitude, and we reintroduce it, in the final result. In the
following, we often add arguments to the integral to denote the
order of photons entering the loop. We give the definition of
those arguments:
\begin{align}
    I_{4}^{n}(abcd) \ & = \ I_{4}^{n}\left(s_{ab},s_{ad} \right)
    \ = \ \parbox{2cm}{\includegraphics[width=2cm]{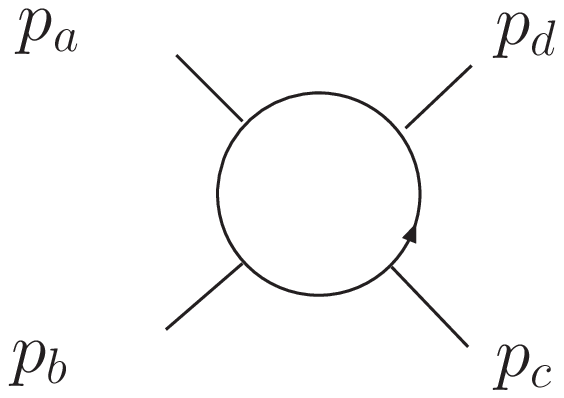}}
    & \nonumber \\
    I_{3}^{n}(s_{cd}) \ & = \
    \parbox{2cm}{\includegraphics[width=2cm]{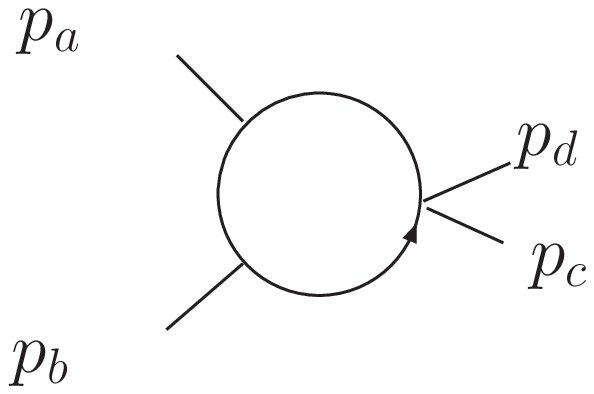}}
    & \nonumber \\
    I_{2}^{n}(s_{cd}) \ & = \ I_{2}^{n}(s_{ab}) \ = \ \parbox{2cm}{\includegraphics[width=2cm]{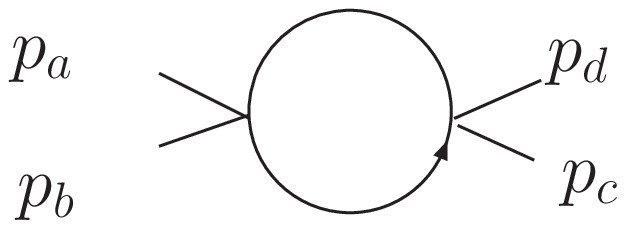}}
\end{align}
\noindent In Appendix $\ref{Masterintegrals}$, we give some
analytic expressions in terms of polylogarithms of those massive
and massless scalar integrals.

\hfil

We decide to express the amplitude $A^{fermion/scalar/{\cal N}
=1}$ as a combination of master integrals $\left(
\ref{decomposition} \right)$. To calculate the different
coefficients in front of each master integral, we use the
generalized unitarity-cut in $n$ dimensions, which we recall in
the next subsection.

\hfil

\subsection{Generalized unitarity-cuts}

\hfil

The unitarity cut method come from the Cutkosky rules
\cite{Cutkosky}. In the last ten years, there has been an intense
development around the unitarity-cuts and several generalizations
have been made. The first generalization of those rules is that we
can cut not only two propagators but also three
\cite{bern3cuts,Mastrolia:2006ki} or four propagators
\cite{Britto:2004nc, Britto:2005ha,Britto:2006sj}. But we will see
that the more we cut propagators, the more we loose information.
However, the more we cut propagators, the more the amplitude is
simple to calculate. The second generalization is to evaluate the
loop integral in $n=4-2 \epsilon$ dimensions
\cite{Bern:massiveloop}, which was improved in \cite{method7}. The
four-dimension cuts are very efficient to calculate coefficients
in front of structures but the extra-dimensional cuts are powerful
to calculate the rational terms. We will see a link between
extra-dimensions and rational terms
\cite{Bern:rational,Papadopoulos:rational}. But the generalization
of this link is not so obvious for a general amplitude. Another
extension, which has recently be done, is the generalization of
the unitarity cuts to massive theories. In
\cite{Bern:massiveloop}, we find the calculation of four-gluon
amplitudes at one loop in massive theory with the two-cut
techniques. And recently, the unitarity-cut techniques in massive
theories was generalized to three and four-cut techniques
\cite{method7bis}. Finally, a few months ago, Papadopoulos and al.
gave a general method to calculate each coefficient in front of
the master integrals \cite{Papa:cut}, extended by Forde
\cite{Forde:2007mi,Kilgore}.

\hfil

\begin{figure}[httb!]
\centering
\includegraphics[width=3cm]{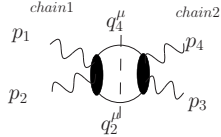}
\caption{\scriptsize \textit{Fermion loop with a cut in the
channel $s_{12}$.}} \label{Figure1}
\end{figure}

The Cutkosky rules \cite{Cutkosky} require to consider an
invariant or a channel, constituted of several consecutive legs.
Consider a loop amplitude, called $A$. One first computes the
discontinuities across branch cuts (imaginary parts) by evaluating
a phase space integral. The imaginary part of the amplitude is the
sum over all the discontinuities:
\begin{equation}
    2 \ \textrm{Im} \left( A_{4}^{scalar}(++++)\right) \ = \ \sum_{i \in
    \textrm{channel}} \textrm{Disc}_{s_{i}} A .
\end{equation}
\noindent The discontinuity in a channel can be computed by
replacing the propagators separating the set of legs by delta
functions:
\begin{equation}
    \frac{i}{D_{i}^{2}} \ \rightarrow  \ 2\pi \ \delta^{(+)} \left( D_{i}^{2}
    \right).
\end{equation}
\noindent The real part is then reconstructed via a dispersion
relation. The existence of a linear combination of scalar
integrals allows us to avoid this reconstruction explicitly. We
perform the cut calculation a bit differently. Consider the loop
amplitude, with a cut in the channel: $s_{12}$ $\left(
\textrm{Fig.} \ref{Figure1}\right)$. We compute the discontinuity
$\textrm{Disc}_{s_{12}}\left(A \right)$. It is convenient to
replace the phase-space integral with an unrestricted loop
momentum integral which has the correct branch cuts $\left(
\ref{feynmancoupure} \right)$. In this integral $\left(
\ref{feynmancoupure} \right)$, the tree-amplitudes are kept
on-shell. Then we decompose the discontinuity
$\textrm{Disc}_{s_{12}} \left(A \right)$ as a linear combination
of scalar cut-integrals in this channel $\left(
\ref{decompositioncoupure} \right)$. The reconstruction of the
amplitude is hidden in the rebuilding of the scalar integrals.

\begin{align}
    \textrm{Disc}_{s_{12}}\left( A \right) & \ = \ \left( 2\pi \right) ^{2} \int
    \frac{d^{n}Q}{\left( 2\pi \right)^{n}} \ \delta^{(+)} ( D_{2}^{2} ) \ A_{tree}(2) \ \delta ^{(+)}( D_{4}^{2} )
    \ A_{tree}(1) \label{coupure} &\\
    & \ = \ \int \frac{d^{n}Q}{\left( 2\pi \right)
    ^{n}} \frac{i}{D_{2}^{2}} A_{tree}(2) \left. \frac{i}{D_{4}^{2}}
    A_{tree}(1)  \right|_{s_{12}} \label{feynmancoupure} &\\
    & \ = \ \sum_{i} c_{i} \left. I_{i}^{n}
    \right|_{s_{12}} = \sum_{i} c_{i} \ \textrm{Disc}_{s_{12}}\left( I_{i}^{n}
    \right).
    \label{decompositioncoupure} &
\end{align}
From this, we can write:
\begin{equation}
    A \ = \ \sum_{i} c_{i} \ I_{i}^{n} + \Delta .
    \label{reconstruction}
\end{equation}
\noindent In the term $\Delta$, there is a combination of scalar
integrals which cannot have a cut in the channel $s_{12}$. To
obtain the coefficient in front of all scalar integrals, we have
to consider cutting amplitudes in all channels
$\left(s_{12},s_{13},s_{14} \right)$. In the following, we note:
\begin{equation}
    2\pi \delta ^{\left( + \right)} \ \equiv \ \delta .
\end{equation}
\noindent We introduce a notation to label the number of cuts and
the channels.``$\textrm{Disc}_{2,s}$'' means cutting the two
internal lines in channel ``$s$''. Cutting a third internal in all
possible ways leads to ``$\textrm{Disc}_{3,s}$'', while cutting
the two remaining lines leads to ``$\textrm{Disc}_{4}$'', in which
there is no need to specify the channel. We denote
$\textrm{Disc}_{N} = \sum_{i \in \textrm{channel}}
\textrm{Disc}_{N,s_{i}}, N=2,3$.

\hfil

We are going to calculate all the helicity amplitudes with two,
three and four-cut techniques in $n=4-2 \epsilon$ dimensions for
massive theories, and then we compare all those techniques. Now,
as we have given the definition of all objects used in this paper,
we are going to calculate the first helicity amplitude $
A_{4}^{scalar}(1^{+},2^{+},3^{+},4^{+})$ in the next section.

\hfil

\section{$ A_{4}^{scalar}$ helicity amplitudes}

\hfil

\subsection{$ A_{4}^{scalar}(1^{+},2^{+},3^{+},4^{+})$ helicity
amplitude} \label{amplitudescalar++++}

\subsubsection{Four-cut technique}

\hfil

The four-cut technique
\cite{Britto:2004nc,Britto:2005ha,Britto:2006sj} says that we cut
all four propagators $D_{i}^{2}, \ i=[1..4]$. We have:
\begin{align*}
\textrm{Disc}_{4} \left(A_{4}^{scalar}(++++) \right) =
\frac{1}{4}\sum_{\sigma(1,2,3,4)}
\parbox{2cm}{\includegraphics[width=2cm]{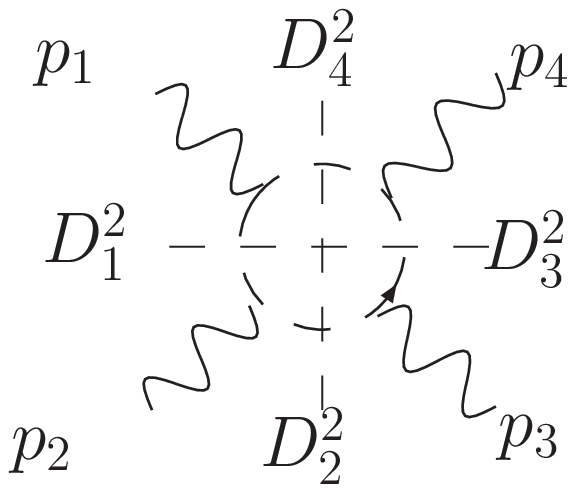}}.
\end{align*}
\noindent We define the loop momenta of propagators as $Q_{i} =
Q_{i-1} +p_{i}$ and $Q_{0} = Q_{4}$. Using Feynman rules, we
compute the discontinuity $\textrm{Disc}
\left(A_{4}^{scalar}(++++) \right)$:
\begin{align}
    \textrm{Disc}_{4} \left(A_{4}^{scalar}(++++) \right) = \dsp \frac{(-2ie)^{4}}{\sqrt{2}^{4}} & \sum_{\sigma(2,3,4)}  \int d^{n}Q \frac{\langle R Q_{1} 1 \rangle}{\langle R1
    \rangle } \frac{\langle R Q_{2} 2 \rangle}{\langle R2 \rangle} & \nonumber \\
    & \frac{\langle R Q_{3} 3 \rangle}{\langle R3 \rangle } \frac{\langle R Q_{4} 4 \rangle}{\langle R4
    \rangle} \delta \left( D_{1}^{2} \right) \delta \left( D_{2}^{2}\right) \delta \left( D_{3}^{2}\right) \delta \left( D_{4}^{2}\right). & \label{M++++scamasse4cuts}
\end{align}
\noindent As the four-dimensional and the $-2\epsilon$-dimensional
spaces are orthogonal, therefore, the spinor product can be
simplified: $\langle R Q_{2} 2 \rangle = \langle R q_{2} 2 \rangle
+ \langle R \mu 2 \rangle = \langle R q_{2} 2 \rangle$. In the
following, we do this simplification each time it is possible and
then we use the fact that all propagators are on-shell to simplify
eq. $\left( \ref{M++++scamasse4cuts} \right)$. We use the first
on-shell tree computed in Appendix $\ref{onshelltrees}$. This tree
has only two photons, so we split the integrand into two groups of
photons $\left( p_{1},p_{2} \right)$ and $\left( p_{3},p_{4}
\right)$, and we apply the formula $\left(\ref{arbre1++} \right)$
for the two groups of photons. The discontinuity $\left(
\ref{M++++scamasse4cuts} \right)$ directly becomes:
\begin{align}
    \textrm{Disc}_{4} \left(A_{4}^{scalar}(++++) \right)& = \dsp (e\sqrt{2})^{4} \sum_{\sigma(2,3,4)}
    \frac{[12][34]}{\langle 12 \rangle \langle 34
    \rangle} \int d^{n}Q \left( \mu^{2} + m^{2} \right)^{2} \delta \left( D_{1}^{2} \right) \delta \left( D_{2}^{2}\right) \delta \left( D_{3}^{2}\right) \delta \left(
    D_{4}^{2}\right)&  \\
    & = \dsp (e\sqrt{2})^{4} \sum_{\sigma(2,3,4)}
    \frac{[12][34]}{\langle 12 \rangle \langle 34
    \rangle} \textrm{Disc}_{4} \left( \int d^{n}Q \frac{\left( \mu^{2} + m^{2} \right)^{2}}{ D_{1}^{2} D_{2}^{2}D_{3}^{2}
    D_{4}^{2}} \right)& \\
    & = \dsp (e\sqrt{2})^{4} \sum_{\sigma(2,3,4)}
    \frac{[12][34]}{\langle 12 \rangle \langle 34
    \rangle} \textrm{Disc}_{4} \left(K_{4}^{n}(1234) \right).& \label{resultat++++4cuts}
\end{align}

\hfil

\subsubsection{Three-cut technique}

\hfil

The three-cut technique imposes three propagators on-shell. So in
the case of the four-photon amplitudes, we have four branch cuts
per four-point diagram, only one branch cut per three-point
diagram and zero branch cut per two-point diagram. But actually,
as the four photons are on-shell, we have only two invariants per
four-point diagrams, and therefore only two independent branch
cuts. So we divide the result by two. Moreover we collect diagrams
to construct set of gauge invariant trees, so we divide again per
two. The discontinuity $ \textrm{Disc}_{3}
\left(A_{4}^{scalar}(++++) \right)$ is:
\begin{align*}
\textrm{Disc}_{3} \left(A_{4}^{scalar}(++++) \right)= & \
\frac{1}{2}\frac{1}{2}\frac{1}{4} \sum_{\sigma(1,2,3,4)} \left(
\parbox{2cm}{\includegraphics[width=2cm]{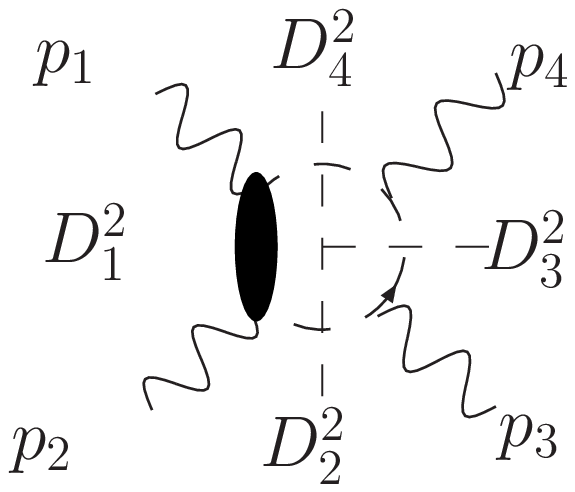}} +
\parbox{2cm}{\includegraphics[width=2cm]{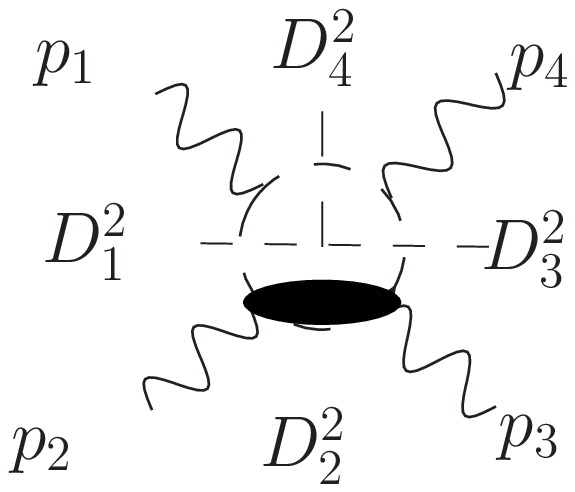}}\right)
& \\
& + \frac{1}{2}\frac{1}{2}\frac{1}{4} \sum_{\sigma(1,2,3,4)}
\left(
\parbox{2cm}{\includegraphics[width=2cm]{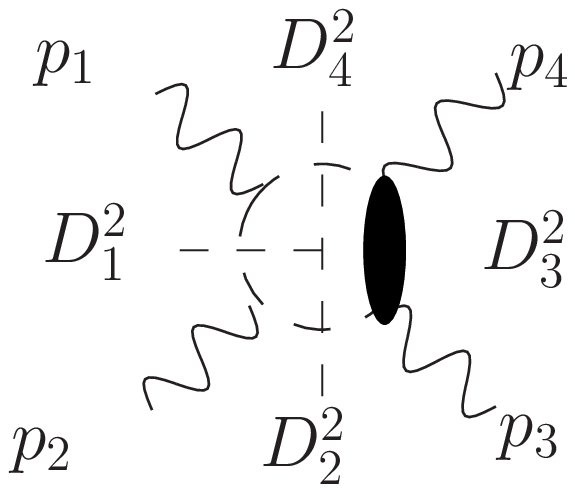}} +
\parbox{2cm}{\includegraphics[width=2cm]{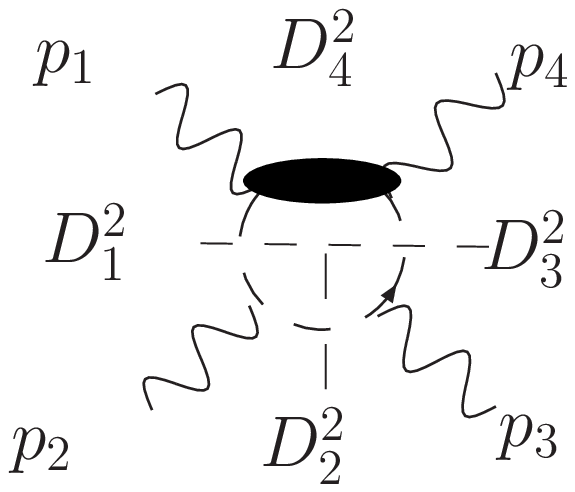}}\right).
\end{align*}
\noindent In the first group, for example, $D_{1}^{2}$ is the
propagator between the two photons $p_{1}$ and $p_{2}$ but we
should not forget the diagram with the four-point vertex. There
are three diagrams in each group. Thanks to permutations, all
groups of cut-diagrams are the same, so the discontinuity becomes:
\begin{align}
\textrm{Disc}_{3} \left(A_{4}^{scalar}(++++) \right) = \frac{1}{4}
\sum_{\sigma(1,2,3,4)}
\parbox{2cm}{\includegraphics[width=2cm]{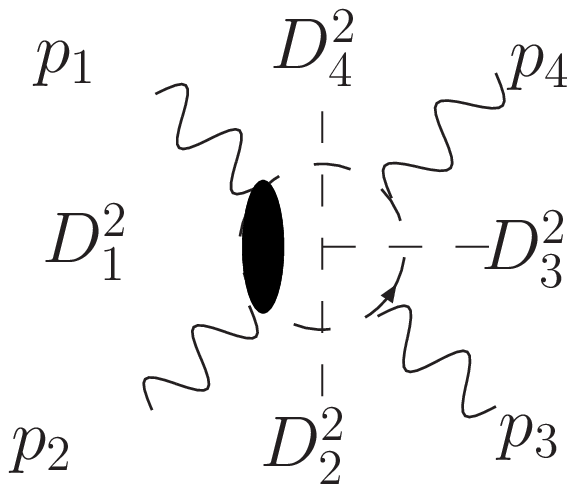}}.
\end{align}
\noindent So using the Feynman rules, the discontinuity
$\textrm{Disc}_{3} \left(A_{4}^{scalar}(++++) \right)$ is:
\begin{align}
    \textrm{Disc}_{3} \left( A_{4}^{scalar}(++++) \right) = \dsp \frac{(-i\sqrt{2}e)^{4}}{4}\sum_{\sigma(1,2,3,4)} &  \int d^{n}Q \left( \sum_{\sigma(1,2)} \frac{\langle R q_{1} 1 \rangle}{\langle R1
    \rangle }\frac{i}{D_{1}^{2}} \frac{\langle R q_{2} 2 \rangle}{\langle R2 \rangle} \right) & \nonumber \\
    & \left( \frac{ \langle R q_{3} 3 \rangle}{\langle R3 \rangle} \frac{\langle R q_{4} 4 \rangle}{\langle R4
    \rangle} \right)\delta \left( D_{2}^{2}\right) \delta \left( D_{3}^{2}\right) \delta \left( D_{4}^{2}\right). & \label{M++++scamasse3cuts}
\end{align}
\noindent We use the expression of the on-shell tree
$\left(\ref{arbre1++} \right) $ for the second group
$(p_{3},p_{4})$. For the first group of photons $(p_{1},p_{2})$,
the propagators around this group are on shell, but the propagator
joining the two photons are not on-shell, so we use $\left(
\ref{arbre2++} \right)$:
\begin{equation}
    \dsp \sum_{\sigma(1,2)} \frac{\langle R q_{1} 1 \rangle}{\langle R1
    \rangle }\frac{i}{D_{1}^{2}} \frac{\langle R q_{2} 2 \rangle}{\langle R2 \rangle} \ = \
    -\left( \mu^{2} + m^{2} \right) \frac{[12]}{ \langle 12 \rangle} \sum_{\sigma(1,2)} \frac{i}{D_{1}^{2}}. \label{arbre2++++scamasse3cuts}
\end{equation}
\noindent Inserting $\left(\ref{arbre1++} \right)$ and $\left(
\ref{arbre2++++scamasse3cuts}\right)$ in $\left(
\ref{M++++scamasse3cuts}\right)$, the discontinuity
$\textrm{Disc}_{3} \left( A_{4}^{scalar}(++++) \right)$ becomes:
\begin{align}
    \textrm{Disc}_{3} \left( A_{4}^{scalar}(++++) \right) & = \frac{(e\sqrt{2})^{4}}{4} \dsp\sum_{\sigma(1,2,3,4)} \frac{[12][34]}{\langle 12 \rangle \langle 34 \rangle}\int
    d^{n}Q \left( \mu^{2} + m^{2} \right)^{2}\sum_{\sigma(1,2)} \frac{i}{D_{1}^{2}} \delta \left( D_{2}^{2}\right) \delta \left( D_{3}^{2}\right) \delta \left( D_{4}^{2}\right) & \\
    &= \dsp (e\sqrt{2})^{4} \sum_{\sigma(2,3,4)}
    \frac{[12][34]}{\langle 12 \rangle \langle 34 \rangle} \left( \textrm{Disc}_{3,s_{12}} \left( K_{4}^{n}\left( 1234 \right) \right) + \textrm{Disc}_{3,s_{14}} \left( K_{4}^{n}(1234) \right) \right)
    &\\
    &= \dsp (e\sqrt{2})^{4} \sum_{\sigma(2,3,4)}
    \frac{[12][34]}{\langle 12 \rangle \langle 34 \rangle}
    \textrm{Disc}_{3}
    \left(K_{4}^{n}\left(1234\right)\right).
    &\label{resultat++++3cuts}
\end{align}
\noindent In the last step we have gathered the two branch cuts of
the scalar integrals $K_{4}^{n}$ to rebuild the entire
discontinuity. Indeed the computation could be more simple.
Consider a discontinuity with several branch cuts. The self
consistency of the unitarity says that we find the same
coefficients in front of all branch cuts of each scalar integral
of this discontinuity. Actually, we need to calculate only the
coefficient in front of one branch cut of each scalar integral of
the discontinuity.

\hfil

\subsubsection{Two-cut technique}

\hfil

This time, only two propagators are on-shell. As we have on-shell
photons, we have only two channels per diagram. Therefore we have
two branch cuts and the imaginary part of the amplitude is
the sum of the discontinuity of the two branch cuts. So, for each
diagram, we decide to cut the propagators $D_{2}^{2},D_{4}^{2}$
and then we cut the propagators $D_{1}^{2},D_{3}^{2}$. Finally
regrouping diagrams with the same cuts to make trees, we obtain:
\begin{align}
2 \ \textrm{Im} \left(A_{4}^{scalar}(++++)\right)=
\textrm{Disc}_{2} \left(A_{4}^{scalar}(++++)\right) =
\frac{1}{4}\sum_{\sigma(2,3,4)} \left(
\parbox{2cm}{\includegraphics[width=2cm]{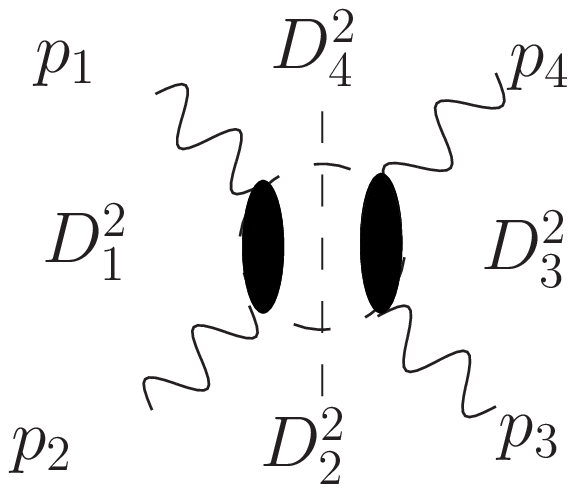}}
+\parbox{2cm}{\includegraphics[width=2cm]{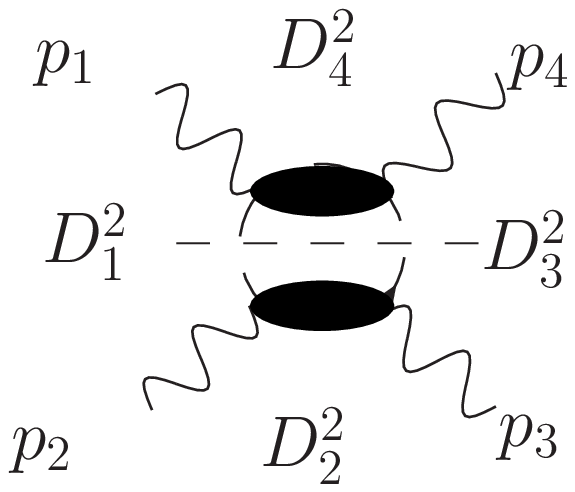}}\right).
\end{align}
\noindent The numerical factor $\frac{1}{4}$ comes from the fact
that we have gathered diagrams to create gauge invariant trees. We
point out, thanks to permutations, the two groups of cut-integrals
are the same. Therefore the discontinuity $\textrm{Disc}_{2}
\left(A_{4}^{scalar}(++++)\right)$ is written:
\begin{align}
    \textrm{Disc}_{2}
    \left(A_{4}^{scalar}(++++)\right) = \dsp
    \frac{1}{2}\sum_{\sigma(2,3,4)} (-i\sqrt{2}e)^{4} \int d^{n}Q &
    \left( \sum_{\sigma(1,2)} \frac{\langle R q_{1} 1 \rangle}{\langle
    R1\rangle }\frac{i}{D_{1}^{2}} \frac{\langle R q_{2} 2 \rangle}{\langle R2 \rangle} \right) &\nonumber \\
    & \dsp *\left( \sum_{\sigma(3,4)} \frac{ \langle R q_{3} 3 \rangle}{\langle R3 \rangle}\frac{i}{ D_{3}^{2}} \frac{\langle R q_{4} 4 \rangle}{\langle R4
    \rangle} \right)\delta \left( D_{2}^{2}\right) \delta \left( D_{4}^{2}\right). & \label{M++++scamasse2cuts}
\end{align}
\noindent Using $\left( \ref{arbre2++} \right)$, to simplify the
two trees in $\left( \ref{M++++scamasse2cuts} \right)$, the
discontinuity $\textrm{Disc}_{2}
\left(A_{4}^{scalar}(++++)\right)$ becomes:
\begin{align}
    \textrm{Disc}_{2}
    \left( A_{4}^{scalar}(++++)\right) & = \frac{(e\sqrt{2})^{4}}{2}
    \dsp\sum_{\sigma(2,3,4)}  \frac{[12][34]}{\langle
    12 \rangle \langle 34 \rangle}\int
    d^{n}Q \left( \mu^{2} + m^{2} \right)^{2}\sum_{\sigma(1,2)}\frac{i}{D_{1}^{2}}
    \sum_{\sigma(3,4)} \frac{i}{D_{3}^{2}} \delta \left( D_{2}^{2}\right)\delta \left( D_{4}^{2}\right) & \\
    & = \dsp  2 (e\sqrt{2})^{4} \sum_{\sigma(2,3,4)}  \frac{[12][34]}{\langle 12 \rangle \langle 34
    \rangle}  \textrm{Disc}_{2,s_{12}} \left( K_{4}^{n}\left(1234\right)\right). &
\end{align}
\noindent We have one branch cut of the scalar integrals. But we
want to reconstruct the entire discontinuity of the scalar
integral $ K_{4}^{n}$. Using the conservation of energy-momentum
of external moments, $ \frac{[12][34]}{\langle 12 \rangle \langle
34 \rangle} $ is invariant by permutation. So we can split the
amplitude in two equal terms. We transform, thanks to
permutations, one of this term to obtain the second branch cut,
and the discontinuity of the scalar integral appears:
\begin{align}
    \textrm{Disc}_{2} \left(A_{4}^{scalar}(++++)\right) & = (e\sqrt{2})^{4}
    \sum_{\sigma(2,3,4)} \frac{[12][34]}{\langle 12 \rangle \langle 34
    \rangle} \left( \textrm{Disc}_{2,s_{12}} \left( K_{4}^{n}\left(1234\right)
    \right) + \textrm{Disc}_{2,s_{14}} \left( K_{4}^{n}\left(1234\right)
    \right) \right) & \\
    & = (e\sqrt{2})^{4} \sum_{\sigma(2,3,4)} \frac{[12][34]}{\langle 12 \rangle \langle 34
    \rangle} \textrm{Disc}_{2} \left( K_{4}^{n}\left(1234\right)
    \right). & \label{resultat++++2cuts}
\end{align}

\hfil

With this last discontinuity we reconstruct directly the helicity
amplitude $A_{4}^{scalar}(++++)$ by transform the discontinuity of
each scalar integral to the scalar integral. We multiply the
result by the factor $K = i (4\pi)^{-n/2}$:
\begin{equation}
    \dsp A^{scalar}_{4}(++++) =  i\frac{(e\sqrt{2})^{4}}{(4\pi)^{n/2}} \sum_{\sigma(2,3,4)} \frac{[12][34]}{\langle 12 \rangle \langle 34
    \rangle} K_{4}^{n}\left(1234\right) = \dsp 4 i \alpha^{2} \sum_{\sigma(2,3,4)} \frac{[12][34]}{\langle 12 \rangle \langle 34
    \rangle} K_{4}^{n}\left(1234\right),
\end{equation}
\noindent where $\alpha = e^{2}/4\pi$. The reconstruction of
scalar integrals is automatic, we don't need dispersive relation.
There is full agreement with \cite{Bern:massiveloop}.

\hfil

\subsection{$ A_{4}^{scalar}(1^{-},2^{+},3^{+},4^{+})$ helicity
amplitude}\label{amplitudescalar-+++}

\subsubsection{Four-cut technique}

\hfil

The four-cut technique sets the four propagators $D_{i}^{2}, \
i=[1..4]$ on-shell:
\begin{align*}
    \textrm{Disc}_{4} \left(A_{4}^{scalar}(-+++)\right) =
    \frac{1}{4}\sum_{\sigma(1,2,3,4)}\parbox{2cm}{\includegraphics[width=2cm]{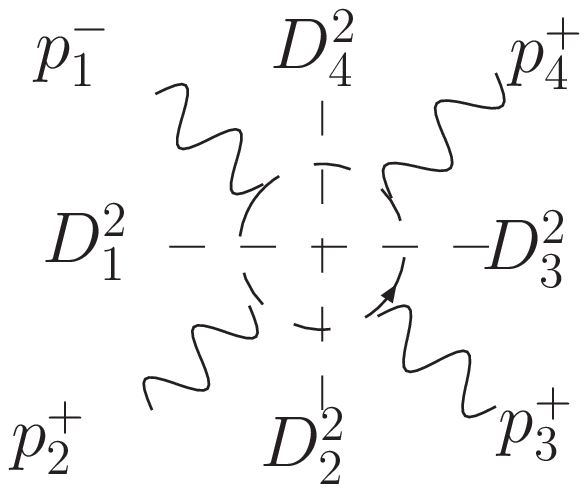}}.
\end{align*}
\noindent So the discontinuity $\textrm{Disc}_{4}
\left(A_{4}^{scalar} (-+++)\right)$ is:
\begin{equation}
    \dsp \textrm{Disc}_{4} \left( A_{4}^{scalar}(-+++) \right) = (-i\sqrt{2}e)^{4} \sum_{\sigma(2,3,4)}  \int d^{n}Q  \frac{[r q_{1} 1 ]}{[1r] }
    \frac{\langle R q_{2} 2 \rangle}{\langle R2 \rangle} \frac{\langle R q_{3} 3 \rangle}{\langle R3 \rangle } \frac{\langle R q_{4} 4 \rangle}{\langle R4
    \rangle} \delta \left( D_{1}^{2} \right) \delta \left( D_{2}^{2}\right) \delta \left( D_{3}^{2}\right) \delta \left( D_{4}^{2}\right). \label{M-+++scamasse4cuts}
\end{equation}
\noindent We split the integrand in two groups of photons: $\left(
p_{1},p_{2} \right) $ and $\left( p_{3},p_{4} \right) $. As the
helicity of the two photons of the first group are different,
therefore we use $\left( \ref{arbre-+} \right)$ in the limit of
all propagators on-shell:
\begin{equation}
\frac{[r q_{1} 1 ]}{[1r] }
    \dsp \frac{\langle R q_{2} 2 \rangle}{\langle R2 \rangle} \ = \ \dsp
    \lim_{D_{1}^{2},D_{3}^{2} \rightarrow 0} \left(
    \ref{arbre-+} \right) \ = \ -\frac{1 }{\langle 2 3 1\rangle}
    \left(  \langle 1 q_{2} 2 3 2\rangle  +\left( \mu^{2} + m^{2} \right) [ 231 ] \right), \label{arbre1-+++scamasse4cuts}
\end{equation}
\noindent and for the second group of photons, we use the relation
$\left( \ref{arbre1++} \right)$. The discontinuity $\left(
\ref{M-+++scamasse4cuts}\right)$ becomes:
\begin{align}
    \textrm{Disc}_{4} & \left( A_{4}^{scalar}(-+++)\right) \  = \dsp \ (e\sqrt{2})^{4} \sum_{\sigma(2,3,4)}  \frac{[34]
    [231]}{ \langle 34 \rangle \langle 231 \rangle} \int d^{n}Q
    \left( \mu^{2}+ m^{2} \right)^{2}
    \delta \left( D_{1}^{2} \right) \delta \left( D_{2}^{2}\right) \delta \left( D_{3}^{2}\right) \delta \left( D_{4}^{2}\right) & \nonumber \\
    & \dsp + \ (e\sqrt{2})^{4}\sum_{\sigma(2,3,4)}  \frac{[34]}{\langle
    34\rangle \langle 231 \rangle} \int d^{n}Q
    \langle 1q_{2}232 \rangle\left( \mu^{2} + m^{2}\right) \delta \left( D_{1}^{2} \right)
    \delta \left( D_{2}^{2}\right) \delta \left( D_{3}^{2}\right) \delta \left(
    D_{4}^{2}\right).
    & \label{M-+++1scamasse4cuts}
\end{align}
\noindent Here, to complete the computation, we have two ways. The
first way is to use the integration formula of the tensor integral
$\left( \ref{annexeintegrationI4} \right)$. The second way is to
use the fact that we have four on-shell propagators, which gives
us four conditions. Those conditions are sufficient to define
exactly the loop momentum. We explain this calculation in Appendix
\ref{calcul-+++}. We find:
\begin{equation}
    \textrm{Disc}_{4}\left( A_{4}^{scalar}(-+++)\right) \ = \ \dsp (e\sqrt{2})^{4} \sum_{\sigma(2,3,4)}  \frac{[34]
    [231]}{ \langle 34 \rangle \langle 231 \rangle} \textrm{Disc}_{4}\left(
     K_{4}^{n} + \frac{ts}{2u} J_{4}^{n} \right). \label{resultat-+++4cuts}
\end{equation}

\hfil

\subsubsection{Three-cut technique}

\hfil

We have a photon with a negative helicity. So not to break the
helicity symmetry, we gather cut-diagrams in three groups rather
than two or four, and we multiply by a factor $1/2$. The
conservation of the symmetry allow us to rebuild easily the
discontinuities with the branch cuts. In terms of cut-diagrams,
the discontinuity $\textrm{Disc}_{3}\left(
A_{4}^{scalar}(-+++)\right)$ is written:
\begin{align*}
\textrm{Disc}_{3}\left( A_{4}^{scalar}(-+++)\right) =
\frac{1}{2}\sum_{\sigma(2,3,4)} \left(
\parbox{2cm}{\includegraphics[width=2cm]{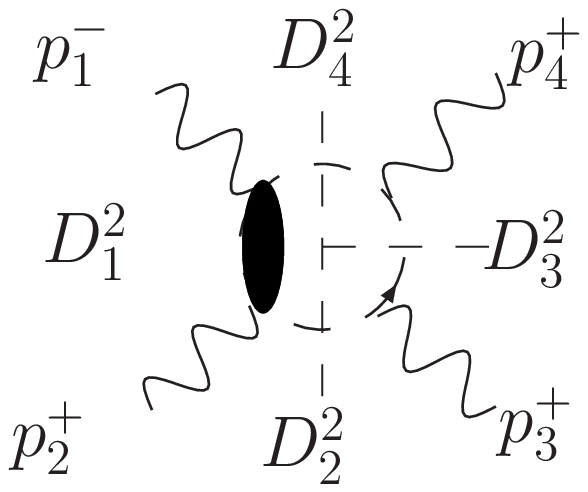}} + \frac{1}{2}\parbox{2cm}{\includegraphics[width=2cm]{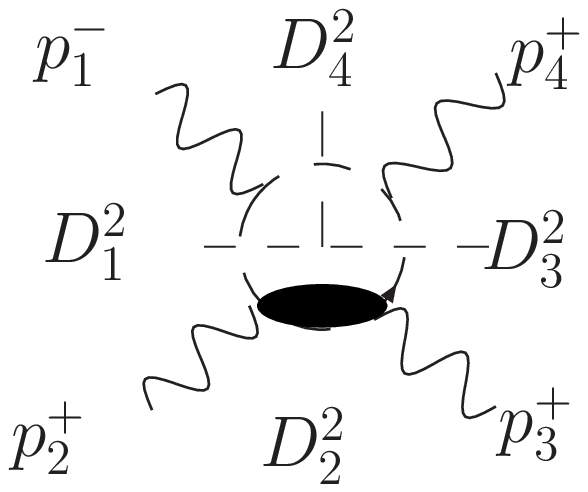}}
 +
 \frac{1}{2}\parbox{2cm}{\includegraphics[width=2cm]{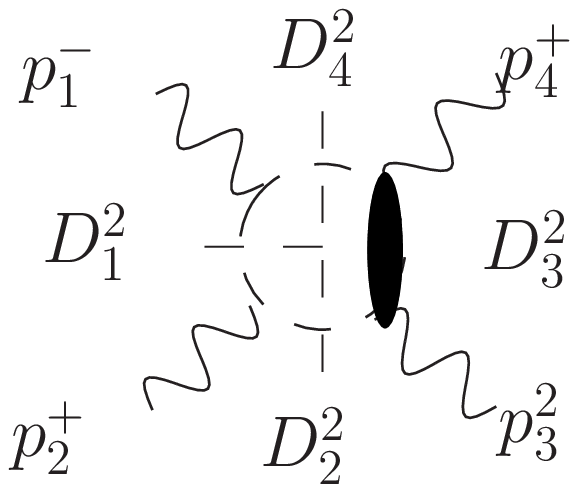}}\right).
\end{align*}
\noindent So, using the Feynman rules, the discontinuity
$\textrm{Disc}_{3}\left( A_{4}^{scalar}(-+++)\right)$ is:
\begin{align}
    \textrm{Disc}_{3}& \left(A_{4}^{scalar}(-+++)\right) = \dsp \ \ \frac{1}{2}\sum_{\sigma(2,3,4)} (-i\sqrt{2}e)^{4} & \nonumber \\
    &\int d^{n}Q \left(- \frac{[r2]\langle R1 \rangle}{[1r] \langle R2 \rangle} + \sum_{\sigma(1,2)} \frac{[r q_{1} 1 ]}{[1r] }
    \frac{i}{D_{1}^{2}} \frac{\langle R q_{2} 2 \rangle}{\langle R2 \rangle} \right) \left( \frac{ \langle R q_{3} 3 \rangle}{\langle R3 \rangle} \frac{\langle R q_{4} 4 \rangle}{\langle R4
    \rangle} \right)\delta \left( D_{2}^{2}\right) \delta\left( D_{3}^{2}\right) \delta \left( D_{4}^{2}\right) & \nonumber \\
    & + \frac{1}{2}\int d^{n}Q \left( \sum_{\sigma(2,3)} \frac{\langle R q_{2} 2 \rangle}{\langle
    R2 \rangle}\frac{i}{D_{2}^{2}}\frac{ \langle R q_{3} 3 \rangle}{\langle R3 \rangle} \right)
    \left( \frac{\langle R q_{4} 4 \rangle}{\langle R4\rangle} \frac{[r q_{1} 1 ]}{[1r] } \right) \delta \left(
    D_{1}^{2}\right) \delta \left( D_{3}^{2}\right) \delta \left(D_{4}^{2}\right) & \nonumber \\
    & + \frac{1}{2} \int d^{n}Q \left( \sum_{\sigma(3,4)} \frac{\langle R q_{3} 3 \rangle}{\langle
    R3 \rangle}\frac{i}{D_{3}^{2}} \frac{ \langle R q_{4} 4 \rangle}{\langle R4 \rangle} \right)
    \left( \frac{[r q_{1} 1 ]}{[1r] } \frac{\langle R q_{1} 2
    \rangle}{\langle R2\rangle} \right) \delta \left( D_{1}^{2}\right) \delta \left(
    D_{2}^{2}\right) \delta \left( D_{4}^{2}\right). &\label{M-+++scamasse3cuts}
\end{align}
\noindent We calculate the first tree $- \frac{[r2]\langle R1
\rangle}{[1r] \langle R2 \rangle} + \sum_{\sigma(1,2)} \frac{[r
q_{1} 1 ]}{[1r] } \frac{i}{D_{1}^{2}} \frac{\langle R q_{2} 2
\rangle}{\langle R2 \rangle}$, imposing $|r\rangle = |2 \rangle$,
$|R\rangle = |1\rangle$, and using the formula $ \left(
\ref{arbre-+} \right)$ in the limit of the propagator $D_{3}^{2}$
on-shell, we obtain:
\begin{align}
    \sum_{\sigma(1,2)} \frac{[r q_{1} 1 ]}{[1r] }
    \frac{i}{D_{1}^{2}} \frac{\langle R q_{2} 2 \rangle}{\langle R2
    \rangle} & = \ \ \lim _{D_{3}^{2} \rightarrow 0} \left( \ref{arbre-+}
    \right) & \nonumber \\
    & \dsp = \ -\frac{i }{ \langle 231 \rangle} \left(  \frac{ \langle 1 q_{2} 232\rangle +\left( \mu^{2} + m^{2}
    \right) [231]}{D_{1}^{2}} + \frac{  \langle 1 q_{2} 31 2\rangle +\left( \mu^{2} + m^{2}
    \right)[231]}{{D_{1}^{'}}^{2}} \right). \label{arbre1-+++scamasse3cuts}
\end{align}
\noindent To evaluate the on-shell trees with the same helicities,
we use $\left( \ref{arbre1++}, \ref{arbre2++} \right)$. For the
two last trees $ \frac{\langle R q_{4} 4 \rangle}{\langle R4
\rangle} \frac{[r q_{1} 1 ]}{[1r]}$ and $ \frac{[r q_{1} 1
]}{[1r]} \frac{\langle R q_{1} 2 \rangle}{\langle R2 \rangle} $,
we can use again the formula $\left( \ref{arbre-+} \right)$, but
with very few spinor manipulations we obtain:
\begin{align}
    \dsp \frac{\langle R q_{4} 4 \rangle}{\langle R4 \rangle} \frac{[r q_{1} 1 ]}{[1r]} & \ =
    \ \left( -s_{12} + D_{2}^{2} \right) \frac{[4q_{4}1]}{\langle 421
    \rangle}  - \left( \mu^{2} +m^{2} \right) \frac{[421]}{\langle 421
    \rangle},& \label{arbre4-+++scamasse3cuts} \\
    \dsp \frac{[r q_{1} 1 ]}{[1r]} \frac{\langle R q_{1} 2 \rangle}{\langle R2 \rangle}  & \ =
    \ \left( -s_{14} + D_{3}^{2} \right) \frac{[2q_{1}1]}{\langle 231
    \rangle}  - \left( \mu^{2} +m^{2} \right) \frac{[231]}{\langle 231 \rangle}.& \label{arbre5-+++scamasse3cuts}
\end{align}
\noindent We gather $\left(
\ref{arbre1-+++scamasse3cuts},\ref{arbre4-+++scamasse3cuts}
,\ref{arbre5-+++scamasse3cuts} \right)$ in the discontinuity
$\left( \ref{M-+++scamasse3cuts} \right)$. We obtain some linear
tensor integrals, which we integrate with $\left(
\ref{annexeintegrationI4},\ref{annexeintegrationJ4} \right)$. Some
three-point linear-tensor integrals appear, but many of them are
zero. Consider the three-point function with one external mass
$s_{23}$: $I_{3}(s_{23})(q_{i}^{\mu})$. This triangle can be
expressed as a linear combination of its two on-shell legs
(Appendix \ref{Masterintegrals}): $I_{3}(s_{23})(q_{i}^{\mu}) = A
\ p_{1}^{\mu} +B \ p_{4}^{\mu}$. As the momenta $p_{1}$ and
$p_{4}$ are light-like vectors, therefore the linear tensor
integral $ I_{3}(s_{23})( \langle 1 q_{i}^{\mu} 4 \rangle )$ is
zero. After integrating, we have a linear combination of
discontinuities of four-point, three-point and two-point scalar
cut-integrals. But some of those discontinuities are spurious.
Consider the invariant $s_{ij}$. $s_{ij}$ is a channel of the one
external mass triangle $I_{3}(s_{kl})$ if the external mass
$s_{kl}$ is equal to the channel $ s_{kl} = s_{ij}$. And there is
the same argument for the bubbles. Integrals which don't respect
this condition are spurious and we drop them. They appeared
because we lifted a cut-integral to a Feynman integral. For the
rational terms there is no problem. The reduction of the Feynman's
integrals gives some extra-scalar integrals. The development of
those scalar integrals in function of $\epsilon$ create the
rational terms. So we keep only the extra-scalar integrals
verifying cuts. Keeping only those integrals with cuts, we rebuild
the discontinuity. The discontinuity of four point scalar
integrals need two branch cuts to be rebuilt, whereas, only one
branch cut is sufficient to rebuild a three or two point scalar
integrals. So, we obtain:
\begin{align}
    \textrm{Disc}_{3}\left( A_{4}^{scalar}(-+++)\right) = & \dsp \  \frac{(e\sqrt{2})^{4}}{2} \sum_{\sigma(2,3,4)}   \frac{[34]
    [231]}{ \langle 34 \rangle \langle 231 \rangle} \textrm{Disc}_{3}\left(
    K_{4}^{n}(1234)
    + \frac{ts}{2u} J_{4}^{n}(1234)+ \frac{t}{u} J_{3}^{n}(s) -\frac{t}{u}  J_{3}^{n}(t) \right)
    & \nonumber \\
    & \dsp \  + \frac{(e\sqrt{2})^{4}}{4} \sum_{\sigma(2,3,4)} \frac{[23] [421]}{ \langle 23
    \rangle \langle 421 \rangle} \textrm{Disc}_{3,s_{23}} \left(  K_{4}^{n}(1234)
     +\frac{ts}{u}J_{4}^{n}(1234) -\frac{2s}{u} J_{3}^{n}(t) \right) & \nonumber \\
    & \dsp \  + \frac{(e\sqrt{2})^{4}}{4} \sum_{\sigma(2,3,4)}  \frac{[34] [231]}{ \langle
    34 \rangle \langle 231 \rangle} \textrm{Disc}_{3,s_{12}} \left(
    K_{4}^{n}(1234) +\frac{ts}{u}J_{4}^{n}(1234) + \frac{2t}{u} J_{3}^{n}(s)\right). & \label{M-+++2scamasse3cuts}
\end{align}
\noindent Pointing out that $ \frac{[34] [231]}{ \langle 34
\rangle \langle 231 \rangle} = \frac{[23] [421]}{ \langle 23
\rangle \langle 421 \rangle} $, we can gather discontinuities. We
symmetrize the coefficient in front of the three-point
extra-dimension integral $J_{3}^{n}$. The discontinuity
$\textrm{Disc}_{3}\left( A_{4}^{scalar}(-+++)\right)$ becomes:
\begin{align}
    \textrm{Disc}_{3}\left( A_{4}^{scalar}(-+++)\right) = \dsp  \ (e\sqrt{2})^{4} \sum_{\sigma(2,3,4)}  \frac{[34]
    [231]}{ \langle 34 \rangle \langle 231 \rangle} & \textrm{Disc}_{3}\left(  K_{4}^{n}(1234) + \frac{st}{2u}
    J_{4}^{n}(1234) \right. &\nonumber \\
    & + \left. \left( \frac{s^{2}+t^{2}+u^{2}}{2tu}\right)
    J_{3}^{n}(s)\right). & \label{resultat-+++3cuts}
\end{align}

\hfil

\subsubsection{Two-cut technique}

\hfil

Each diagram has to be cut in the two channels, corresponding to
the two branch cuts. In terms of cut-diagrams, the discontinuity
$\textrm{Disc}_{2}\left( A_{4}^{scalar}(-+++)\right)$ is:
\begin{align*}
\textrm{Disc}_{2}\left( A_{4}^{scalar}(-+++)\right) =
\frac{1}{4}\sum_{\sigma(2,3,4)} \left(
\parbox{2cm}{\includegraphics[width=2cm]{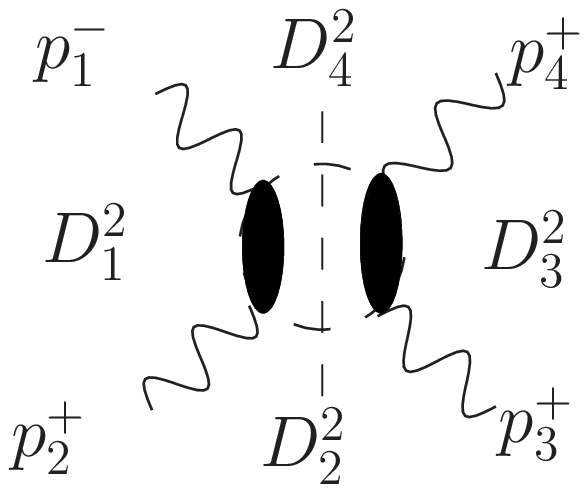}}
+\parbox{2cm}{\includegraphics[width=2cm]{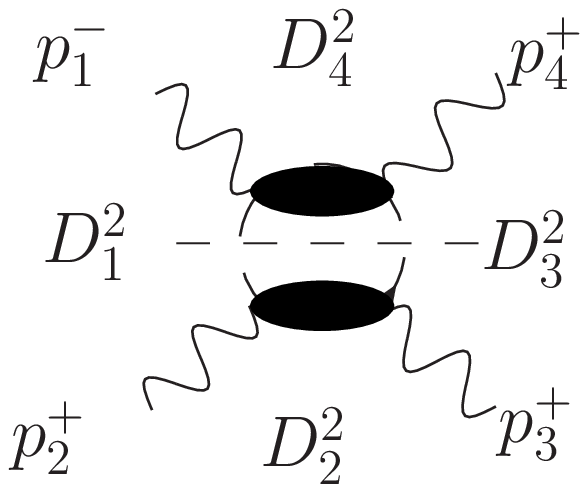}}\right).
\end{align*}
\noindent If we do permutations, we can see easily that all
cut-diagrams are doubled. We gather them and the discontinuity
$\textrm{Disc}_{2}\left( A_{4}^{scalar}(-+++)\right)$ is:
\begin{align}
    \textrm{Disc}_{2}\left( A_{4}^{scalar}(-+++)\right)  = \dsp \frac{1}{2}\sum_{\sigma(2,3,4)} (-i\sqrt{2}e)^{4} \int d^{n}Q & \left(- \frac{[r2]\langle R1
\rangle}{[1r] \langle R2 \rangle} + \sum_{\sigma(1,2)} \frac{[r
q_{1} 1 ]}{[1r]}\frac{i}{ D_{1}^{2}}
    \frac{\langle R q_{2} 2 \rangle}{\langle R2 \rangle} \right) &
    \nonumber \\
    & \ \ \ * \left( \sum_{\sigma(3,4)} \frac{ \langle R q_{3} 3 \rangle}{\langle R3 \rangle}\frac{i}{ D_{3}^{2}} \frac{\langle R q_{4} 4 \rangle}{\langle R4
    \rangle} \right)\delta \left( D_{2}^{2}\right)\delta \left(D_{4}^{2}\right). & \label{M-+++scamasse2cuts}
\end{align}
\noindent We use $\left( \ref{arbre-+} \right)$ and $\left(
\ref{arbre2++} \right)$ to calculate the trees of the
discontinuity. We obtain some tensor triangles, which we integrate
with the formulas $\left( \ref{annexeintegrationI4},
\ref{annexeintegrationJ4} \right)$. We keep only cut-integrals
which are not spurious, and we rebuild the discontinuities.
Finally, the discontinuity $\left( \ref{M-+++scamasse2cuts}
\right)$ is:
\begin{align}
    \textrm{Disc}_{2}\left( A_{4}^{scalar}(-+++)\right) = \dsp \ (e\sqrt{2})^{4} \sum_{\sigma(2,3,4)} \frac{[34]
    [231]}{ \langle 34 \rangle \langle 231 \rangle} & \textrm{Disc}_{2}\left(  K_{4}^{n}(1234) + \frac{st}{2u}
    J_{4}^{n}(1234) \right. &\nonumber \\
    & + \left. \left( \frac{s^{2}+t^{2}+u^{2}}{2tu}\right)
    J_{3}^{n}(s)\right).& \label{resultat-+++2cuts}
\end{align}

\hfil

The two-cut technique gives all information to reconstruct the
helicity amplitude. We find straightforwardly:
\begin{equation}
    A_{4}^{scalar}(-+++) \ = \  \dsp  4 i \alpha^{2}  \ \sum_{\sigma(2,3,4)} \frac{[34]
    [231]}{ \langle 34 \rangle \langle 231 \rangle} \left( K_{4}^{n}(1234) + \frac{st}{2u}  J_{4}^{n}(1234) + \left(
    \frac{s^{2}+t^{2}+u^{2}}{2tu}\right)J_{3}^{n}(s)\right).
\end{equation}

\hfil

\subsection{$ A_{4}^{scalar}(1^{-},2^{-},3^{+},4^{+})$ helicity
amplitude} \label{amplitudescalar--++}

\hfil

This helicity amplitude is usually called the MHV (Maximal
Helicity Violating) amplitude.

\hfil

\subsubsection{Four-cut technique}

\hfil

One of the difficulty of this helicity amplitude, is that we have
two kinds of topologies of helicities. The helicities are either
alternate or they are paired as shown eq. $\left(
\ref{--++sca4cut} \right)$. We group diagrams according the
topology of helicities and the four propagators $D_{i}^{2}, \
i=[1..4]$ are on-shell so the MHV discontinuity is:
\begin{align}
\textrm{Disc}_{4} \left( A_{4}^{scalar}(--++) \right) =
\sum_{\sigma(1,2)}\sum_{\sigma(3,4)}
\parbox{2cm}{\includegraphics[width=2cm]{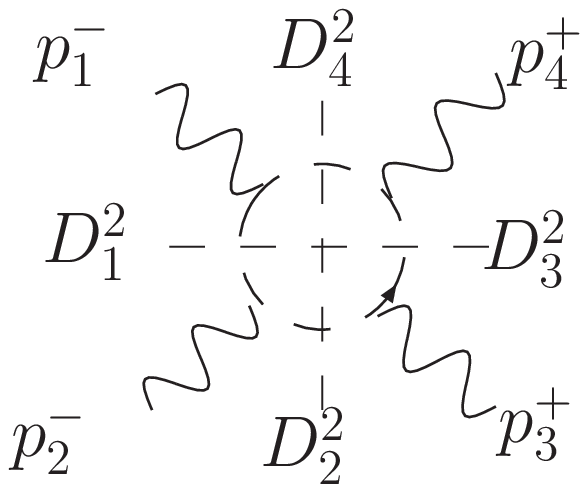}} + \sum_{\sigma(1,2)}
\parbox{2cm}{\includegraphics[width=2cm]{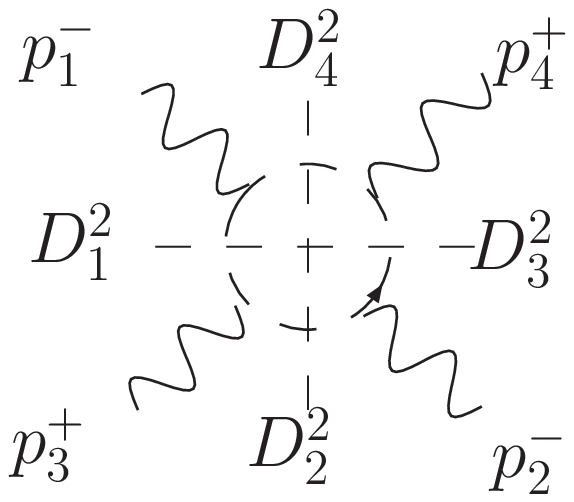}}.
\label{--++sca4cut}
\end{align}
\noindent We obtain:
\begin{align}
    & \quad \quad \quad  \textrm{Disc}_{4} \left( A_{4}^{scalar}(--++) \right) & \nonumber \\
    = \ & \dsp  \ (e\sqrt{2})^{4} \sum_{\sigma(1,2)}\sum_{\sigma(3,4)}\int d^{n}Q  \frac{ [ r q_{1} 1
    ]}{[1r]} \frac{ [r q_{2} 2]}{[2r]}  \frac{\langle R q_{3} 3 \rangle}{\langle R3 \rangle} \frac{\langle
    R q_{4} 4 \rangle}{\langle R4 \rangle}  \delta \left( D_{1}^{2}\right) \delta \left( D_{2}^{2}\right) \delta \left( D_{3}^{2}\right)\delta \left(D_{4}^{2}\right) & \nonumber \\
    & \dsp +   (e\sqrt{2})^{4} \sum_{\sigma(1,2)} \int d^{n}Q  \frac{[ r q_{1} 1
    ]}{[1r]}\frac{\langle R q_{2} 3 \rangle}{\langle R3 \rangle} \frac{ [r q_{3} 2]}{[2r]}  \frac{\langle
    R q_{4} 4 \rangle}{\langle R4 \rangle} \delta \left( D_{1}^{2}\right) \delta \left( D_{2}^{2}\right)
    \delta \left( D_{3}^{2}\right)\delta \left(D_{4}^{2}\right) &\\
    = \ & \dsp \ \ I_{1} + I_{2}. & \label{M--++1}
\end{align}
\noindent We have split the two topologies into two integrals:
$I_{1}$ and $I_{2}$. Applying two times the tree formulas $\left(
\ref{arbre1++},\ref{arbre1--} \right)$, the first topology $I_{1}$
is directly:
\begin{equation}
    I_{1} \ = \ (e\sqrt{2})^{4} \sum_{\sigma(1,2)} \sum_{\sigma(3,4)} \frac{\langle 12 \rangle}{[12]}\frac{[34]}{\langle 34 \rangle}
    \textrm{Disc}_{4} \left( K_{4}^{n}(1234) \right). \label{I1}
\end{equation}
\noindent For the second topology $I_{2}$, instead of using
$\left( \ref {arbre-+} \right)$, we gather photon with the same
helicity and we reduce directly:
\begin{align}
    \frac{[ r q_{1} 1]}{[1r]}\frac{ [r q_{3} 2]}{[2r]} &  = -
    \frac{\langle 1q_{1}q_{3}2 \rangle}{[12]} = - \frac{\langle 12 \rangle \left( \mu^{2} +m^{2} \right) + \langle 14q_{3}2
    \rangle}{[12]}, \label{arbre1--++scamasse4cuts}&\\
    \frac{\langle R q_{2} 3 \rangle}{\langle R3 \rangle}\frac{\langle
    R q_{4} 4 \rangle}{\langle R4 \rangle} & = -
    \frac{[ 3q_{2}q_{4}4 ]}{\langle 34\rangle} = -\frac{ [34] \left( \mu^{2} +m^{2} \right) + [3q_{2}24]
    }{\langle 34 \rangle}. \label{arbre2--++scamasse4cuts}&
\end{align}
\noindent Therefore with $\left(
\ref{arbre1--++scamasse4cuts},\ref{arbre2--++scamasse4cuts}
\right)$, $I_{2}$ becomes a sum of four terms which we develop and
integrate with the formulas $\left(
\ref{annexeintegrationI4},\ref{annexeintegrationJ4} \right)$. Here
as we have four cuts, all triangles and bubbles are spurious,
because they don't have four cuts. So $I_{2}$ is spelt:
\begin{equation}
    I_{2} = 2 ( e\sqrt{2})^{4} \frac{\langle 12 \rangle [34]}{[12] \langle 34 \rangle}
    \left( \textrm{Disc}_{4} \left( K_{4}^{n}(1324) \right) + \frac{2ut}{s}\textrm{Disc}_{4} \left( J_{4}^{n}(1324) \right) +
    \frac{u^{2}t^{2}}{2s^{2}}\textrm{Disc}_{4} \left( I_{4}^{n}(1324) \right)\right).
\end{equation}
\noindent The amplitude contains a four-point scalar integral in
$n$ dimensions $I_{4}^{n}$. This integral $\left( I_{4}^{n}
\right)$ in a massless theory has IR divergences. Each diagram of
the four-photon amplitudes has no IR divergence. So those
divergences should be compensated by other divergent integrals
like three-point scalar integrals. If we have three-point
integrals in massless theory, we have probably the same in a
massive theory. To simplify the problem it is better to transform
the $n$-dimensional four-point scalar integral into a
$(n+2)$-dimensional integral, which is no longer IR divergent.
This transformation $ \left( I_{4}^{n} \rightarrow I_{4}^{n+2}
\right) $ is given by the formula $\left( \ref{passageJaI}
\right)$. Keeping only integrals with four cuts, we have:
\begin{equation}
    I_{2} \ = \ 2 ( e\sqrt{2})^{4} \frac{\langle 12 \rangle [34]}{[12] \langle 34 \rangle}\left( \textrm{Disc}_{4} \left( K_{4}^{n}(1324) \right)
    - \frac{tu}{s}\textrm{Disc}_{4} \left( I_{4}^{n+2}(1324) \right) \right). \label{I2}
\end{equation}
\noindent So as the discontinuity $\textrm{Disc}_{4} \left(
A^{scalar}(--++) \right)$ is the addition of the integral $I_{1}$,
given in $\left( \ref{I1} \right)$ and the integral $I_{2}$, given
in $\left( \ref{I2} \right)$, therefore we obtain:
\begin{equation}
    \textrm{Disc}_{4} \left(A^{scalar}(--++) \right) \ = \ \dsp  (e\sqrt{2})^{4} \frac{\langle
    12 \rangle}{[12]}\frac{[34]}{\langle 34 \rangle} \left( -
    \frac{2tu}{s}  \textrm{Disc}_{4} \left(I_{4}^{n+2}(1324) \right) +\sum_{\sigma(2,3,4)}
    \textrm{Disc}_{4} \left( K_{4}^{n}(1234) \right)   \right). \label{resultat--++4cuts}
\end{equation}

\hfil

\subsubsection{Three-cut technique}

\hfil

The discontinuity, after grouping together diagrams with the same
cuts is:
\begin{align*}
    \textrm{Disc}_{3} \left(A^{scalar}(--++) \right) \ = \ & \frac{1}{4} \sum_{\sigma(1,2)}\sum_{\sigma(3,4)} \left(\parbox{2cm}{\includegraphics[width=2cm]{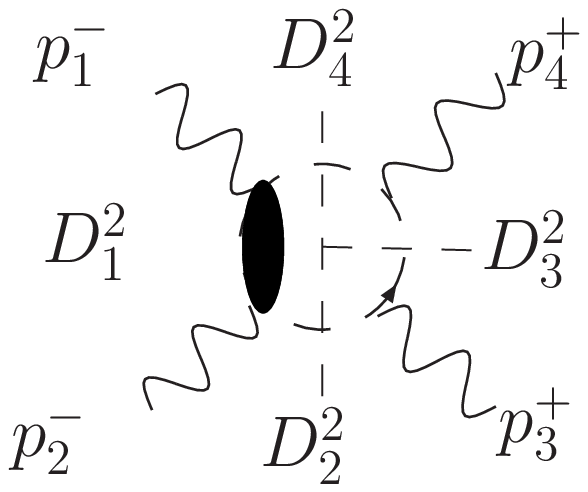}} +
    \parbox{2cm}{\includegraphics[width=2cm]{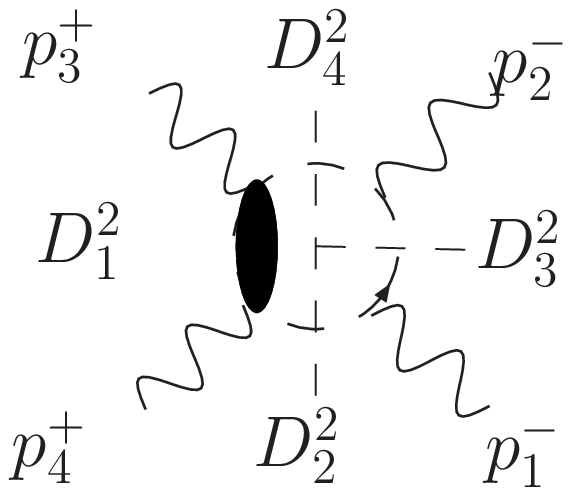}}\right)&\\
    & + \frac{1}{4} \sum_{\sigma(1,2)} \sum_{\sigma(1,3)}\sum_{\sigma(2,4)}
    \left(
    \parbox{2cm}{\includegraphics[width=2cm]{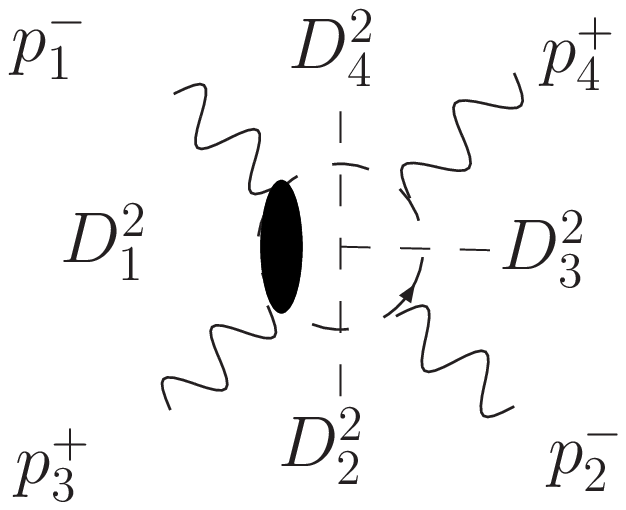}} +
    \parbox{2cm}{\includegraphics[width=2cm]{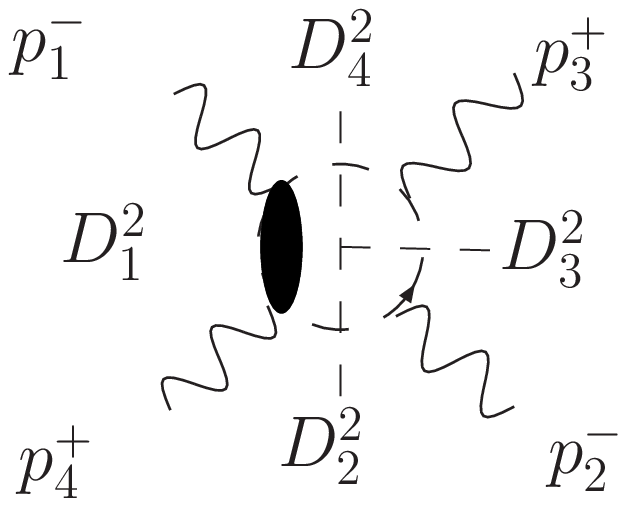}}\right).&
\end{align*}
\noindent Using Feynman rules, the discontinuity is:
\begin{align}
    & \textrm{Disc}_{3} \left(A^{scalar}(--++) \right) = \dsp \frac{(-i\sqrt{2}e)^{4}}{4}\Bigg[  & \nonumber \\
    & \sum_{\sigma(1,2)}\sum_{\sigma(3,4)} \int d^{n}Q \sum_{\sigma(1,2)}\left( \frac{ [ r q_{1} 1
    ]}{[1r]} \frac{i}{D_{1}^{2}}\frac{ [r q_{2} 2]}{[2r]} \right)
    \frac{\langle R q_{3} 3 \rangle}{\langle R3 \rangle} \frac{\langle R q_{4} 4 \rangle}{\langle R4 \rangle}
    \delta \left( D_{2}^{2}\right)\delta \left( D_{3}^{2}\right) \delta \left(D_{4}^{2}\right) & \nonumber \\
    & \dsp + \sum_{\sigma(1,2)}\sum_{\sigma(3,4)} \int d^{n}Q \sum_{\sigma(3,4)} \left( \frac{\langle R q_{1} 3 \rangle}
    {\langle R3 \rangle} \frac{i}{D_{1}^{2}} \frac{\langle R q_{2} 4 \rangle}{\langle R4 \rangle}\right) \frac{ [ r q_{3} 1]}{[1r]} \frac{ [r q_{4} 2]}{[2r]}
    \delta \left( D_{2}^{2}\right)\delta \left( D_{3}^{2}\right) \delta \left(D_{4}^{2}\right) & \nonumber \\
    & \dsp + \sum_{\sigma(1,2)} \sum_{\sigma(1,3)}\sum_{\sigma(2,4)} \int d^{n}Q
    \left(-\frac{[r3] \langle R1\rangle}{ [1r] \langle R3 \rangle} + \sum_{\sigma(1,3)}\frac{ [ r q_{1} 1 ]}{[1r]} \frac{i}{D_{1}^{2}} \frac{\langle R q_{2} 3
    \rangle}{\langle R3 \rangle}\right) \frac{ [r q_{3} 2]}{[2r]}
    \frac{\langle R q_{4} 4 \rangle}{\langle R4 \rangle}
    \delta \left( D_{2}^{2}\right)\delta \left( D_{3}^{2}\right) \delta \left(D_{4}^{2}\right) & \nonumber \\
    & \dsp + \left. \sum_{\sigma(1,2)} \sum_{\sigma(1,3)}\sum_{\sigma(2,4)} \int d^{n}Q \sum_{\sigma(1,4)}
    \left( \frac{ [ r q_{1} 1 ]}{[1r]} \frac{i}{D_{1}^{2}} \frac{\langle R q_{2}4
    \rangle}{\langle R4 \rangle}\right) \frac{ [r q_{3} 2]}{[2r]}
    \frac{\langle R q_{4} 3 \rangle}{\langle R3 \rangle}
    \delta \left( D_{2}^{2}\right)\delta \left( D_{3}^{2}\right) \delta
    \left(D_{4}^{2}\right)\right].
    &
\end{align}
\noindent We are not going to develop all the computation because
there is no difficulty and all trees have already been calculated
in this paper. It remains some tensor integrals, which are reduced
with the formulas $\left( \ref{annexeintegrationI4},
\ref{annexeintegrationJ4} \right)$. Then we use the formula
$\left( \ref{passageJaI} \right)$ to transform the $n$-dimensional
boxes into  $(n+2)$-dimensional boxes. We find:
\begin{align}
    \textrm{Disc}_{3} \left(A^{scalar}(--++) \right) \ = \ & \dsp (e\sqrt{2})^{4} \frac{\langle 12 \rangle}{[12]}\frac{[34]}{\langle 34 \rangle}
    \left[ - \frac{2tu}{s} \textrm{Disc}_{3} \left(I_{4}^{n+2}(1324) \right)  +
    \sum_{\sigma(2,3,4)} \textrm{Disc}_{3} \left(K_{4}^{n}(1234) \right) \right. & \nonumber \\
    & + \left. \sum_{\sigma(1,2)} \left( \frac{ t-u}{s}
    \textrm{Disc}_{3} \left(I_{2}^{n}(u) \right)+ 4\frac{u}{s} \textrm{Disc}_{3} \left(J_{3}^{n}(u) \right)
    \right)\right].
    &\label{resultat--++3cuts}
\end{align}

\hfil

\subsubsection{Two-cut technique} \label{twocut--++}

\hfil

We have again two kinds of topologies. The discontinuity, in terms
of cut-diagrams, is:
\begin{align*}
    \textrm{Disc}_{2} \left(A^{scalar}(--++) \right) = & \frac{1}{2}\left(\parbox{2cm}{\includegraphics[width=2cm]{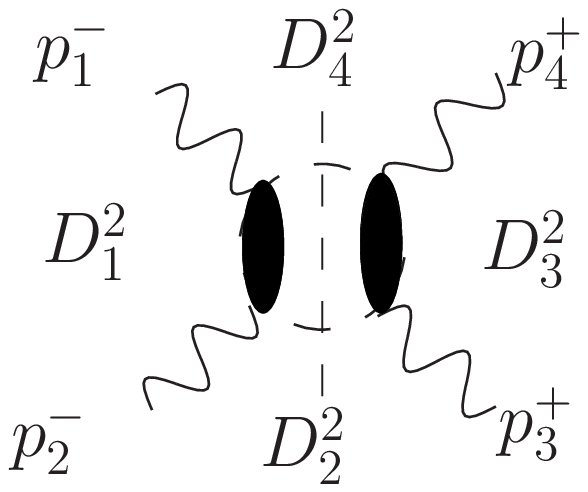}} +
    \parbox{2cm}{\includegraphics[width=2cm]{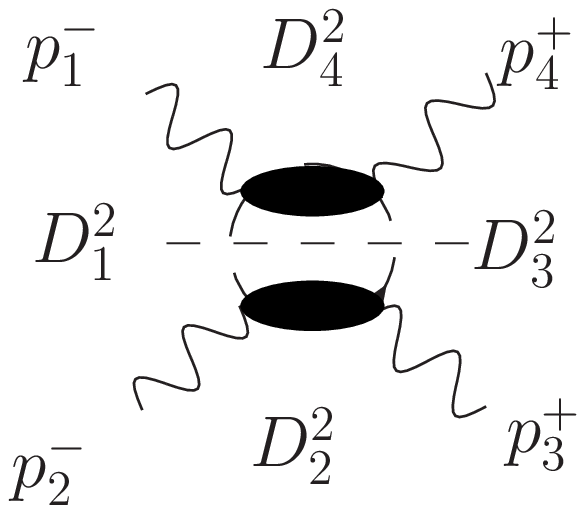}}\right) + \frac{1}{2} \sum_{\sigma(1,2)}
    \left(
    \parbox{2cm}{\includegraphics[width=2cm]{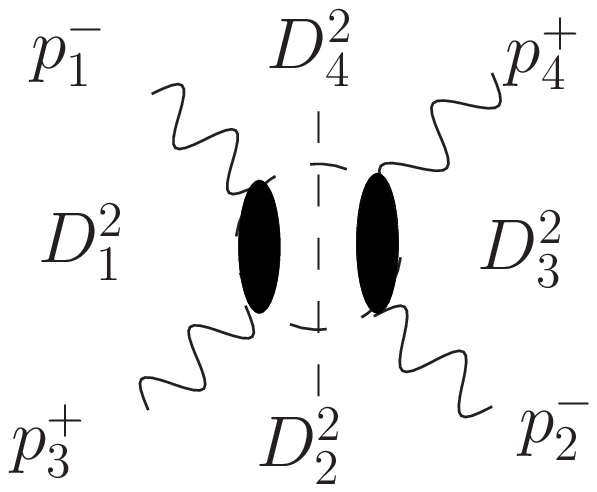}} +
    \parbox{2cm}{\includegraphics[width=2cm]{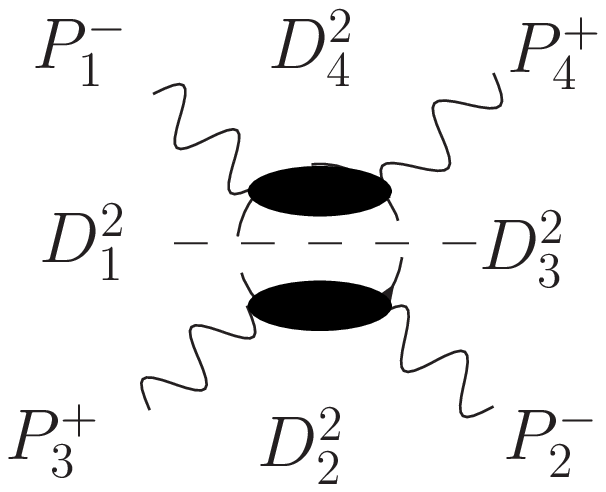}}\right).&
\end{align*}
\noindent Here, thanks to the permutations, all diagrams are
doubled. So we gather diagrams and the discontinuity becomes:
\begin{align}
    \textrm{Disc}_{2} & \left(A^{scalar}(--++) \right) = & \nonumber \\
    & \dsp \ (e\sqrt{2})^{4} \int d^{n}Q \sum_{\sigma(1,2)}\left( \frac{ [ r q_{1} 1
    ]}{[1r]} \frac{i}{D_{1}^{2}}\frac{ [r q_{2} 2]}{[2r]} \right) \sum_{\sigma(3,4)} \left( \frac{\langle R q_{3} 3 \rangle}{\langle R3 \rangle}
    \frac{i}{D_{3}^{2}} \frac{\langle R q_{4} 4 \rangle}{\langle R4 \rangle} \right)
    \delta \left( D_{2}^{2}\right) \delta \left(D_{4}^{2}\right) & \nonumber \\
    & \dsp + (e\sqrt{2})^{4} \sum_{\sigma(1,2)}  \int d^{n}Q  \left( -\frac{[r3] \langle R1\rangle}{ [1r] \langle R3 \rangle} + \sum_{\sigma(1,3)}  \frac{[ r q_{1} 1
    ]}{[1r]} \frac{i}{D_{1}^{2}}\frac{\langle R q_{2} 3 \rangle}{\langle R3 \rangle}
    \right)  & \nonumber \\
    & \quad \quad \quad \quad \quad \quad \quad  * \left( -\frac{[r4] \langle R2\rangle}{ [2r] \langle R4 \rangle} + \sum_{\sigma(2,4)}\frac{ [r q_{3}
    2]}{[2r]}\frac{i}{D_{3}^{2}} \frac{\langle R q_{4} 4 \rangle}{\langle R4 \rangle} \right) \delta \left(
    D_{2}^{2}\right)\delta \left(D_{4}^{2}\right) &\\
    = & \ \dsp I_{1} + I_{2}. & \label{M--++8}
\end{align}
\noindent The collection of diagrams, to create gauge invariant
trees, has mixed the topologies. We express the trees of $I_{1}$
with the formulas $\left( \ref{arbre2++},\ref{arbre2--} \right)$.
For the on-shell trees of $I_{2}$, we can use the formula $\left(
\ref{arbre-+} \right)$, but it is not the best way. We obtain
directly:
\begin{align}
    -\frac{[r3] \langle R1\rangle}{ [1r] \langle R3 \rangle} + \sum_{\sigma(1,3)} \frac{[ r q_{1} 1]}{[1r]} \frac{i}{D_{1}^{2}}\frac{\langle R q_{2} 3 \rangle}{\langle R3
    \rangle} & = -\frac{i}{u} [3q_{4}1]^{2}\sum_{\sigma(1,3)} \frac{1}{D_{1}^{2}}, & \label{arbre3--++scamasse2cuts} \\
    -\frac{[r4] \langle R2\rangle}{ [2r] \langle R4 \rangle} + \sum_{\sigma(2,4)} \frac{ [r q_{3} 2]}{[2r]} \frac{i}{D_{3}^{2}} \frac{\langle
    R q_{4} 4 \rangle}{\langle R4 \rangle} & = - \frac{i}{u} [4q_{2}2]^{2}\sum_{\sigma(2,4)}
    \frac{1}{D_{3}^{2}}.
    &\label{arbre4--++scamasse2cuts}
\end{align}
\noindent So the discontinuity $\textrm{Disc}_{2}
\left(A^{scalar}(--++) \right)$ becomes:
\begin{align}
    \textrm{Disc}_{2} & \left(A^{scalar}(--++) \right) \ = \dsp \ (e\sqrt{2})^{4} \sum_{\sigma(1,2)}\sum_{\sigma(3,4)} \frac{\langle 12 \rangle}{[12]}\frac{[34]}{\langle 34 \rangle}
    \textrm{Disc}_{2,s_{12}} \left( K_{4}^{n}(1234) \right) & \nonumber \\
    & \dsp +(e\sqrt{2})^{4} \sum_{\sigma(1,2)}\frac{i^{2}}{u^{2}} \int d^{n}Q  \left( [3q_{4}1]^{2} \sum_{\sigma(1,3)} \frac{1}{D_{1}^{2}} \right)
    \left( [4q_{2}2]^{2}\sum_{\sigma(2,4)} \frac{1}{D_{3}^{2}} \right) \delta \left( D_{2}^{2}\right)\delta
    \left( D_{4}^{2}\right)
    & \\
    = & \ N_{1} + N_{2}. & \label{M--++9}
\end{align}
\noindent Now we simplify the second integral $N_{2}$. We first
introduce two spinors $\langle 34 \rangle$ and $ [21]$ to build
the numerator as one product of spinors. We obtain four integrals
with the same numerators:
\begin{align}
    N_{2} \  = \ & (e\sqrt{2})^{4} \sum_{\sigma(1,2)}\frac{i^{2}}{u^{2}} \int d^{n}Q  \left( [3q_{4}1]^{2} \sum_{\sigma(1,3)} \frac{1}{D_{1}^{2}} \right)
     \left( [4q_{2}2]^{2}\sum_{\sigma(2,4)} \frac{1}{D_{3}^{2}} \right) \delta \left( D_{2}^{2}\right)\delta
    \left( D_{4}^{2}\right)  & \nonumber\\
    = \ & (e\sqrt{2})^{4} \sum_{\sigma(1,2)} \frac{ i^{2} \langle 12 \rangle [34]}{u^{2} s^{2} [12] \langle 34 \rangle }
    \int d^{n}q  \langle 1q_{1}34q_{3}21q_{1}34q_{3}21 \rangle
    \left( \frac{1}{D_{1}^{2}}+ \frac{1}{{D_{1}^{'}}^{2}} \right)
    \left( \frac{1}{D_{3}^{2}}+ \frac{1}{{D_{3}^{'}}^{2}} \right)
    \delta \left( D_{2}^{2}\right)\delta
    \left( D_{4}^{2}\right). \label{I2--++} &
\end{align}
\noindent We have four integrals of rank four. We can use standard
reduction techniques to integrate them, but it is not the most
efficient method. It is better to use the property of the "axis of
cut", which is an axis of symmetry. The distribution of helicity
is symmetric in relation with this axis. To simplify the
expression of the numerator of $I_{2}$, we use the on-shell
conditions of external photons and we note $q_{2} = q $. We write
the numerator as the product of two equal scalars, named $P$,
according to the symmetry of the cut axis:
\begin{equation}
    \langle 1q_{1}34q_{3}21q_{1}34q_{3}21 \rangle \ = \ \langle 1q34q21q34q21 \rangle \ = \ \langle 1q34q21 \rangle \langle 1q34q21 \rangle \ = \
    P^{2}.
\end{equation}
\noindent We want to decrease the rank of $P$ and to introduce
$D_{i}^{2}$ in the numerator. Using gamma matrix relations, we
obtain:
\begin{equation}
    \langle 1q34q21 \rangle \ = \ 2(q.4)\langle 1q321 \rangle - 2(q.3)\langle 1q421 \rangle + \left( \mu^{2} + m^{2}
    \right)\langle 13421 \rangle . \label{developpement}
\end{equation}
\noindent Now, to continue the simplification of $P$, we have to
know the distribution of photons around the loop. The scalar
products $ 2(p_{j}.q_{i})$ can be expressed as a sum of
denominators and Mandelstam variables. Many tensor triangle
integrals can be eliminated. A one external mass triangle integral
with rank one or two can be expressed as a linear combination of
one external mass scalar triangle and scalar bubbles, according to
the formula $\left( \ref{reductiontriangle} \right)$. However
permutations allow us to simplify many tensors. The integral
$N_{2}$ $\left( \ref{I2--++} \right)$ becomes:
\begin{align}
    N_{2} = (-i\sqrt{2}e)^{4} & \sum_{\sigma(1,2)} \frac{i^{2} \langle 12 \rangle [34]}{u s^{2} [12] \langle 34 \rangle }
    \int d^{n}Q \ \ \ \delta \left( D_{2}^{2}\right)\delta
    \left( D_{4}^{2}\right) & \nonumber \\
    & \left( \frac{ \langle 1q34q213q21\rangle - s\left( \mu^{2}+m^{2} \right) \langle 1q34q21
    \rangle}{D_{1}^{2}D_{3}^{2}} + \frac{ - s\left( \mu^{2}+m^{2} \right) \langle 1q34q21
    \rangle}{{D_{1}^{'}}^{2}D_{3}^{2}} \right. & \nonumber \\
    & \left. \ \ + \frac{ - s\left( \mu^{2}+m^{2} \right) \langle 1q34q21
    \rangle}{D_{1}^{2}{D_{3}^{'}}^{2}} + \frac{ \langle 1q34q213q21\rangle - s \left( \mu^{2}+m^{2} \right)\langle 1q34q21
    \rangle}{{D_{1}^{'}}^{2}{D_{3}^{'}}^{2}}  \right) &
    \label{LLLL}
\end{align}
\noindent We apply, again, the development $\left(
\ref{developpement} \right)$ in each terms of $ \left( \ref{LLLL}
\right)$ to reduce the rank of each integral. When we have only
rank one terms we integrate with the formulas $\left(
\ref{annexeintegrationI4},\ref{annexeintegrationJ4} \right)$. And
the integral $I_{2}$ becomes:
\begin{align}
    N_{2} = (e\sqrt{2})^{4} \sum_{\sigma(1,2)} \frac{ \langle 12
    \rangle [34]}{  [12] \langle 34 \rangle } \textrm{Disc}_{2,s_{13}} & \left(  -2\frac{tu}{s}I_{4}^{n+2}(1324)  + 4 \frac{u}{s}
    J_{3}(u)\right. & \nonumber \\
    & \left. + 2\left( K_{4}^{n}(1324)+K_{4}^{n}(3124)\right)  - \frac{ u-t}{s} I_{2}^{n}(u)\right) .
\end{align}
\noindent We gather $N_{1}$ and $N_{2}$ to rebuild the
discontinuity $\textrm{Disc}_{2} \left(A^{scalar}(--++) \right)$
and we obtain:
\begin{align}
    \textrm{Disc}_{2} \left( A^{scalar}(--++) \right) = & \dsp (e\sqrt{2})^{4} \frac{\langle 12 \rangle}{[12]}\frac{[34]}{\langle 34 \rangle}
    \left[ - \frac{2tu}{s} \textrm{Disc}_{2} \left(I_{4}^{n+2}(1324) \right)  +
    \sum_{\sigma(2,3,4)} \textrm{Disc}_{2} \left(K_{4}^{n}(1234) \right) \right. & \nonumber \\
    & + \left. \sum_{\sigma(1,2)} \left( \frac{ t-u}{s}
    \textrm{Disc}_{2} \left(I_{2}^{n}(u) \right)+ 4\frac{u}{s} \textrm{Disc}_{2} \left(J_{3}^{n}(u) \right)
    \right)\right].
    &\label{resultat--++2cuts}
\end{align}

\hfil

The reconstruction of the helicity amplitude is easy and we find:
\begin{align}
    A_{4}^{scalar} (--++) = & \dsp 4 \ i \alpha^{2} \frac{\langle 12 \rangle}{[12]}\frac{[34]}{\langle 34 \rangle}
    \left( - \frac{2tu}{s} I_{4}^{n+2}(1324)  \right. & \nonumber \\
    & \left. + \sum_{\sigma(1,2)} \left( \frac{ t-u}{s}
    I_{2}^{n}(u)+ 4\frac{u}{s} J_{3}^{n}(u) \right) + \sum_{\sigma(2,3,4)} K_{4}^{n}(1234) \right). &
\end{align}
\noindent This expression is valid to all orders in $\epsilon$.
One of the reason of the compactness of the result is that we have
a symmetry of the helicity structure. Moreover, thanks to the fact
that we use a four-point scalar integral in $n+2$ dimensions
rather a four-point scalar integral in $n$ dimensions, we don't
have any triangle except the scalar integral $J_{3}^{n}$.

\hfil

\subsection{Collection of the main result of the four-photon
helicity amplitudes in massive scalar QED.} \label{summaryscalar}

\hfil

The helicity amplitudes of four-photon scattering are:
\begin{align}
    A^{scalar}_{4}(++++)  = \ \dsp 4 \ i \ \alpha^{2} & \sum_{\sigma(2,3,4)} \frac{[12][34]}{\langle 12 \rangle \langle 34
    \rangle} K_{4}^{n}\left(1234\right),  &\\
    A_{4}^{scalar}(-+++) =  \ \dsp 4 \ i \ \alpha^{2} & \sum_{\sigma(2,3,4)}  \frac{[34]
    [231]}{ \langle 34 \rangle \langle 231 \rangle} \left( K_{4}^{n} +\frac{st}{2u}  J_{4}^{n}(1234) + \left( \frac{s^{2}+t^{2}+u^{2}}{2tu}\right)
    J_{3}^{n}(s)\right),& \\
    A_{4}^{scalar}(--++) = \ \dsp 4 \ i \ \alpha^{2} &\frac{\langle 12 \rangle}{[12]}\frac{[34]}{\langle 34 \rangle}
    \left(  - \frac{2tu}{s} I_{4}^{n+2}(1324) + \sum_{\sigma(2,3,4)} K_{4}^{n}(1234) \right. & \nonumber \\
    & \left. \quad  \quad \quad \ \ + \sum_{\sigma(1,2)} \left( \frac{ t-u}{s}
    I_{2}^{n}(u)+ 4\frac{u}{s} J_{3}^{n}(u) \right) \right). &
\end{align}
\noindent If we compare the different cut techniques, we see that
the four-cut technique is very powerful to calculate the
coefficients in front the boxes. But we cannot obtain the
coefficients in front of bubbles and triangles. We can point out
that the three-cut technique is sufficient to reconstruct all the
amplitude. I explain this fact in the subsection
$\ref{multicutexplanation}$. In the leading order in $\epsilon$,
the extra-scalar integrals are purely rational (Appendix
$\ref{integraldimsup}$), this is the origin of the rational terms.
Thanks to the spinor formalism, the results are more compact than
all results of four-photon amplitude obtained in the past. In the
Appendix $\ref{knownresult}$, I give the massless limit in the
leading order in $\epsilon$ and I find the known results.
Therefore, we point out that, in massless theory, only the MHV
amplitude $\left( A^{scalar}_{4}(--++) \right)$ has a
polylogarithm structure in the leading order in $\epsilon$. The
two helicity amplitudes $A^{scalar}_{4}\left( \pm+++ \right)$ are
only rational terms.

\section{Four-photon helicity amplitudes in QED: $A_{4}^{fermion}$} \label{amplitudefermions}

\subsection{$ A_{4}^{fermion}(1^{\pm},2^{+},3^{+},4^{+})$ helicity amplitudes}

\hfil

The two helicity amplitudes $
A_{4}^{fermion}(1^{\pm},2^{+},3^{+},4^{+})$ in QED, are directly
related to the scalar QED helicity amplitude $
A_{4}^{scalar}(1^{\pm},2^{+},3^{+},4^{+})$ in massless and massive
theories:
\begin{equation}
    A_{4}^{fermion}(1^{\pm},2^{+},...,N^{+}) \ = \ -2 \ A_{4}^{scalar}(1^{\pm},2^{+},...,N^{+}). \label{supersymmetricdecompostion-+++}
\end{equation}
\noindent This result is true diagram per diagram. To proof this
result, we consider a fermion loop with N photons entering the
loop. We impose, first, that all photons have a positive helicity
and the same reference vector $ |R \rangle $. Therefore we have $
\forall (i,j) \in [1..N], \ \ \varepsilon_{i}.\varepsilon_{j} =
0$. Now we develop a one-loop diagram, called $D^{fermion}$, in
QED:
\begin{equation}
    D^{fermion} \ = \ - e^{N} \int d^{n} Q \frac{tr \left( \varepssla_{1} \left( \qsla_{1} + m \right) ... \varepssla_{N} \left( \qsla_{N} + m
    \right)\right)
    }{D_{1}^{2}...D_{N}^{2}} \ = \ - e^{N} \int d^{n} Q \frac{tr \left( \varepssla_{1} \qsla_{1}
    ... \varepssla_{N} \qsla_{N}\right) }{D_{1}^{2}...D_{N}^{2}}. \label{III}
\end{equation}
\noindent All terms proportional to $m^{2}$ are proportional to
$\varepsilon_{i}.\varepsilon_{j} = 0$, and so vanish. Now if we
put the explicit formula of the polarisation vectors of each
photon $ \left( \ref{vectorpolarisation} \right)$ in $ \left(
\ref{III} \right)$, then we obtain directly that the amplitude of
the scalar QED diagram corresponding $D^{fermion} = -2
D^{scalar}$. We add all diagram and we obtain straightforward
$\left( \ref{supersymmetricdecompostion-+++} \right)$. Secondly,
for $ A_{4}^{fermion}(-+++)$, as we have one negative-helicity
photon, we impose the reference vector of the positive-helicity
photons equal to the momentum of the negative-helicity photon. In
this case, we have $ \forall (i,j) \in [1..N], \ \
\varepsilon_{i}.\varepsilon_{j} = 0$ too, and the proof is the
same as the first case.

\hfil

\subsection{Relation between the QED theories}

\hfil

We can relate the different QED theories with the Gordon relation.
A development of this link was initiated in \cite{SecondOrder}.
Currents, in QED, are charged whereas in scalar QED, currents are
not charged. So to relate the two theories, we have to separate
the QED in an uncharged part and a charged part, which is the
magnetic moment of a gauge field.

\hfil

We define the magnetic momenta of a gauge field, with a momentum
$p$ and the helicity $\sigma$, by:
\begin{equation}
    {\mbox{$\not{\! \! M_{p}^{\sigma}}$}} \ = \ e\sigma^{\mu\nu} p_{\nu}{\varepsilon_{p}}_{\mu}^{\sigma} \ = \ \frac{i e}{2} \left[
    \varepssla_{p}^{\sigma} , \psla \right] \ = \ \frac{i e}{2} \left( \varepssla_{p}^{\sigma}\psla - \psla \varepssla_{p}^{\sigma}\right). \label{magnetic}
\end{equation}
\noindent The spinor formulas of the polarisation vectors $\left(
\ref{vectorpolarisation} \right)$, give us:
\begin{equation}
    {\mbox{$\not{\! \! M_{p}^{+}}$}} \ = \ i \ e \sqrt{2} |p- \rangle \langle p+| \ \ \ \textrm{and} \ \ \ \
    {\mbox{$\not{\! \! M_{p}^{-}}$}} \ = \ - i \ e \sqrt{2} |p+ \rangle \langle p-| . \label{moment}
\end{equation}

We consider a vertex in QED between an ingoing photon with a
momentum $p$ and two fermions with the momenta $k$ and $k+p$. We
decomposed the sum over the two ingoing and outgoing currents of
fermions in the vertex as the simple vertex of the scalar QED plus
another vertex called "the magnetic term", just with some gamma
matrix relations:
\begin{align}
    -ie\frac{\ksla+\psla}{\left(k+p\right)^{2}}\varepssla_{p} - ie \varepssla_{p}
    \frac{\ksla}{\left(k+p\right)^{2}}
    & = -ie \left( 2 \ksla + \psla + \psla \right) \frac{\varepssla_{p}}{2 \left(k+p\right)^{2}} -
    ie \varepssla_{p} \frac{\ksla}{\left(k+p\right)^{2}} &\\
    & = \frac{-ie}{\left(k+p\right)^{2}} \left( \left( 2k+p \right).\varepsilon_{p} - \frac{\varepssla_{p}}{2}
    \left( 2 \ksla +\psla \right) + \psla \frac{\varepssla_{p}}{2} + \varepssla_{p} \ksla \right) & \\
    & = \frac{-ie}{\left(k+p\right)^{2}} \left( \left( 2k+p \right).\varepsilon_{p}  + i \sigma^{\mu\nu}{\varepsilon_{p}}_{\mu}p_{\nu} \right) &\\
    & = \frac{-ie}{\left(k+p\right)^{2}}\left( 2 k^{\mu} +i\sigma^{\mu\nu}p_{\nu} \right) \varepsilon_{p}^{\mu}.\label{Gordon} &
\end{align}
\noindent In the left hand side of this relation, we have the QED
vertex with two currents, which look like the currents of
fermions. In the right hand side, we recognize the simple scalar
QED vertex and the magnetic moment of the photon entering the
vertex. So we are going to define an effective interaction, which
describe the QED.

\hfil

We define an effective interaction, described by the vertex
$U_{p}$ between a photon, with the momentum $p$ and a fermion,
with a momentum $k$:
\begin{equation}
    U_{p} \ = \ -ie \left( 2k^{\mu}+p^{\mu} +i\sigma^{\mu\nu}p_{\nu} \right)
    \varepsilon_{p}^{\mu} \ = \ -ie \left( 2k^{\mu}+p^{\mu} \right)\varepsilon_{p}^{\mu} + {\mbox{$\not{\! \! M_{p}}$}}.  \label{newvertex}
\end{equation}

\hfil

The Gordon relation is written, with this effective interaction:
\begin{equation}
    -ie\frac{\ksla+\psla}{\left(k+p\right)^{2}}\varepssla_{p} - ie\varepssla_{p}
    \frac{\ksla}{\left(k+p\right)^{2}} \ = \ \frac{U_{p}}{\left(k+p\right)^{2}}. \label{Gordon2}
\end{equation}
\noindent We are going to show that the relation between QED and
scalar QED is complete for a loop of fermion with ingoing photons.

\hfil

Consider the amplitude $ \gamma_{1}+ \gamma_{2}+ \gamma_{3}+
\gamma_{4} \rightarrow 0 $ in QED theory. If we note $ B_{S}^{12}
= 2ie^{2} \eta
^{\mu\nu}{\varepsilon_{1}}_{\mu}{\varepsilon_{2}}_{\nu} $ the
double vertex in scalar QED, then the amplitude in QED becomes:
\begin{align}
    A_{4}^{fermion} =  - \frac{1}{8} \sum_{\sigma(1..4)} \int d^{n}Q & \ i^4 \frac{tr\left(U_{1}
    U_{2} U_{3}U_{4} \right) }{D_{1}^{2}D_{2}^{2}D_{3}^{2}D_{4}^{2}} + i^{3} \frac{B_{S}^{12} tr\left(U_{3}U_{4}\right)}{D_{2}^{2}D_{3}^{2}D_{4}^{2}}
    + i^{3}\frac{B_{S}^{23}tr\left(U_{1}U_{4}\right)}{D_{1}^{2}D_{3}^{2}D_{4}^{2}}
    + i^{3} \frac{B_{S}^{34}tr\left(U_{1}U_{2}\right)}{D_{1}^{2}D_{2}^{2}D_{4}^{2}}& \nonumber \\
    & + i^{3}\frac{B_{S}^{41}tr\left(U_{2}U_{3}\right)}{D_{1}^{2}D_{2}^{2}D_{3}^{2}}
    + i^{2} \frac{B_{S}^{12}B_{S}^{34}}{D_{2}^{2}D_{4}^{2}}
    + i^{2} \frac{B_{S}^{23}B_{S}^{14}}{D_{1}^{2}D_{3}^{2}}.
    &\label{amplitudenew}
\end{align}

To proof this relation , we consider the four-photon QED
amplitude:
\begin{equation}
    A_{4}^{fermion} = - \frac{1}{4} \sum_{\sigma(1..4)} \int d^{n}Q \ i^4 \frac{tr\left(\varepssla_{1} \left(
    \qsla_{1}+m \right)
    \varepssla_{2} \left( \qsla_{2}+m \right) \varepssla_{3} \left( \qsla_{3}+m \right) \varepssla_{4} \left( \qsla_{4}
     +m \right) \right) }{D_{1}^{2}D_{2}^{2}D_{3}^{2}D_{4}^{2}}.
\end{equation}
\noindent First we begin to develop the amplitude
$A_{4}^{fermion}$ in $2^{4}=16$ terms. Then we apply the relation
linking QED vertex and the effective interaction $\left(
\ref{Gordon2} \right)$ to eliminate all $\qsla_{i}$ in the
numerator of each term. In the next step we use again the relation
$\left( \ref{Gordon2} \right)$, which inverts the rotating
direction in the loop. So, to we find the initial direction, we
apply the gamma matrix relation : $ tr \left( \gamma_{1}....
\gamma_{N} \right) = tr \left( \gamma_{N} .... \gamma_{1}
\right)$. At the end we restore the symmetry to create double
vertex $B_{S}^{ij}$.

\hfil

This result is remarquable. QED is written like scalar QED except
for the fact that the simple vertex $ -ie \left( 2k^{\mu}+p^{\mu}
\right) $ becomes the effective interaction $ U_{p} = -ie \left(
2k^{\mu}+p^{\mu} +i\sigma^{\mu\nu}p_{\nu} \right)$. This result
can be extended to the N-photon one-loop amplitudes. Using this
trick, we are going to calculate the last helicity amplitude $
A_{4}^{fermion}(1^{-},2^{-},3^{+},4^{+})$.

\hfil

\subsection{$ A_{4}^{fermion}(1^{-},2^{-},3^{+},4^{+})$ helicity amplitude with four-cut technique}

\hfil

The four-cut technique assumes that all the propagators $
D_{1}^{2},D_{2}^{2},D_{3}^{2},D_{4}^{2}$ are on-shell. Using the
formula $\left( \ref{amplitudenew} \right)$, the discontinuity
$\textrm{Disc}_{4}\left( A_{4}^{fermion}(--++) \right)$ is:
\begin{equation}
    \textrm{Disc}_{4}\left( A_{4}^{fermion}(--++)
    \right) = - \frac{1}{2} \sum_{\sigma(2,3,4)} \int d^{n}Q \
    tr\left(U_{1}
    U_{2} U_{3}U_{4} \right) \delta \left( D_{1}^{2}\right)\delta \left( D_{2}^{2}\right)\delta
    \left( D_{3}^{2}\right)\delta \left(D_{4}^{2}\right).
\end{equation}
\noindent The effective interaction $ U_{p} = -ie \left(
2k^{\mu}+p^{\mu} +i\sigma^{\mu\nu}p_{\nu} \right)
\varepsilon_{p}^{\mu} $ is a sum of two terms, so the development
of the discontinuity gives us $2^{4}=16$ terms. But, a moment
magnetic $\left( \ref{moment} \right)$ is a commutator, therefore,
a trace of it is zero $tr(M_{i})=0$ and a trace with two magnetic
moments with two different helicities is zero too. So the
development of $\textrm{Disc}_{4}\left( A_{4}^{fermion}(--++)
\right)$ has only five terms. The one, with only the QED scalar
vertices, is the scalar discontinuity with the factor ``-2'':
\begin{align}
    \textrm{Disc}_{4}& \left( A_{4}^{fermion}(--++)
    \right) \ = \  -2 \ \textrm{Disc}_{4}\left( A_{4}^{scalar}(--++)\right) & \nonumber\\
    & \ + \sum_{\sigma(1,2)}\sum_{\sigma(3,4)}
    -\frac{1}{2} (-2ie)^{2}i^{4}  \int d^{n}Q \ tr \left(
    M_{3}M_{4}\right) \ \varepsilon_{1}.q_{1} \ \varepsilon_{2}.q_{2} \  \delta \left( D_{1}^{2}\right)\delta \left( D_{2}^{2}\right)\delta
    \left( D_{3}^{2}\right)\delta \left(D_{4}^{2}\right) & \nonumber\\
    & \ + \sum_{\sigma(1,2)}\sum_{\sigma(3,4)}
    -\frac{1}{2} (-2ie)^{2} i^{4} \int d^{n}Q \ tr \left(
    M_{1}M_{2}\right) \ \varepsilon_{3}.q_{3} \ \varepsilon_{4}.q_{4} \ \delta \left( D_{1}^{2}\right)\delta \left( D_{2}^{2}\right)\delta
    \left( D_{3}^{2}\right)\delta \left(D_{4}^{2}\right) & \nonumber\\
    & \ + \frac{1}{2} \sum_{\sigma(1,2)}\sum_{\sigma(3,4)}
    -\frac{1}{2} (-2ie)^{2} i^{4} \int d^{n}Q \ tr \left(
    M_{3}M_{4}\right) \ \varepsilon_{1}.q_{1} \ \varepsilon_{2}.q_{3} \ \delta \left( D_{1}^{2}\right)\delta \left( D_{2}^{2}\right)\delta
    \left( D_{3}^{2}\right)\delta \left(D_{4}^{2}\right) & \nonumber \\
    & \ + \frac{1}{2} \sum_{\sigma(1,2)}\sum_{\sigma(3,4)}
    -\frac{1}{2} (-2ie)^{2} i^{4} \int d^{n}Q \ tr \left(
    M_{1}M_{2}\right) \ \varepsilon_{3}.q_{2} \ \varepsilon_{4}.q_{4} \ \delta \left( D_{1}^{2}\right)\delta \left( D_{2}^{2}\right)\delta
    \left( D_{3}^{2}\right)\delta \left(D_{4}^{2}\right) & \nonumber \\
    = & \ -2 \  \textrm{Disc}_{4}\left( A_{4}^{scalar}(--++)
    \right) + I_{1} + I_{2} + I_{3} + I_{4}. & \label{M--++spimasse4cuts}
\end{align}
\noindent The factor ``2'' in front of the scalar discontinuity,
comes from the fact that we need two complex scalars to build a
fermion and the sign ``-'' comes from the fact that we change a
fermion-loop into a boson-loop. We find this factor in the
supersymmetric decomposition $\left(
\ref{supersymmetricdecomposition} \right)$. Now we have just to
calculate all trees containing in $\left( \ref{M--++spimasse4cuts}
\right)$ to obtain the discontinuity. We, first, compute the
traces of magnetic moments using the definition $\left(
\ref{moment} \right)$ and then we simplify the formula of trees in
$\left( \ref{M--++spimasse4cuts} \right)$ with $\left(
\ref{arbre1++} \right)$. We obtain:
\begin{align}
    \textrm{Disc}_{4}\left( A_{4}^{fermion}(--++)
    \right) = & \dsp   \ -2 \  \textrm{Disc}_{4}\left( A_{4}^{scalar}(--++)\right) & \nonumber \\
    & +(e\sqrt{2})^{4}\frac{\langle 12\rangle [34]}{[12] \langle
    34\rangle} s \left(\sum_{\sigma(2,3,4)} \textrm{Disc}_{4}\left( J_{4}^{n}(1234) \right) - 2 \textrm{Disc}_{4}\left(I_{4}^{n+2}
    (1324)\right)\right). &
\end{align}
\noindent Now we use the two-cut technique to calculate the same
amplitude.

\hfil

\subsection{$ A_{4}^{fermion}(1^{-},2^{-},3^{+},4^{+})$ helicity amplitude with the two-cut technique}

\hfil

The QED discontinuity $\textrm{Disc}_{2} \left(
A_{4}^{fermion}(--++) \right)$ is:
\begin{align}
    \textrm{Disc}_{2} \left( A_{4}^{fermion}(--++) \right) = & - \frac{1}{2} \sum_{\sigma(2,3,4)}  \int d^{n}Q \ i^4 \frac{tr\left(U_{1}
    U_{2} U_{3}U_{4} \right) }{D_{1}^{2}D_{3}^{2}} \delta\left( D_{2}^{2}\right) \delta\left( D_{4}^{2}\right) +
    i^{4}\frac{tr\left(U_{1} U_{2} U_{3}U_{4} \right) }{D_{2}^{2}D_{4}^{2}} \delta\left( D_{1}^{2}\right) \delta\left( D_{3}^{2}\right) & \nonumber \\
    & +  \left(  i^{3}\frac{B_{S}^{34}tr\left(U_{1}U_{2}\right)}{D_{1}^{2}}
    + i^{3}\frac{B_{S}^{12} tr\left(U_{3}U_{4}\right)}{D_{3}^{2}}
    + i^{2} \frac{B_{S}^{12}B_{S}^{34}}{1} \right)
    \delta\left( D_{2}^{2}\right) \delta\left( D_{4}^{2}\right) & \nonumber \\
    & + \left( i^{3}\frac{B_{S}^{14}tr\left(U_{2}U_{3}\right)}{D_{2}^{2}}
    + i^{3} \frac{B_{S}^{23}tr\left(U_{1}U_{4}\right)}{D_{4}^{2}}
    + i^{2} \frac{B_{S}^{14}B_{S}^{23}}{1}\right)
    \delta\left(D_{1}^{2}\right) \delta\left( D_{3}^{2}\right). &
\end{align}
\noindent The computation give us directly:
\begin{align}
    \textrm{Disc}_{2}\left( A_{4}^{fermion}(--++)
    \right) = & \dsp  -2 \ \textrm{Disc}_{2}\left( A_{4}^{scalar}(--++)\right) & \nonumber \\
    & +(e\sqrt{2})^{4}\frac{\langle 12\rangle [34]}{[12] \langle
    34\rangle} s \left(\sum_{\sigma(2,3,4)} \textrm{Disc}_{2}\left( J_{4}^{n}(1234) \right) - 2 \textrm{Disc}_{2}\left(I_{4}^{n+2}
    (1324)\right)\right). &
\end{align}

\hfil

\subsection{Collection of the main result of the four-photon
helicity amplitudes in massive QED.}

\hfil

With the two-cut technique or the four-cut technique, we find the
same discontinuity of the MHV four-photon amplitude in massive
QED. The reconstruction gives us:
\begin{equation}
    A_{4}^{fermion}(--++) \ = \ \dsp  -2 \ A_{4}^{scalar}(--++) + 4 \ i\ \alpha^{2} \frac{\langle 12\rangle [34]}{[12] \langle 34\rangle} s
    \left(\sum_{\sigma(2,3,4)} J_{4}^{n}(1234) - 2 \ I_{4}^{n+2} (1324) \right). \label{supersymmetricdecompostion--++}
\end{equation}
\noindent If we take the formula of scalar amplitude already
calculated $\left( \ref{resultat--++2cuts} \right)$, the QED
amplitude is :
\begin{align}
    A^{fermion} (--++) \ = \ \dsp -8 \ i \ \alpha^{2} \frac{\langle 12 \rangle}{[12]}\frac{[34]}{\langle 34 \rangle}
    & \left( \frac{t^{2}+u^{2}}{s} I_{4}^{n+2}(1324) + \sum_{\sigma(1,2)} \left( \frac{ t-u}{s}
    I_{2}^{n}(u)+ 4\frac{u}{s} J_{3}^{n}(u) \right) \right. & \nonumber \\
    & \left. + \sum_{\sigma(2,3,4)} \left(K_{4}^{n}(1234)-\frac{s}{2} J_{4}^{n}(1234)\right)
    \right).
    &
\end{align}
\noindent The known massless limit is given in the Appendix
$\ref{knownresult}$.

\hfil

\section{Discussions} \label{discussions}

\subsection{On the analytical structures}

\hfil

At one loop order, the helicity amplitudes of the four-photon
process have the same structure for QED or scalar QED theories.
The first two are only a rational term whereas the MHV amplitude
has a polylogarithm structure, carried by the scalar integrals
$I_{4}^{n+2}$ and $I_{2}^{n}$.

\hfil

Now we are going to prove that in massless QED and in massless
scalar QED, and therefore in massless supersymmetric
$\textrm{QED}^{{\cal N} =1}$, four-photon amplitudes could be
decomposed without triangle. It comes from the fact in the
decomposition $\left( \ref{decomposition} \right)$, the infrared
divergence are carried out by scalar triangles. Consider one
diagram, named $L$ of a four-photon amplitude in QED (we have
exactly same proof for the scalar QED). The numerator of fermion's
propagators implies that all IR singularities vanish. Now consider
a sub-diagram of $L$ by removing propagators. We have four
sub-diagrams. After removing a propagator that connects two
photons with on-shell momenta $p_{1}$ and $p_{2}$, the amplitude
depends only on the sum of the two momenta $g = p_1 + p_2$ with
non zero $g^{2} \neq 0$. The two photons around the pinched
propagators have the behavior of one massive photon. However, the
mass of a massive entering particle regularizes IR divergences.
Therefore all sub-diagrams of $L$ are not IR divergent. So as each
sub-diagram is a tensor triangle and doesn't have any IR
divergences therefore each sub-diagram cannot be decomposed with
scalar triangle. Finally we can not have any triangle in massless
theories. But in massive theory, this argument is not exact,
because, the infrared divergences exist only in massless theories.
However no triangle is expected, except some extra-scalar
triangles with the extra dimension term $\mu^{2} +m^{2}$.

\hfil

The bubbles have two origins. The first origin is the
decomposition of three-point tensors integrals and the second
origin is the UV divergences of the loop. Here we have no
triangle, so the contribution of bubbles coming from the reduction
of triangles is zero. Now we study the UV limit of one diagram in
scalar QED. We use the notations introduce in subsection
$\ref{schemeregularization}$. For example, we have:
\begin{align}
     \dsp \int d^{n} Q \frac{\varepsilon_{1}.q_{1}\varepsilon_{2}.q_{2}\varepsilon_{3}.q_{3}\varepsilon_{4}.q_{4}}
    {D_{1}^{2}D_{2}^{2}D_{3}^{2}D_{4}^{2}} & \propto \dsp
    \epsilon_{1}^{\mu}\epsilon_{2}^{\nu}\epsilon_{3}^{\rho}\epsilon_{4}^{\sigma}\int d^{n}Q
    \frac{{q_{1}}_{\mu}{q_{2}}_{\nu}{q_{3}}_{\rho}{q_{4}}_{\sigma}}{D_{1}^{2}D_{2}^{2}D_{3}^{2}D_{4}^{2}}&
    \\
    & \xrightarrow[UV]{} \ \dsp \epsilon_{1}^{\mu}\epsilon_{2}^{\nu}\epsilon_{3}^{\rho}\epsilon_{4}^{\sigma}
    \int d^{n}Q
    \frac{{q}_{\mu}{q}_{\nu}{q}_{\rho}{q}_{\sigma}}{D_{1}^{2}D_{2}^{2}D_{3}^{2}D_{4}^{2}}&\\
    & \xrightarrow[UV]{} \ \dsp \epsilon_{1}^{\mu}\epsilon_{2}^{\nu}\epsilon_{3}^{\rho}\epsilon_{4}^{\sigma}
    \left(\eta^{\mu\nu} \eta^{\rho\sigma} + \eta^{\mu\rho} \eta^{\nu\sigma} + \eta^{\mu\sigma} \eta^{\nu\rho} \right) \int d^{n}Q
    \frac{{q^{2}}^{2}}{D_{1}^{2}D_{2}^{2}D_{3}^{2}D_{4}^{2}}. &
\end{align}
\noindent The tensor integral is UV divergent, therefore the
reduction creates bubbles. But, whatever the helicity amplitudes,
the contraction of tensors $\epsilon_{1}^{\mu}\epsilon_{2}^{\nu}
\epsilon_{3}^{\rho}\epsilon_{4}^{\sigma} \left(\eta^{\mu\nu}
\eta^{\rho\sigma} + \eta^{\mu\rho} \eta^{\nu\sigma} +
\eta^{\mu\sigma} \eta^{\nu\rho} \right)$ is zero. So we conclude
that each diagram of the four-photon amplitude is UV finite,
thanks to the gauge invariant. The UV finiteness express by
compensations of the divergences of the bubbles. We clearly
observe this phenomenon in the MHV amplitudes. But, for the first
two helicity amplitude $A \left( \pm +++ \right)$, we have at
least three positive-helicity photon, so all the discontinuities,
in four dimensions, are zero and it implies that the coefficients
in front of each bubble for each diagram is zero.

\hfil

\subsection{On the origin of the rational terms} \label{rationnalterms}

\hfil

In the very simple example of the four-photon amplitudes, we point
out that the rational terms come from the extra-dimension
integrals $J_{i}$ and $K_{i}$. We can discuss the origin of those
integrals. Consider a one loop diagram of the four photon
amplitude in the four-dimensional helicity scheme, described in
the paragraph $\ref{schemeregularization}$. We can write a diagram
as:
\begin{equation}
    A_{4}^{s} \ = \ \int d^{n}Q
    \frac{\textrm{Num}(Q)}{D_{1}^{2}D_{2}^{2}D_{3}^{2}D_{4}^{2}},
\end{equation}
\noindent where $\textrm{Num}(Q) = \varepsilon_{1}.Q_{1}
\varepsilon_{2}.Q_{2} \varepsilon_{3}.Q_{3}
\varepsilon_{4}.Q_{4}$. However the regularization scheme imposes
that the vertices are in four dimensions. And the $4$-dimensional
space and the $-2\epsilon$-dimensional space are orthogonal
therefore, $ \varepsilon_{1}.Q_{1} = \varepsilon_{1}.q_{1}$. The
numerator is actually a function of the four dimensional part of
the loop momentum: $\textrm{Num}(Q) \rightarrow \textrm{Num}(q)$.
Moreover the denominator of a propagator is spelt: $D_{i}^{2} =
q_{i}^{2} -\mu^{2} -m^{2}$. During the reduction some squares of
momenta appear, like $q_{i}^{2}$, in the numerator. To rebuild a
denominator, we subtract and add the mass: $m^{2}+\mu^{2}$:
\begin{equation}
    q_{i}^{2} \ = \ D_{i}^{2} + \left( m^{2}+\mu^{2} \right).
\end{equation}
\noindent The integrals with $\mu^{2}$ are only rational, this is
the origin of the rational terms. In this case, it is very simple
to find the rational terms. We have just to shift the mass of the
scalar:
\begin{equation}
    m^{2} \ \rightarrow \ m^{2}+\mu^{2}.
\end{equation}
\noindent The four-photon amplitudes are a special case where the
particle in the loop is a scalar or a fermion. Consider the case
where we have a photon propagator in the loop. The internal photon
is in $n$ dimensions and its propagator, proportional to the
metric $\eta^{\mu\nu}$. The contraction of this metric in $n$
dimensions with loop momenta $Q_{\mu}$ creates some $\mu^{2}$
terms because there are not enough vertices to reduce all the
propagators in four dimensions. In this case, we cannot associate
the $\mu^{2}$ terms with a mass.

\hfil

The four-photon amplitudes have no ultraviolet and infrared
divergences so the amplitudes don't depend on the scheme of
regularization. In the case where the amplitude have divergences,
which are not regularized, then the rational terms depend on the
scheme. In \cite{Catani1}, we have methods to change the scheme of
regularization without much effort, just by adding or subtracting
a rational term.

\hfil

\subsection{On the multi-cut techniques} \label{multicutexplanation}

\hfil

In this work we apply three kind of unitarity-cut techniques with
two, three or four cut propagators. We first note that, the more
there are cuts, the more we have on-shell conditions and the
simpler is the computation. But the more we have cuts, less
coefficients in front of the scalar integrals could be calculated.
Actually the four-cut technique is very powerful to calculate the
coefficient in front of the four-point scalar integrals. The
three-cut technique is sufficient to calculate the coefficient in
front of scalar triangles, scalar boxes and scalar bubbles.

\hfil

However, the fact that we can calculate the coefficient in front
of bubbles with three-cut technique, is a peculiarity of the
four-on-shell-photon amplitudes. Consider one four-photon diagram
which is made by four photons ingoing into a fermion loop. We
assume that we apply the two-cut technique to it. We call the axis
of cut, the main cut. This axis shares the four photons in two
trees of two photons. Each group is the same invariant. Then we
add a cut. So we cut one tree with two photons into two trees with
one photon. As one on-shell photon cannot constitute an invariant,
therefore when we cut, we don't divide an invariant into two
invariants. So it remains only one invariant, which is the one of
the two cut technique. Therefore, we don't touch the analytic
information contents in the branch cut and don't loose information
when we extend the two-cut technique to the three cut technique.
This fact explains why the two-point functions, which respect the
main cut, are not spurious in the three-cut technique.

\hfil

\section{Supersymmetric amplitude $A_{4}^{{\cal N} =1}$} \label{amplitudesuper}

\hfil

We use the supersymmetric decomposition $\left(
\ref{supersymmetricdecomposition} \right)$ to extract directly the
supersymmetric amplitude $A_{4}^{ {\cal N} =1}$. Since the
$A_{4}^{fermion}$ obeys to the supersymmetric decomposition
$\left( \ref{supersymmetricdecomposition} \right)$, that means
that we can identify the $A_{4}^{{\cal N} =1}$, without computing
all diagrams. So, with the formula $\left(
\ref{supersymmetricdecompostion-+++} \right)$, we can identify the
$A_{4}^{{\cal N} =1}(\pm +++)$ and, with the formula $ \left(
\ref{supersymmetricdecompostion--++} \right)$, we can identify
$A_{4}^{ {\cal N} =1}(--++)$. We obtain:
\begin{align}
    A_{4}^{{\cal N} =1}(++++) & \ = \ 0, &\\
    A_{4}^{{\cal N} =1}(-+++) & \ = \ 0, &\\
    A_{4}^{{\cal N} =1}(--++) & \ = \ 4 \ i\ \alpha^{2} \frac{\langle 12\rangle [34]}{[12] \langle 34\rangle} s
    \left(\sum_{\sigma(2,3,4)} J_{4}^{n}(1234) - 2 I_{4}^{n+2} (1324) \right). &
\end{align}
\noindent In massless case, in the limit $\epsilon \ \rightarrow \
0$, we have:
\begin{align}
    A_{4}^{{\cal N} =1}(++++) & \ = \ 0, &\\
    A_{4}^{{\cal N} =1}(-+++) & \ = \ 0, &\\
    A_{4}^{{\cal N} =1}(--++) & \ = \ -8 \ i\ \alpha^{2}\frac{\langle 12\rangle [34]}{[12] \langle 34\rangle} s I_{4}^{n+2} (1324) + O\left(\epsilon \right). &
\end{align}
\noindent There is full agreement with \cite{gamsusy}. With a
massless or massive loop momentum, the supersymmetric amplitudes
have no rational term, no bubble, and no triangle, only boxes.

\hfil

We are going to prove that diagrams of the four-photon amplitudes
in supersymmetric QED : ${\cal N} =1$ are UV finite. We identify
the decomposition of a fermion loop $\left( \ref{amplitudenew}
\right)$ and the formula of the supersymmetric decomposition
$\left( \ref{supersymmetricdecomposition} \right)$. As the trace
of one magnetic moment is zero, therefore we see that all the
terms belonging to the supersymmetric amplitude have at least two
magnetic moments. So we can do the power counting of one of those
terms. We define $r$ the power of the loop momentum of the
$N$-photons amplitude. We have:
\begin{equation}
    r \ = \ n-1 +N-2-2N \ = \ 1-2\epsilon - N.
\end{equation}
\noindent So, if $N\geq3$, therefore the loop is not UV divergent
and so diagrams of the four-photon amplitude in supersymmetric
$\textrm{QED}^{{\cal N} =1}$ are UV finite.

\hfil

In the last section, we show that the diagrams in
$\textrm{QED}^{{\cal N}=1}$ have no IR divergence and therefore no
triangle. The bubbles in the supersymmetric amplitude could come
only from the UV structure. But we see that each diagram has no UV
divergence, so they are no bubble in each diagram. We can observe
it with standard reduction of the four-photon amplitude. There are
some interferences in the loop between bosons and fermions, which
reduce the UV power and eliminate all bubbles. The interferences
create magnetic moments, which are gauge invariants. Interferences
increase the power of the gauge invariance.

\hfil

In the next section, we calculate, the most simple helicity
configuration of six-photon amplitude in massive theories.

\hfil

\section{The first helicity amplitude $A_{6}(++++++)$} \label{amplitude6photons}

\hfil

In \cite{Papa:6gamma,nagy}, the six-photon helicity amplitudes in
massless and massive loop were numerically computed. Here we
obtain an analytic expression of the most simple helicity
amplitude, all the six photons have a positive helicity for a
massive loop.

\hfil

A six-photon one-loop diagram is not IR/UV divergent, so in this
part, the dimensional regularization is not needed here and the
integrals are in four dimensions. With standard techniques, we
show that this amplitude has neither bubble, nor rational term and
nor triangle with one and two external mass. We verify it
explicitly by the computation.

\hfil

\noindent Thanks to the supersymmetric decomposition $\left(
\ref{supersymmetricdecompostion-+++} \right)$, the scalar
amplitude gives us directly the fermionic amplitude and the
supersymmetric amplitude. Now consider a diagram and we apply the
two-cut method. There are two kinds of discontinuity. The first
kind of discontinuities separate the six photons in two groups of
three photons, whereas the second kind of discontinuities separate
the six photons in a group with four photons and a group with two
photons. We don't have bubble contributions so the second kind of
discontinuity is better because with only one cut we can have the
coefficient in front of all kind of the scalar integrals. The
problem of the first kind of discontinuities is that we cannot
have the coefficient in front of triangle with three external
mass. Let us cut the diagrams in the channel $s_{56}$:
\begin{equation}
    \textrm{Disc}_{2,s_{56}} \left(  A^{scalar}_{6}(++++++) \right) = (-2ie)^{6}i^{6} \int d^{4}q \ A_{tree}\left( 1^{+},2^{+},3^{+},4^{+}
    \right) A_{tree}\left( 5^{+},6^{+} \right) \delta\left( D_{4}^{2}
    ,D_{6}^{2} \right). \label{M++++++scamasse2cuts}
\end{equation}
\noindent In this part, to reduce, we note $\delta\left( D_{4}^{2}
,D_{6}^{2} \right) = \delta\left( D_{4}^{2} \right) \delta
\left(D_{6}^{2} \right)$. We need on-shell trees with two photons
and four photons. The computation give us:
\begin{align}
    A_{tree}\left( 5^{+},6^{+} \right)& = \dsp
    \sum_{\sigma(5,6)}\frac{\varepsilon_{5}^{+} .q_{5} \varepsilon_{6}^{+} .q_{6}}{
    D_{5}^{2}} = - m^{2} \frac{[56]}{ 2 \langle
    56 \rangle} \sum_{\sigma(5,6)} \frac{ 1 }{D_{5}^{2}},
    &\label{arbre1++++++scamasse2cuts} &\\
    A_{tree}\left( 1^{+},2^{+},3^{+},4^{+} \right)& = \dsp
    \sum_{\sigma(1,2,3,4)}\frac{\varepsilon_{1}^{+} .q_{1} \varepsilon_{2}^{+} .q_{1} \varepsilon_{3}^{+} .q_{3}\varepsilon_{4}^{+} .q_{4}}{
    D_{1}^{2}D_{2}^{2}D_{3}^{2}} &\\
    & = \dsp  m^{4} \sum_{\sigma(1,2,3,4)}\frac{[12] [34]}{ 4 \langle 12 \rangle \langle 34 \rangle}
    \frac{1}{D_{1}^{2}D_{2}^{2}D_{3}^{2}} + m^{4}
    \sum_{\sigma(1,2,3,4)}\frac{[4(4+3+2+1)q_{0}1]}{ 4 \langle 12 \rangle \langle 23
    \rangle \langle 34 \rangle} \frac{1}{D_{1}^{2}D_{2}^{2}D_{3}^{2}}.& \label{arbre2++++++scamasse2cuts}
\end{align}
\noindent We can find some équivalent formulae in \cite{arbre}. We
put those trees $\left( \ref{arbre1++++++scamasse2cuts}
,\ref{arbre2++++++scamasse2cuts} \right)$ in the amplitude $\left(
\ref{M++++++scamasse2cuts} \right)$. We have to integrate tensor
hexagons rank one, which give only four point scalar integrals. So
we are going to calculate the coefficient in front of each scalar
box. We begin to calculate the discontinuity of the two adjacent
mass four point scalar function :
\begin{align*}
     \parbox{5cm}{\includegraphics[width=5cm]{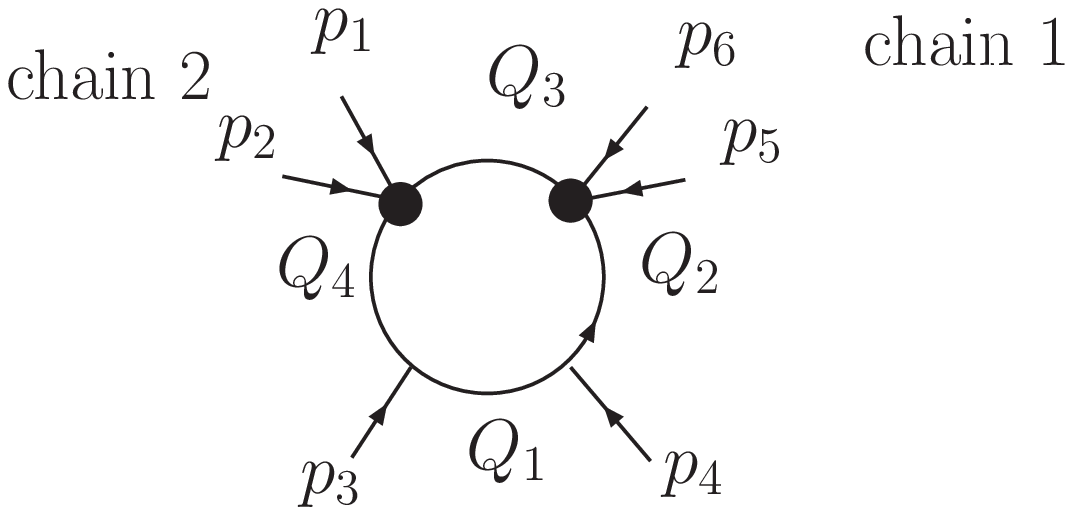}},
\end{align*}
\noindent where we have two invariants $s_{12}, s_{56}$ :
\begin{align}
    \textrm{Disc}_{4,s_{12},s_{56}} \left(  A^{scalar}_{6}\right) & \ = \ \int d^{4}q \ A_{tree}\left( 1^{+},2^{+}\right)
    \prod_{i=3}^{4} \varepsilon_{i}.q_{i} \ A_{tree}\left( 5^{+},6^{+}\right)
    \delta\left( D_{2}^{2} , D_{3}^{2}, D_{4}^{2}, D_{6}^{2} \right) & \\
    & \ = \ -(e\sqrt{2})^{6} m^{6} \frac{[12][34][56]}{\langle 12 \rangle \langle 34 \rangle \langle 56
    \rangle} \textrm{Disc}_{4,s_{12},s_{56}} \left( I_{6}^{n} \right). &
\end{align}
\noindent We use the trees $\left( \ref{arbre1++++++scamasse2cuts}
\right)$, the computation is straightforward and the
reconstruction too. All information of the two adjacent mass four
point functions is hold in the hexagon. Now, we compute the
coefficient in front of the one mass four point function:
\begin{align*}
     \parbox{5cm}{\includegraphics[width=5cm]{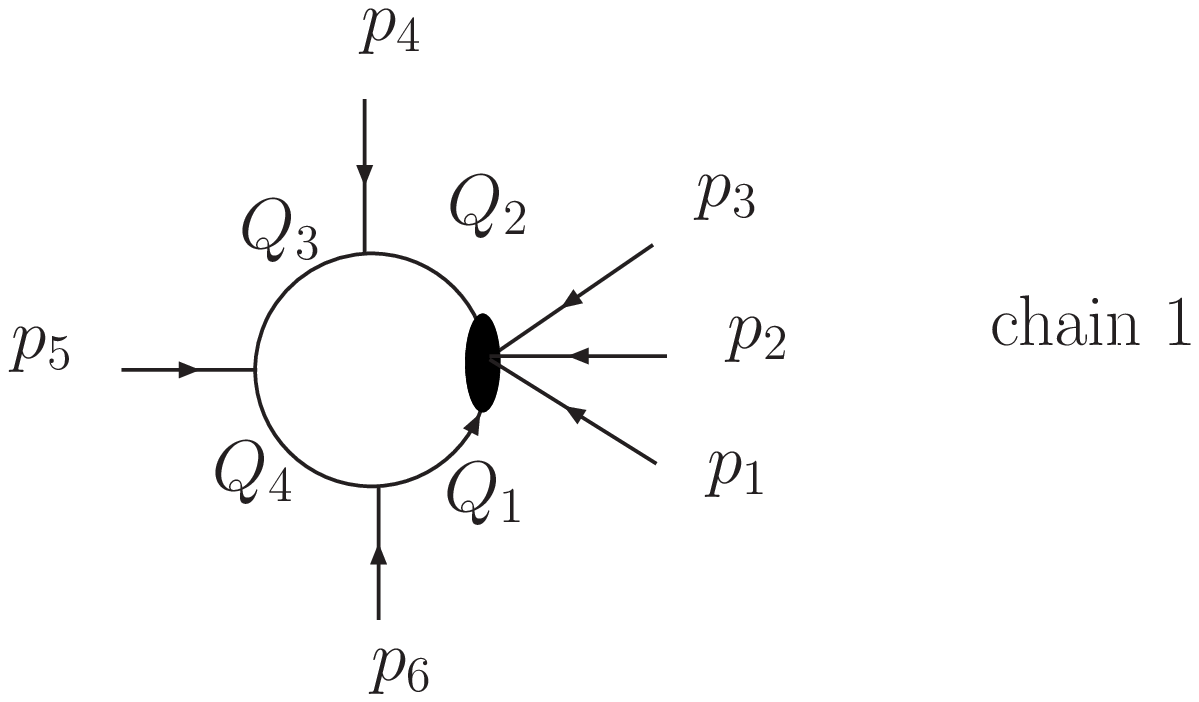}}.
\end{align*}
\noindent The discontinuity is:
\begin{equation}
    \textrm{Disc}_{4,s_{123}} \left( A^{scalar}_{6}\right) \ = \ \int d^{4}q \ A_{arbre}\left( 1^{+},2^{+},3^{+}\right)
    \varepsilon_{4}^{+}.q_{4} \varepsilon_{5}^{+}.q_{5}\varepsilon_{6}^{+}.q_{6} \ \delta\left( D_{3}^{2},D_{4}^{2},D_{5}^{2}, D_{6}^{2} \right).
\end{equation}
\noindent To not break the symmetry, we use the four-photon scalar
trees $\left( \ref{arbre2++++++scamasse2cuts} \right)$. The trees
with a even number of photons are simpler than the trees with a
odd number of photons. We simplify thanks to the permutations and
after reductions, we obtain:
\begin{align}
    \sum_{\sigma(1,2,3,4)}\textrm{Disc}_{4,s_{123}} \left(  A^{s}_{6}\right) \ = \ & -(e \sqrt{2})^{6} m^{6} \frac{[12][34][56]}{\langle 12 \rangle \langle 34 \rangle \langle 56
    \rangle} \sum_{\sigma(1,2,3,4)} \textrm{Disc}_{4,s_{123}} \left( I_{6}^{n} \right) & \nonumber \\
    & -(e\sqrt{2} )^{6} m^{4} \frac{[56]}{\langle 56 \rangle}\sum_{\sigma(1,2,3,4)} \int d^{4}q\frac{[4 \Msla q_{6}1]}
    { \langle 12 \rangle \langle 23 \rangle \langle 34 \rangle}
    \frac{1}{D_{1}^{2}} \delta\left(
    D_{3}^{2},D_{4}^{2},D_{5}^{2},D_{6}^{2}\right),
\end{align}
\noindent where $M = p_{1}+p_{2}+p_{3}$ is the momentum of the
external mass. Rotating the gammas matrix, and using the on-shell
conditions, the numerator becomes:
\begin{equation}
    \langle 54 \Msla q_{6}165 \rangle \ = \  D_{1}^{2} \langle 54 \Msla 65 \rangle + M^{2} \langle 54165 \rangle
    - \langle 54 \Msla \qsla _{3}1 65 \rangle.
\end{equation}
\noindent So the discontinuity is :
\begin{align}
    \textrm{Disc}_{4,s_{123}} \left(  A^{scalar}_{6}\right) \ = \ & \ \ -(e\sqrt{2})^{6} m^{6} \frac{[12][34][56]}{\langle 12 \rangle \langle 34 \rangle \langle 56
    \rangle} \textrm{Disc}_{4,s_{123}} \left( I_{6}^{n} \right) & \nonumber \\
    & \ \ + (e\sqrt{2} )^{6} m^{4} \frac{\langle 54 \Msla 65 \rangle}{2\langle 12 \rangle \langle 23 \rangle \langle 34 \rangle \langle 45 \rangle \langle 56 \rangle \langle 61 \rangle }
    \textrm{Disc}_{4,s_{123}} \left(I_{4,1}^{n} \left( s_{123} \right) \right).
\end{align}
\noindent Finally the compute the discontinuity of the two
opposite mass four point scalar integral:
\begin{align*}
     \parbox{5cm}{\includegraphics[width=5cm]{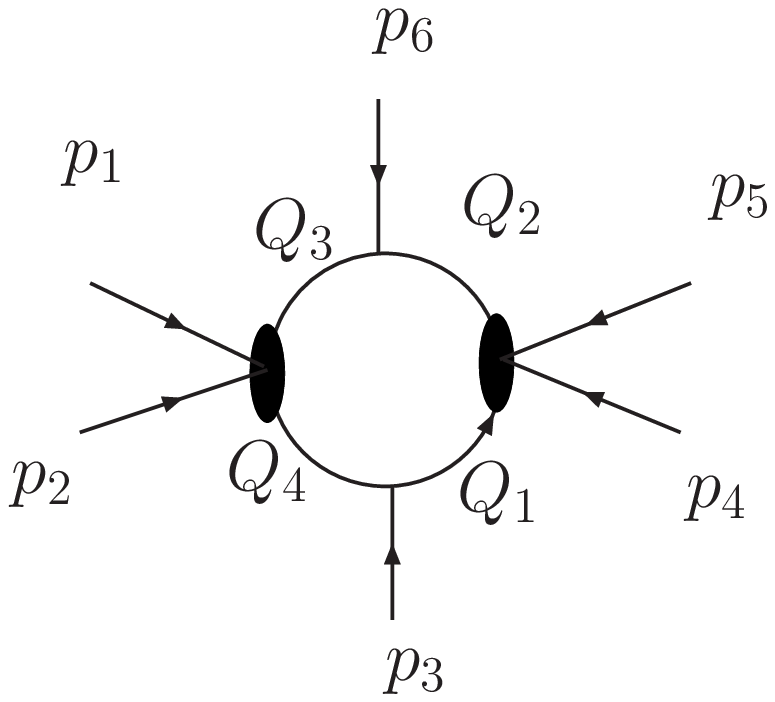}},
\end{align*}
\noindent and the discontinuity is:
\begin{equation}
    \textrm{Disc}_{4,s_{12}, s_{45}} \left(  A^{scalar}_{6}\right) \ = \ \int d^{4}q \ A_{tree}\left( 1^{+},2^{+}\right)
    \varepsilon_{3}^{+}.q_{3} A_{tree}\left( 4^{+},5^{+}\right) \varepsilon_{6}^{+}.q_{6} \ \delta\left( D_{2}^{2}, D_{3}^{2}, D_{5}^{2},D_{6}^{2}
    \right).
\end{equation}
\noindent To not break the symmetry of the scalar integral we use
the three photon scalar trees:
\begin{align}
    A_{tree}\left( 1^{+},2^{+},3^{+} \right) & \ = \ - \frac{m^{2}}{\sqrt{2}^{3}} \sum_{\sigma(1..3)}
    \frac{1}{\langle 12 \rangle \langle 23 \rangle}
    \frac{[3q_{3}(1+2+3)1]}{ D_{1}^{2}D_{2}^{2}}. &
\end{align}
\noindent We apply twice this formula and, thanks to the
permutations, the discontinuity gives:
\begin{align}
    & \ \sum_{\sigma(1,2,3)\sigma(4,5,6)} \textrm{Disc}_{4,s_{12}, s_{45}} \left(  A^{scalar}_{6}\right) & \nonumber \\
    = & \ (e \sqrt{2} )^{6} m^{4} \sum_{\sigma(1,2,3)\sigma(4,5,6)} \int
    d^{4} q \frac{[3\qsla_{3}21][6\qsla_{6}54]}{\langle 12 \rangle \langle 23\rangle \langle 45
    \rangle \langle 56\rangle D_{1}^{2}D_{4}^{2}}\delta\left( D_{2}^{2}, D_{3}^{2}, D_{5}^{2},D_{6}^{2} \right).
\end{align}
\noindent We simplify the numerator by rotating the gammas matrix:
\begin{align}
    [3\qsla_{3}21][6\qsla_{6}54] \ = & \ -\frac{[3216543]}{\langle 16 \rangle \langle 43
    \rangle} + D_{1}^{2} D_{4}^{2}  \frac{[3(1+2)5(5+4)3]}{\langle 16\rangle \langle
    43\rangle} & \nonumber \\
    & + D_{1}^{2} s_{12} \frac{[36543]}{\langle 16\rangle \langle
    43\rangle} -D_{1}^{2} \frac{[3(1+2) \qsla_{2}6543]}{\langle 16\rangle \langle
    43\rangle},
\end{align}
\noindent and we obtain directly:
\begin{align}
    \textrm{Disc}_{4,s_{12}, s_{45}} \left(A^{scalar}_{6}\right) \ =
    & \ -(e\sqrt{2})^{6} m^{6} \frac{[12][34][56]}{\langle 12 \rangle \langle 34 \rangle \langle 56
    \rangle} \textrm{Disc}_{4,s_{12},s_{45}} \left( I_{6}^{n} \right)
    & \nonumber \\
    &  \ +(e\sqrt{2})^{6} m^{4} \frac{[3(1+2)6(5+4)3]}{2\langle 12 \rangle \langle 23 \rangle \langle 34
    \rangle \langle 45\rangle \langle 56 \rangle \langle 61\rangle} \textrm{Disc}_{4,s_{12}, s_{45}} \left(
    I_{4,2B}^{n}(s_{12},s_{45})\right).
\end{align}
\noindent In each discontinuity, we have the trace of the scalar
hexagon. The entire reconstruction is immediate:
\begin{align}
    A_{6}^{scalar} \ = & \ -i\frac{(e\sqrt{2})^{6} m^{6} }{96 \pi^{2}}
    \sum_{\sigma(1,2,3,4,5,6)} \frac{[12][34][56]}{\langle 12 \rangle \langle 34 \rangle \langle 56
    \rangle} I_{6}^{n} & \nonumber \\
    & \ \ +i\frac{(e\sqrt{2})^{6} m^{4} }{96 \pi^{2}}
    \sum_{\sigma(1,2,3,4,5,6)}
    \frac{[3(1+2)6(5+4)3]}{2\langle 12 \rangle \langle 23 \rangle \langle 34
    \rangle \langle 45\rangle \langle 56 \rangle \langle 61\rangle} I_{4,2B}^{n}(s_{12},s_{45})& \nonumber \\
    & \ \ + i\frac{(e\sqrt{2})^{6} m^{4} }{96 \pi^{2}}
    \sum_{\sigma(1,2,3,4,5,6)}
    \frac{\langle 54 \Msla 65 \rangle}{2\langle 12 \rangle \langle 23 \rangle \langle 34 \rangle \langle 45 \rangle \langle 56 \rangle \langle 61 \rangle }
    I_{4,1}^{n} ( s_{123} ) &
\end{align}
\noindent The coefficient in front of each four point scalar
integral could be written as:
\begin{align}
    \frac{\det\left( S \right)}{\langle 12 \rangle \langle 23 \rangle \langle 34 \rangle \langle 45 \rangle \langle 56 \rangle \langle 61 \rangle
    },
\end{align}
\noindent where $S$ is the kinematical matrix of the massless
scalar integrals. If we put all the fonctions in $n+2$ dimensions,
therefore all three points functions vanish. Thanks to the
supersymmetric decomposition $\left(
\ref{supersymmetricdecompostion-+++} \right)$, we have:
\begin{align}
    & A^{fermion}_{6}(++++++) \ = \ -2 A^{scalar}_{6}(++++++), &\\
    & A^{{\cal N} = 1}_{6}(++++++). \ = \ 0
\end{align}

\hfil

\section{Conclusion}

\hfil

In this paper, we have calculated all the four-photon helicity
amplitudes in massive and massless QED, scalar QED and
supersymmetric $\textrm{QED}^{{\cal N}=1}$. To compute them, we
use two very powerful methods: the unitarity-cuts and helicity
amplitudes accompanied with the spinor formalism. So we don't need
any standard reduction method. Thanks to four cuts in four
dimensions, we obtain the coefficients in front of boxes, thanks
to three cuts we obtain the coefficients in front of triangles and
bubbles. The extension in $4-2\epsilon$ dimensions of the
unitarity-cuts allows us to calculate straightforward the rational
terms.

\hfil

In this example, we have simplifications because we have only four
massless external legs. So we don't need the two cuts methods to
compute the coefficients in front of bubbles.

\hfil

Moreover, the transformation of the four point scalar integrals in
$n$ dimensions in four point scalar integrals in $n+2$ dimensions
reduces strongly the final result. Indeed this transformation
allows us to separate the infrared structure of the amplitude.
This separation generates many explicite compensations.

\hfil

As we have some very compact expressions of the six-photon
amplitudes in the massless theories \cite{bernicot}, we hope to
obtain expressions in massive theories, by understanding of the
origin of the rational terms. Thanks to the two-cut techniques, we
could calculate the first of the four six-photon helicity
amplitudes. In a next paper, we develop the calculation of the
next six-photon helicity amplitudes.

\hfil

\section*{Acknowledgments}

I would like to thank J.P Guillet for his explanations on IR
divergences and the rational terms. I also would like to thank P.
Aurenche for a careful reading of the manuscript.


\hfil

\hfil

\begin{appendix}

{\LARGE \bf APPENDIX}

\renewcommand{\theequation}{\Alph{section}.\arabic{equation}}
\setcounter{equation}{0}

\hfil

\hfil

We give, for sake of completeness, the vertices in QED and scalar
QED, and then the known results on the four-photon amplitudes are
given in the second appendix. The third appendix recall the
reduction of tensor integrals and extra-dimension scalar
integrals. Moreover, in the fourth appendix, we give the
definition of the master integrals used in this paper is recalled
and in the next appendix the reduction of the pentagon. Finally,
we give the proof of the amplitude of each chain of photons used
in this paper.

\hfil

\section{Vertices} \label{vertex}

\noindent The QED vertex is:
\begin{align}
    \parbox{2cm}{\includegraphics[width=2cm]{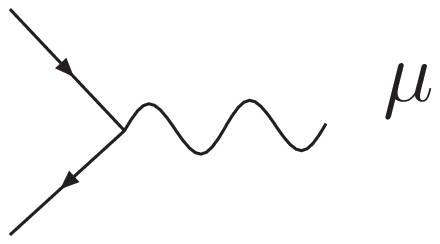}} =
    -ie \gamma^{\mu},
\end{align}
\noindent whereas the two scalar QED vertices are:
\begin{align}
    \parbox{2.5cm}{\includegraphics[width=2.5cm]{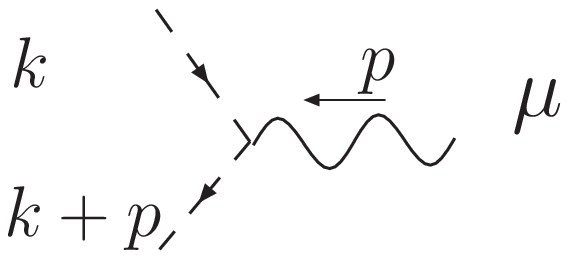}} =
    -ie \left\{ k + (p+k) \right\}^{\mu} \ \ \ \parbox{2cm}{\includegraphics[width=2cm]{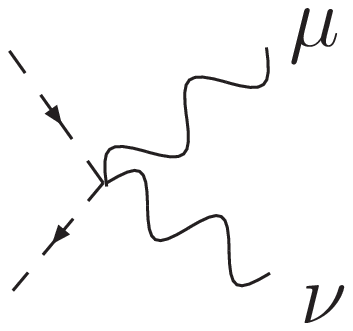}} =
    2ie^{2}\eta^{\mu\nu}.
\end{align}

\hfil

\section{Massless limit of the four photons amplitudes} \label{knownresult}

\hfil

We take the massless limit in the leading order in $\epsilon$ of
the results in the section $\ref{summaryscalar}$. We use the
Schouten Identity, and we find that $ \frac{[12][34]}{\langle 12
\rangle \langle 34 \rangle} $, $ \frac{[34] [231]}{ \langle 34
\rangle \langle 231 \rangle} $ are invariant by permutation. So
the helicity amplitudes are:
\begin{align}
    \dsp A^{scalar}_{4}(++++) \ & \dsp \xrightarrow[m^{2}\rightarrow 0]{} -4 i \alpha^{2} \frac{[12][34]}{\langle 12 \rangle \langle 34
    \rangle} + O(\epsilon), &\\
    \dsp A_{4}^{scalar}(-+++) \ & \dsp \xrightarrow[m^{2}\rightarrow 0]{} -4 i \alpha^{2} \frac{[34]
    [231]}{ \langle 34 \rangle \langle 231 \rangle} + O(\epsilon),
    & \\
    \dsp A_{4}^{scalar}(--++) \ & \dsp \xrightarrow[m^{2}\rightarrow 0]{} 4 i \alpha^{2} \frac{\langle 12 \rangle}{[12]}\frac{[34]}{\langle 34
    \rangle}\left\{ 1 - \frac{2ut}{s}I_{4}^{n+2}(1324) + \frac{t-u}{s}
    \left( I_{2}^{n}(u) - I_{2}^{n}(t)\right) \right\} + O
    (\epsilon), &\\
    \dsp A^{fermion}_{4}(\pm+++) \ & \dsp \xrightarrow[m^{2}\rightarrow 0]{} \ -2
    A^{scalar}_{4}(\pm+++), & \\
    \dsp A^{fermion}_{4} (--++) \ & \dsp \xrightarrow[m^{2}\rightarrow 0]{} -8 i \alpha^{2} \frac{\langle 12 \rangle}{[12]}\frac{[34]}{\langle 34 \rangle}
     \left\{ 1 + \frac{t^{2}+u^{2}}{s} I_{4}^{n+2}(1324) +  \frac{ t-u}{s}\left(
    I_{2}^{n}(u)-I_{2}^{n}(t) \right) \right\} + O\left( \epsilon
    \right).
\end{align}
\noindent We find the results known in the massless limit
\cite{4gammabern:2001,gamsusy}.

\hfil

\section{Definition of the master integrals} \label{Masterintegrals}

\hfil

In this appendix, we give the definition of the master integrals
used in this paper. We write $G_{i}$ the Gram determinant and
$S_{i}$ the kinematical S-matrix. Consider a scalar integral:
\begin{equation*}
\parbox{2cm}{\includegraphics[width=2cm]{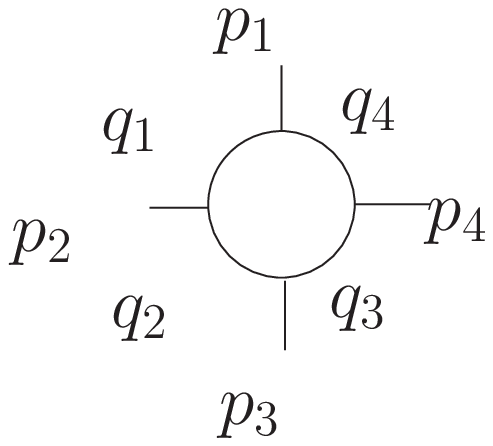}}
\end{equation*}
\noindent We define the Gram and kinematical S-matrix by:
\begin{align}
    G_{ij} & = 2 p_{i}.p_{j}, &\\
    S_{ij} & = \left( q_{j} - q_{i}
    \right)^{2}-m_{i}^{2}-m_{j}^{2},
\end{align}
\noindent and we note the spatial integral:
\begin{equation}
    r_{\Gamma} \ = \ \frac{\Gamma \left( 1 +\epsilon \right) \Gamma^{2} \left( 1 - \epsilon \right)}{\Gamma \left( 1 - 2\epsilon
    \right)}.
\end{equation}
\noindent Finally, in all formulae, the analytic prescriptions
are:
\begin{equation}
     s \rightarrow s + i\lambda \quad \quad m^{2} \rightarrow m^{2}-
     i\lambda.
\end{equation}

\hfil

\subsection{Two-point functions}

\begin{equation*}
\parbox{2cm}{\includegraphics[width=2cm]{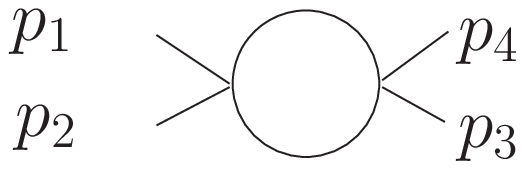}}
\quad \quad \quad \quad \quad \quad \quad \quad
    s = s_{12}= s_{34}
\end{equation*}

\noindent In a massless theory, the two-point scalar integral in
$n$ dimensions is:
\begin{equation}
    I_{2}^{n} \left( s \right) \ = \ \dsp \frac{r_{\Gamma}}{\epsilon (1-2\epsilon)} \left( -s \right)
    ^{-\epsilon} \ = \ \frac{1}{\epsilon} - \ln \left( -s \right) +2 + O\left( \epsilon \right).
\end{equation}
\noindent and in $n+2$ dimensions, the two-point scalar integral
is:
\begin{equation}
    I_{2}^{n+2} \left( s \right) \ = \ \dsp -\frac{r_{\Gamma}}{2\epsilon (1-2\epsilon)(3-2\epsilon)} \left( -s \right)
    ^{1-\epsilon}.
\end{equation}
\noindent In a massive theory, the two-point scalar integral in
$n$ dimensions, in the leading order in $\epsilon$ is:
\begin{equation}
    I_{2}^{n} (s) \ = \ m^{-2\epsilon} \frac{\Gamma ( 1+\epsilon
    )}{\epsilon}+2 + f\left( \sqrt{\rho} \right)
    + O(\epsilon).
\end{equation}
\noindent where $ \rho = 1-\frac{4m^{2}}{s}$. Thanks to the small
imaginary part: $ s \rightarrow s + i\lambda, m^{2} \rightarrow
m^{2}- i\lambda$ and $m^{2} > 0$, we have:
\begin{align}
    &\textrm{if} \ s < 0 \ \textrm{then} \ f\left( \sqrt{\rho} \right) = \sqrt{\rho} \ \ln \left( \frac{\sqrt{\rho}-1}{\sqrt{\rho}+1}\right)- i\lambda &\\
    &\textrm{if} \ s > 4m^{2} \ \textrm{then} \ f\left( \sqrt{\rho}
    \right)= \sqrt{\rho} \ \left( \ln \left( \frac{1- \sqrt{\rho}}{\sqrt{\rho}+1}\right) - i \pi \right) &\\
    &\textrm{if} \ s \in[0,4m^{2}] \ \textrm{then} \ f\left( \sqrt{\rho}
    \right)= -i\sqrt{-\rho} \ \ln \left( \frac{-i\sqrt{-\rho}-1}{-i\sqrt{-\rho}+1}\right).&
\end{align}
The determinants are given by:
\begin{align}
    \det \left( S_{2} \right) & = s \ \left(s-4 m^{2} \right), &\\
    \det \left( G_{2} \right) & = s. &
\end{align}
\noindent Most of those results comes from
\cite{Bern:massiveloop,method8}.

\hfil

\subsection{One external mass three-point functions}

\hfil

\begin{equation*}
\parbox{2cm}{\includegraphics[width=2cm]{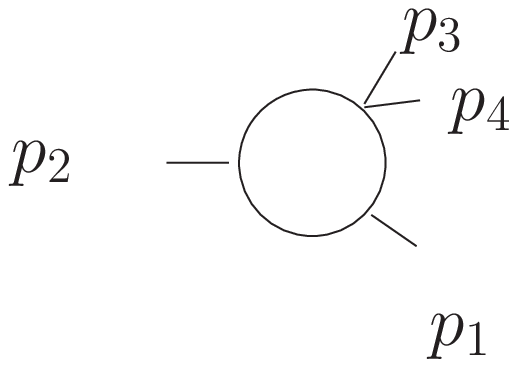}}
\quad \quad \quad \quad \quad \quad \quad \quad
    s = s_{34}
\end{equation*}

\noindent In a massless theory, the one external mass scalar
triangle in $n$ dimensions is:
\begin{equation}
    I_{3}^{n} \left( s \right) \ = \ \dsp \frac{r_{\Gamma}}{\epsilon ^{2}} \frac{\left( -s
    \right)^{-\epsilon}}{s} \ = \ \frac{1}{s} \left( \frac{1}{\epsilon^{2}} - \ln \left( -s \right)
    + \frac{\ln \left( -s \right) ^{2}}{2} \right)+ O\left( \epsilon
    \right),
\end{equation}
\noindent and in $n+2$ dimensions, this triangle is:
\begin{equation}
    I_{3}^{n+2} \left( s \right) \ = \ \dsp \frac{r_{\Gamma}}{2 \epsilon (1-\epsilon)(1-2\epsilon)} \left( -s
    \right)^{-\epsilon}.
\end{equation}
\noindent In a massive theory, the one external mass scalar
triangle in $n$-dimensions, in the leading order in $\epsilon$ is:
\begin{equation}
    I_{3}^{n}(s) \ = \ - \frac{1}{2s} \ln ^{2} \left( \frac{\rho -1}{\rho + 1}
    \right) + O \left( \epsilon \right),
\end{equation}
\noindent where $ \rho = \sqrt{1-\frac{4m^{2}}{s}}$. The
determinants are given by:
\begin{align}
    \det \left( S_{3} \right) & \ = \ 2 s^{2}m^{2}, &\\
    \det \left( G_{3} \right) & \ = \ - s^{2}. &
\end{align}

\hfil

\subsection{No external mass scalar four-point function}

\begin{equation*}
\parbox{2cm}{\includegraphics[width=2cm]{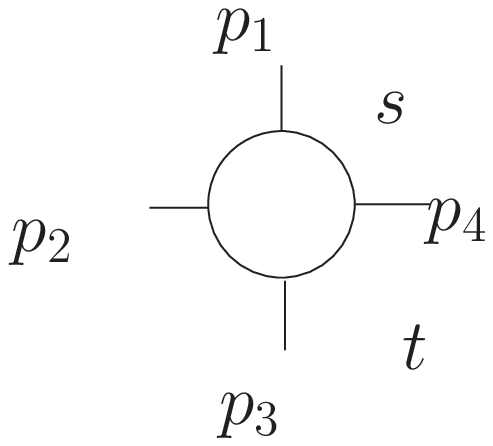}}
\quad \quad \quad \quad \quad \quad \quad \quad
\left\{\begin{array}{l}
    s = s_{12} \\
    t = s_{14} \\
    u = s_{13}
\end{array} \right.
\end{equation*}

\noindent In a massless theory, the no-external-mass scalar box in
$n$ dimensions is:
\begin{equation}
    \dsp I_{4}^{n} \left( s,t \right) \ = \  \frac{2}{st} \frac{r_{\Gamma}}{\epsilon
    ^{2}}\left\{ (-s)^{-\epsilon} + (-t)^{-\epsilon} \right\} -\frac{2}{ st} F_{0}
    (s,t),
\end{equation}
\noindent where $F_{0} (s,t) = \dsp \frac{1}{2}\left\{  \ln ^{2}
\left( \frac{s}{t} \right) + \pi^{2} \right\}$. In $n+2$
dimensions, this box is:
\begin{equation}
    I_{4}^{n+2} \left( s,t \right) \ = \ \frac{2 \sqrt{det(S_{4})}}{(n-3)det(G_{4})}
    F_{0}(s,t) \ = \ \frac{u}{2}\left\{  \ln ^{2} \left( \frac{s}{t} \right) + \pi^{2}
    \right\} + O \left( \epsilon \right).
\end{equation}
\noindent In a massive theory, the no-external-mass scalar box in
$n$ dimensions, in the leading order in $\epsilon$, is:
\begin{equation}
    I_{4}^{n}(s,t) \ = \ - \frac{1}{st} \left\{ H\left(
    -\frac{um^{2}}{st}, \frac{m^{2}}{s} \right) + H\left(
    -\frac{um^{2}}{st}, \frac{m^{2}}{t} \right) \right\} + O \left( \epsilon
    \right),
\end{equation}
\noindent where:
\begin{align}
    H(X,Y) \ = \ &\frac{2}{x_{+}-x_{-}} \left\{ \ln \left( 1 -
    \frac{X}{Y} \right) \ln \left( -\frac{x_{-}}{x_{+}} \right)
    \right. & \nonumber \\
    & \left. - Li_{2} \left( \frac{x_{-}}{y-x_{+}} \right) - Li_{2} \left( \frac{x_{-}}{x_{-}-y}
    \right) + Li_{2} \left( \frac{x_{+}}{x_{+}-y} \right) + Li_{2} \left( \frac{x_{+}}{y-x_{-}}
    \right)\right\},
\end{align}
\noindent with $ \dsp x_{\pm} = \frac{1}{2} \left( 1 \pm
\sqrt{1-4X} \right)$ and $ \dsp y = \frac{1}{2} \left( 1 +
\sqrt{1-4Y} \right) $. The standard definition of the dilogarithm
is $ Li_{2}(x) = - \int_{0}^{1}dt \  \frac{\ln(1-xt)}{t}$. The
determinants are given by:
\begin{align}
    \det \left( S_{4} \right) & \ = \ ts \left[ -4 \ m^{2} (t+s) + ts
    \right],
    &\\
    \det \left( G_{4} \right) & \ = \ -2st(s+t) = 2stu. &
\end{align}

\hfil

\section{Reduction of integrals}

\hfil

\subsection{Reduction of tensors integrals}

\hfil

In this appendix, we give the reduction of the linear-tensor
two-point integrals, linear-tensor one-mass three-point integrals
and linear-tensor no-mass four-point integrals. In the argument of
each integral, we give the numerator of the tensor integrals. We
can find techniques of reduction in \cite{method1}.
\begin{figure}[httb!]
\centering
\includegraphics[width=9cm]{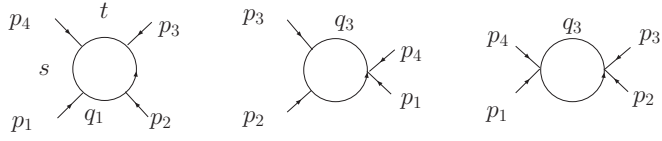}\\
\caption{\scriptsize \textit{Kinematics of bubbles, one external
mass triangle and no external mass box.}}
\end{figure}

\begin{align}
    I_{2}^{n} \left( q_{3}^{\mu} \right) & = \frac{1}{2} \left(
    p_{2}+p_{3} \right)^{\mu}I_{2}^{n}, & \label{reductionbubble}\\
    I_{3}^{n} \left( q_{3}^{\mu} \right) & = \frac{1}{t}I_{2}^{n}
    p_{2}^{\mu} - \left(I_{3}^{n} + \frac{1}{t}I_{2}^{n} \right)p_{3}^{\mu}, & \label{reductiontriangle} \\
    I_{4}^{n}(q_{1}^{\mu}) & = -\frac{1}{2u}\left( t I_{4}^{n}(1234)
    -2I_{3}^{n}(s) +2I_{3}^{n}(t) \right) \left( p_{1}^{\mu}+ p_{3}^{\mu} \right)
    +\frac{\left(p_{1}^{\mu}+p_{4}^{\mu} \right)}{2}I_{4}^{n}(1234),  &
    \label{annexeintegrationI4}\\
    I_{4}^{n}\left( \left( \mu^{2}+m^{2} \right) l_{1}^{\mu} \right) & = -\frac{1}{2u}\left( t J_{4}^{n}(1234)
    -2J_{3}^{n}(s) +2J_{3}^{n}(t) \right) \left( p_{1}^{\mu}+ p_{3}^{\mu} \right)
    +\frac{\left(p_{1}^{\mu}+p_{4}^{\mu} \right)}{2}J_{4}^{n}(1234).  & \label{annexeintegrationJ4}
\end{align}

\hfil

\subsection{Reduction of extra-dimension scalar integrals} \label{integraldimsup}

\hfil

Here we give several formulas to calculate extra-dimension scalar
integrals. Most of those results come
from~\cite{Bern:massiveloop}. We can relate the extra scalar
integral and the scalar integral in $n+2t$ dimensions:
\begin{equation}
    \int \frac{d^{n}Q}{\left( 2\pi\right) ^{n}}
    \frac{(\mu^{2})^{t}}{D_{1}^{2}...D_{N}^{2}} \ = \ - \epsilon ( 1 -\epsilon) .... (t-1-\epsilon) (4\pi)^{t} \int
    \frac{d^{n +2t}Q}{\left( 2\pi\right) ^{n +
    2t}}\frac{1}{D_{1}^{2}...D_{N}^{2}}.
\end{equation}
\noindent Using this formula we can calculate easily $J_{N}$ and
$K_{N}$ function. Here we give some formula useful for this paper:
\begin{align}
    (-\epsilon) I_{4}^{n+2} \ & = \ 0 + O \left( \epsilon \right), &\\
    (-\epsilon)(1- \epsilon)I_{4}^{n+4} \ & = \  -\frac{1}{6} + O \left(\epsilon
    \right).
\end{align}
\noindent We have:
\begin{align}
    J_{N}^{n} \ & = \ m^{2} I_{N}^{n} + (-\epsilon) I_{N}^{n+2}, &\\
    K_{N}^{n} \ & = \ m^{4} I_{N}^{n} -2m^{2} \epsilon I_{N}^{n+2} + (-\epsilon)(1- \epsilon)I_{N}^{n+4}. &
\end{align}
\noindent And the following relation holds between the $J_{4}$ and
$I_{4}^{n+2}$ functions:
\begin{equation}
    J_{4}^{n}(1234) \ = \ -\frac{st}{4u} I_{4}^{n}(1234)+\frac{s}{2u} I_{3}^{n}(s) +\frac{t}{2u}I_{3}^{n}(t) - \frac{1}{2}
    I_{4}^{n+2}(1234). \label{passageJaI}
\end{equation}

\hfil

\section{Reduction of the pentagons and scalar hexagons}
\label{HexaPenta}

\hfil

The method of reduction comes from \cite{method1}. The explicit
reduction of massless scalar pentagon and massless scalar hexagon
is in \cite{Binoth:2001vm,Bern:pentagon,reduction}.

\hfil

\noindent We note $I_{6}^{n}$ the hexagon in $n$ dimensions and
$I_{5}^{n} \left( s_{ij} \right)$ the n-dimensional one-mass
scalar pentagon. This pentagon is obtain by removing the
propagator, of the scalar hexagon, between the external momentum
$p_{i}$ and $p_{j}$. We note ${S_{6}}_{ij}$ the kinematical matrix
of the hexagon and ${S_{5}}_{ij}$ the kinematical matrix of the
pentagon.

\hfil

\noindent According to \cite{Bern:pentagon,method1}, the exact
reduction of the hexagon is:
\begin{equation}
    I_{6}^{n} \ = \ \sum_{k,l=1}^{6} {S_{6}}_{kl}^{-1} I_{5}^{n} \left( s_{l \ l+1}
    \right),
\end{equation}
\noindent where the kinematical matrix is:
\begin{align}
{S^{6}}_{kl} =
\begin{pmatrix}
    0 & 0 & s_{23} & s_{234} & s_{16} & 0 \\
    0 & 0 & 0 & s_{34} & s_{345} & s_{12} \\
    s_{23} & 0 & 0 & 0 & s_{45} & s_{456} \\
    s_{234} & s_{34} & 0 & 0 & 0 & s_{56} \\
    s_{16} & s_{345} & s_{45} & 0 & 0 & 0 \\
    0 & s_{12} & s_{456} & s_{56} & 0 & 0
\end{pmatrix} -2 m^{2} \begin{pmatrix}
    1 & 1 & 1 & 1 & 1 & 1 \\
    1 & 1 & 1 & 1 & 1 & 1 \\
    1 & 1 & 1 & 1 & 1 & 1 \\
    1 & 1 & 1 & 1 & 1 & 1 \\
    1 & 1 & 1 & 1 & 1 & 1 \\
    1 & 1 & 1 & 1 & 1 & 1
\end{pmatrix}.
\end{align}
\noindent The reduction, in the leading order in $\epsilon$, of
the pentagon is:
\begin{align}
    I_{5}^{n}(s) = \sum_{k,l=1}^{5} {S_{5,s}}_{kl}^{-1}
    I_{4}^{n} \left( s,s_{l \ l+1} \right) + O \left(
    \epsilon \right).
\end{align}
\noindent Where $I_{4}^{n} \left( s,s_{l \ l+1} \right)$ is the
four-point scalar integral, obtain by removing the propagateur
between the legs with the momentum $p_{l}$ and $p_{l+1}$ of the
pentagon $I_{5}^{n}(s)$. The kinematical matrix for the pentagon
$I_{5}(s_{12})$ is:
\begin{align}
{S_{5,s_{12}}}_{kl} =
\begin{pmatrix}
    0 & 0 & s_{34} & s_{345} & s_{12}\\
    0 & 0 & 0 & s_{45} & s_{456} \\
    s_{34} & 0 & 0 & 0 & s_{56}  \\
    s_{345} & s_{45} & 0 & 0 & 0 \\
    s_{12} & s_{456} & s_{56} & 0 & 0
\end{pmatrix} -2 m^{2} \begin{pmatrix}
    1 & 1 & 1 & 1 & 1  \\
    1 & 1 & 1 & 1 & 1  \\
    1 & 1 & 1 & 1 & 1  \\
    1 & 1 & 1 & 1 & 1  \\
    1 & 1 & 1 & 1 & 1
\end{pmatrix}.
\end{align}
\noindent For another pentagon, we permute the labels. The
hexagons and pentagons have no infrared divergences. We keep only
the "finite parts" of box functions. This part is usually called
$F_{i}$, where the subscripts $i$ represents the number of
external legs. We found those definitions in \cite{Binoth:2001vm}.

\hfil


\section{Computation of the on-shell trees using in the paper} \label{onshelltrees}

\hfil

\subsection{On-shell trees with two positive-helicity or two
negative-helicity photons}

\hfil

{\bf Proposition 1:} \\
Consider two positive-helicity photons with
the momentum $p_{1}$ and $p_{2}$, join and connect them by an
on-shell propagator, therefore we have:
\begin{equation}
    \dsp \frac{\langle R q_{1} 1 \rangle}{\langle R1
    \rangle} \frac{\langle R q_{2} 2 \rangle}{\langle R2
    \rangle} =  -\left( \mu^{2} + m^{2} \right) \frac{[1 2]}{ \langle 12 \rangle}. \label{arbre1++}
\end{equation}
\noindent In the case of two negative-helicity photons, we have:
\begin{equation}
    \dsp \frac{[ R q_{1} 1 ]}{[1R]} \frac{[ R q_{2} 2 ]}{[2R]} = -\left( \mu^{2} + m^{2} \right) \frac{ \langle 12 \rangle}{[1 2]}. \label{arbre1--}
\end{equation}

\dem We assume to have two photons with a positive helicity. We
note $q_{0} = q_{1}-p_{1}$. And all propagators are on-shell,
therefore $ 2(p_{1}.q_{0}) = 2(p_{2}.q_{2}) = 0 $. Using this
trick the amplitude is spelt:
\begin{align}
    \dsp \frac{\langle R q_{1} 1 \rangle}{\langle R1
    \rangle} \frac{\langle R q_{2} 2 \rangle}{\langle R2
    \rangle}& = \dsp  \frac{ \langle R q_{0} 12 q_{2}
    R\rangle}{\langle R1 \rangle \langle R2 \rangle \langle 12 \rangle}  = \dsp \frac{ 2(p_{2}.q_{2})\langle Rq_{0} 1R\rangle
    -2(p_{1}.q_{0})\langle Rq_{2} 2R\rangle + \langle R 1 q_{0} q_{2} 2 R\rangle}{\langle R1 \rangle \langle R2 \rangle \langle 12 \rangle} & \\
    & \dsp = \frac{  q_{1}^{2} \langle R 12 R\rangle}{\langle R1 \rangle \langle R2 \rangle \langle 12 \rangle} =
    \frac{  \left( \mu^{2} + m^{2} \right)\langle R 12 R\rangle}{\langle R1 \rangle \langle R2 \rangle \langle 12 \rangle} =
    -\left( \mu^{2} + m^{2} \right) \frac{[1 2]}{ \langle 12 \rangle}. &
\end{align}
\noindent There is the same demonstration with negative-helicities
photons.

\hfil

{\bf Proposition 2:}\\
Consider a chain of two positive-helicity
photons with the momentum $p_{1}$ and $p_{2}$ surround with two
on-shell propagators, therefore, we have:
\begin{equation}
    \dsp \sum_{\sigma(1,2)} \frac{\langle R q_{1} 1 \rangle}{\langle R1
    \rangle}\frac{i}{ D_{1}^{2}} \frac{\langle R q_{2} 2 \rangle}{\langle R2
    \rangle} = -\left( \mu^{2} + m^{2} \right) \frac{[1 2]}{ \langle 12 \rangle} \sum_{\sigma(1,2)} \frac{ i }{D_{1}^{2}}. \label{arbre2++}
\end{equation}
\noindent If we have two negative-helicity photons, the chain is:
\begin{equation}
    \dsp \sum_{\sigma(1,2)} \frac{[ R q_{1} 1 ]}{[1R]}\frac{i}{ D_{1}^{2}} \frac{[ R q_{2} 2 ]}{[2R]} =
    -\left( \mu^{2} + m^{2} \right) \frac{ \langle 12 \rangle}{[1 2]} \sum_{\sigma(1,2)} \frac{ i }{D_{1}^{2}}. \label{arbre2--}
\end{equation}
\noindent If the joined propagator is put on-shell therefore we
find formula $\left( \ref{arbre1++},\ref{arbre1--} \right)$.

\dem
\begin{align}
    \dsp \sum_{\sigma(1,2)} \frac{\langle R q_{1} 1 \rangle}{\langle R1
    \rangle}\frac{i}{ D_{1}^{2}} \frac{\langle R q_{2} 2 \rangle}{\langle R2
    \rangle}& = \dsp \sum_{\sigma(1,2)} \frac{i}{\langle R1 \rangle \langle R2 \rangle \langle 12 \rangle}\frac{ \langle R q_{0} 12 q_{2}
    R\rangle}{D_{1}^{2}} &\\
    & = \dsp \sum_{\sigma(1,2)} \frac{i}{\langle R1 \rangle \langle R2 \rangle \langle 12 \rangle}\frac{ 2(p_{2}.q_{2})\langle Rq_{0} 1R\rangle
    -2(p_{1}.q_{0})\langle Rq_{2} 2R\rangle + \langle R 1 q_{0} q_{2} 2 R\rangle}{D_{1}^{2}} &\\
    & = \dsp \sum_{\sigma(1,2)} \frac{i}{\langle R1 \rangle \langle R2 \rangle \langle 12 \rangle}\frac{ \left( D_{2}^{2}- D_{1}^{2}\right)
    \langle Rq_{0} 1R\rangle
    -\left( D_{1}^{2}-D_{4}^{2}\right)\langle Rq_{2} 2R\rangle + q_{1}^{2}\langle R 1 2 R\rangle}{D_{1}^{2}} &\\
    & = \dsp \sum_{\sigma(1,2)} \frac{i}{\langle R1 \rangle \langle R2 \rangle \langle 12 \rangle}\frac{  \left( \mu^{2} + m^{2} \right)\langle R 1 2 R\rangle}{D_{1}^{2}} &\\
    & = \dsp   -\left( \mu^{2} + m^{2} \right) \frac{[1 2]}{ \langle 12 \rangle} \sum_{\sigma(1,2)} \frac{ i }{D_{1}^{2}}. &
\end{align}
\noindent Most of the reduction come from the permutation. For
photons with a negative helicity, the proof is the same.

\hfil

\subsection{ On-shell trees with one positive-helicity photon
and one negative-helicity photon}

\hfil

{\bf Proposition 3:}\\
Consider a diagram with a scalar line. On this scalar line with
have first a chain with one on-shell negative-helicity photon and
one on-shell positive-helicity photon. We assume that the momentum
of negative-helicity photon (respectively positive-helicity
photon) is: $p_{1}$ (respectively $p_{2}$). Moreover we assume
that a third on-shell photon, with the momentum $p_{3}$, is
ingoing in the scalar line just after this chain (fig.
\ref{proposition2}). The propagators around the chain are on-shell
$ D_{0}^{2} =0$ and $ D_{2}^{2}=0$. The amplitude of this diagram
is:

\begin{figure}[httb!]
\centering
\includegraphics[width=2.7cm]{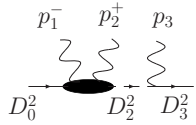}\\
\caption{\scriptsize \textit{ Chain composed with two photons,
with two different helicity, following with one photon.}}
\label{proposition2}
\end{figure}
\begin{align}
    \dsp \frac{[r2]\langle R1\rangle}{2 \langle R2 \rangle [1r]} + \sum_{\sigma(1,2)} \frac{[r q_{1} 1 ]}{[1r] }\frac{i}{D_{1}^{2}}
    \frac{\langle R q_{2} 2 \rangle}{\langle R2 \rangle} = & \dsp
    -\frac{i }{ \langle 231 \rangle}
    \left(  \frac{ \langle 1 q_{2} 232\rangle -D_{3}^{2}\langle 1 q_{2} 2\rangle  +\left( \mu^{2} + m^{2} \right) [231]}{D_{1}^{2}}
    \right) & \nonumber \\
    &-\frac{i }{\langle 231 \rangle} \left(  \frac{  -D_{3}^{2}\langle 1 q_{2} 2\rangle + \langle 1
    q_{2} 31 2\rangle +\left( \mu^{2} + m^{2} \right)[231]}{{D_{1}^{'}}^{2}} \right). \label{arbre-+}
\end{align}

\dem We choose $|R\rangle = |1\rangle $ and $|r\rangle =
|2\rangle$. The computation gives us:
\begin{align}
    & \dsp \sum_{\sigma(1,2)} \frac{[r q_{1} 1 ]}{[1r]} \frac{i}{ D_{1}^{2}}
    \frac{\langle R q_{2} 2 \rangle}{\langle R2 \rangle} =
    -\frac{i \langle 1 q_{2} 2\rangle ^{2}}{s_{12}}
    \sum_{\sigma(1,2)} \frac{1}{D_{1}^{2}} = \dsp
    -\frac{i }{s_{12} \langle 231 \rangle}
    \left(  \frac{ \langle 1 q_{2} 231 q_{2} 2\rangle}{D_{1}^{2}} + \frac{ \langle 1 q_{2} 231 q_{2} 2\rangle}{{D_{1}^{'}}^{2}} \right) &\\
    = & \dsp
    -\frac{i }{s_{12} \langle 231 \rangle}
    \left(  \frac{ 2(p_{1}. q_{2})\langle 1 q_{2} 23 2\rangle -2(p_{3}. q_{2})\langle 1 q_{2} 21
    2\rangle +2(p_{2}. q_{2})\langle 1 q_{2} 31 2\rangle
    - q_{2}^{2} \langle 12312 \rangle}{D_{1}^{2}}
    \right)& \nonumber \\
    & \dsp -\frac{i }{s_{12} \langle 231 \rangle}
    \left(  \frac{ 2(p_{1}. q_{2})\langle 1 q_{2} 23 2\rangle -2(p_{3}. q_{2})\langle 1 q_{2} 21 2\rangle
    +2(p_{2}. q_{2})\langle 1 q_{2} 31 2\rangle - q_{2}^{2} \langle 12312 \rangle}{{D_{1}^{'}}^{2}}
    \right)&\\
    = & \dsp
    -\frac{i }{ \langle 231 \rangle}
    \left(  \frac{ \langle 1 q_{2} 232\rangle -D_{3}^{2}\langle 1 q_{2} 2\rangle  +\left( \mu^{2} + m^{2} \right) [231]}{D_{1}^{2}}
    \right)-\frac{i }{\langle 231 \rangle}
    \left(  \frac{  -D_{3}^{2}\langle 1 q_{2} 2\rangle + \langle 1 q_{2} 31 2\rangle +\left( \mu^{2} + m^{2} \right) [231]}{{D_{1}^{'}}^{2}}
    \right).
\end{align}


\section{Four-cut technique for $A_{4}^{scalar}(-+++)$} \label{calcul-+++}

\hfil

In this appendix, we want to prove that, the integral $\left(
\ref{M-+++1scamasse4cuts} \right)$ could be calculate without
integration formulas, like $\left( \ref{annexeintegrationI4}
\right)$. We use just the four on-shell conditions. We want to
calculate the tensor integral:
\begin{equation}
    (e\sqrt{2})^{4} \sum_{\sigma(2,3,4)}  \frac{[34]}{\langle
    34\rangle \langle 231 \rangle} \int d^{n}Q
    \langle 1q_{2}232 \rangle\left( \mu^{2} + m^{2}\right) \delta \left( D_{1}^{2} \right)
    \delta \left( D_{2}^{2}\right) \delta \left( D_{3}^{2}\right) \delta \left(
    D_{4}^{2}\right) \label{LLL}
\end{equation}
\noindent corresponding to the graph: $
\parbox{3cm}{\includegraphics[width=3cm]{4-+++4cuts.eps}}$.

\hfil

\noindent The four-cut technique
\cite{Britto:2004nc,Britto:2005ha,Britto:2006sj} says that the
integrals $\left( \ref{LLL} \right)$ is:
\begin{align}
    &(e\sqrt{2})^{4}\sum_{\sigma(2,3,4)} \frac{[34]}{\langle
    34\rangle \langle 231 \rangle} \frac{1}{2} \sum_{i} \langle 1{q_{2}}_{i}232 \rangle \textrm{Disc}_{4} \int
    d^{n}Q \frac{ \mu^{2} + m^{2}}{ D_{1}^{2}  D_{2}^{2} D_{3}^{2} D_{4}^{2}} &\\
    = \ & (e\sqrt{2})^{4}\sum_{\sigma(2,3,4)} \frac{[34]}{\langle
    34\rangle \langle 231 \rangle} \frac{1}{2}  \langle 1 \sum_{i}{q_{2}}_{i}232 \rangle \textrm{Disc}_{4} \int
    d^{n}Q \frac{ \mu^{2} + m^{2}}{ D_{1}^{2}  D_{2}^{2} D_{3}^{2} D_{4}^{2}}, & \label{LLL1}
\end{align}
\noindent where ${q_{2}}_{i}$ are the solutions of:
\begin{align}
    D_{1}^{2}  &= 0 \ \Leftrightarrow \  \left(q_{2}-p_{2} \right)^{2}- \left( m^{2} +\mu^{2}\right) = 0 \label{equation1} \\
    D_{2}^{2}  &= 0 \ \Leftrightarrow \  q_{2}^{2} - \left( m^{2} +\mu^{2}\right)  = 0 \label{equation2} \\
    D_{3}^{2}  &= 0 \ \Leftrightarrow \  \left(q_{2}+p_{3} \right)^{2} - \left( m^{2} +\mu^{2}\right) = 0 \label{equation3} \\
    D_{4}^{2}  &= 0 \ \Leftrightarrow \  \left(q_{2}+p_{3}+p_{4} \right)^{2} - \left( m^{2} +\mu^{2}\right) = 0. \label{equation4}
\end{align}

\hfil

To solve this system of equations, we choose a basis of the
four-dimension Minkowski space: $ B = \left\{ p_{2}^{\mu},
p_{3}^{\mu}, \langle 2 \gamma^{\mu} 3 \rangle , \langle 3
\gamma^{\mu} 2 \rangle \right\}$. In our case, ${q_{2}^{\mu}}_{i}$
is a four-dimension vector, therefore, we can project it on the
base $ B $:
\begin{equation}
    {q_{2}^{\mu}}_{i} \ = \ a_{i} \ p_{2}^{\mu} + b_{i} \ p_{3}^{\mu} + \frac{c_{i}}{2} \langle 2 \gamma^{\mu}
    3 \rangle + \frac{d_{i}}{2} \langle 3 \gamma^{\mu} 2 \rangle .
    \label{decompositionq}
\end{equation}
\noindent So to know the vector ${q_{2}^{\mu}}_{i}$, we have to
calculate, the four coefficients $a_{i},b_{i},c_{i}$ and $d_{i}$.
The conditions $\left( \ref{equation1} \right)$ and $\left(
\ref{equation2} \right)$ impose:
\begin{equation}
    \left( {q_{2}}_{i} - p_{2} \right)
    ^{2} = m^{2} + \mu^{2}  \Leftrightarrow 2 \left( p_{2}.{q_{2}}_{i} \right) = 0  \Leftrightarrow b_{i} =
    0.
\end{equation}
\noindent The conditions $\left( \ref{equation2} \right)$ and
$\left( \ref{equation3} \right)$ impose:
\begin{equation}
    \left( {q_{2}}_{i} + p_{3} \right)
    ^{2} = m^{2}+\mu^{2}  \Leftrightarrow 2  \left(p_{2}.{q_{2}}_{i}\right) = 0  \Leftrightarrow a_{i} =
    0.
\end{equation}
\noindent The second condition $\left( \ref{equation2} \right)$
imposes:
\begin{equation}
    {q_{2}}_{i} ^{2} = m^{2} +\mu^{2}  \Leftrightarrow c_{i} d_{i} =
    -\frac{m^{2}+\mu^{2}}{s_{23}}. \label{condition1}
\end{equation}
\noindent And finally the two conditions $\left( \ref{equation3}
\right)$ and $\left( \ref{equation4} \right)$ impose:
\begin{equation}
    \left( {q_{2}}_{i} + p_{3} + p_{4} \right) ^{2} = m^{2}+\mu^{2} \Leftrightarrow
    c_{i}\langle 243 \rangle+ d_{i} \langle 342\rangle = -s_{34}. \label{condition2}
\end{equation}
\noindent So we have $a_{i}=b_{i}=0$ and the two equations $\left(
\ref{condition1},\ref{condition2} \right)$ gives us $c_{i}$ and
$d_{i}$. ${q_{2}}_{i}$ is totally define. Now we insert the
decomposition $\left( \ref{decompositionq} \right)$ in the
equation $\left( \ref{LLL1} \right)$ and we have:
\begin{equation}
     \langle 1 \sum_{i}{q_{2}}_{i}232 \rangle = -t \langle 123
     \rangle\sum_{i} c_{i} = st \frac{\langle 123
     \rangle}{\langle 243\rangle}.
\end{equation}
\noindent We input the last result in the integral $\left(
\ref{LLL1} \right)$, and we obtain:
\begin{align}
    & (e\sqrt{2})^{4}  \sum_{\sigma(2,3,4)} \frac{[34]}{\langle
    34\rangle \langle 231 \rangle} \int d^{n}Q
    \langle 1q_{2}232 \rangle\left( \mu^{2} + m^{2}\right) \delta \left( D_{1}^{2} \right)
    \delta \left( D_{2}^{2}\right) \delta \left( D_{3}^{2}\right) \delta \left(
    D_{4}^{2}\right)  & \nonumber \\
    = \ & (e\sqrt{2})^{4}  \sum_{\sigma(2,3,4)} \frac{[34][231]}{\langle
    34\rangle \langle 231 \rangle} \frac{ts}{2u} \textrm{Disc}_{4}
    J_{4}^{n}(1234). &
\end{align}
\noindent This result is the result which we have obtain with the
classical integration $\left( \ref{resultat-+++4cuts} \right)$.

\end{appendix}


\begin{thebibliography}{99}

\bibitem{karplus}
    R.\ Karplus and M.\ Neumann, \ Phys.\ Rev.\ {\bf 80} (1950) 380,
    {\bf 83} (1951) 776.

\bibitem{tollis}
    B.\ De Tollis, \ Nuovo \ Cim.\ {\bf 32} (1964) 757, {\bf 35} (1965) 1182.

\bibitem{Bern:massiveloop}
    Z.Bern and  A.G.Morgan,
    Nucl. \ Phys. \ B  {\bf 467} (1996) 479-509,
    [arXiv:hep-ph/9511336].

\bibitem{4gammabern:2001}
  Z.Bern, A. De Freitas, L.Dixon, A.Ghinculov and H.L.Wong,
  JHEP \ {\bf 0111} (2001) 031,
  [arXiv:hep-ph/0109079].

\bibitem{gamsusy}
  T.~Binoth,  E.~W.~N.~Glover, P.~Marquard and J.~J.~van der Bij,
  JHEP {\bf 0205} (2002) 060,
  [arXiv:hep-ph/0202266].

\bibitem{method8}
  A.~Brandhuber,  S.~McNamara, B.~J.~Spence and G.~Travaglini,
  JHEP {\bf 0510} (2005) 011,
  [arXiv:hep-th/0506068].

\bibitem{Binoth:2001vm}
    T.~Binoth,  J.~P.~Guillet, G.~Heinrich and C.~Schubert,
    Nucl. \ Phys. \  B {\bf 615} (2001) 385,
    [arXiv:hep-ph/0106243].

\bibitem{Bern:1993kr}
  Z.~Bern,  L.~J.~Dixon and D.~A.~Kosower,
  Nucl. \ Phys. \ B {\bf 412} (1994) 751,
  [arXiv:hep-ph/9306240].

\bibitem{Bern:super}
    Z. Bern, L. Dixon and D. A. Kosower,
    Phys. \ Rev. \ Lett. \ {\bf 70} (1993) 2677-2680,
    [arXiv:hep-ph/9302280v1].


\bibitem{Dixon:TASI}
    L. Dixon,
    [arXiv:hep-ph/9601359].

\bibitem{spinor:chinois}
    Z.~Xu, D.H~Zhang and L.~Chang,
    \ Nucl. \ Phys. \ B {\bf 291} (1987) 392-428.

\bibitem{Cutkosky}
    R.E. Cutkosky,
    \ J. \ Math. \ Phys \ {\bf 1} (1960) 429.


\bibitem{bern3cuts}
    Z. Bern, L. Dixon and D.A. Kosower,
    \ Nucl.Phys. \ B {\bf 513} (1998) 3-86,
    [arXiv:hep-ph/9708239v1].

\bibitem{Mastrolia:2006ki}
    P.~Mastrolia,
    Phys. \ Lett. \ B {\bf 644} (2007) 272,
    [arXiv:hep-th/0611091].

\bibitem{Britto:2004nc}
    R.~Britto,  F.~Cachazo and B.~Feng,
    Nucl.\ Phys. \ B {\bf 725} (2005) 275,
    [arXiv:hep-th/0412103].

\bibitem{Britto:2005ha}
    R.~Britto,  E.~Buchbinder, F.~Cachazo and B.~Feng,
    Phys.\ Rev.\ D {\bf 72} (2005) 065012,
    [arXiv:hep-ph/0503132].

\bibitem{Britto:2006sj}
    R.~Britto,  B.~Feng and P.~Mastrolia,
    Phys. \ Rev. \ D {\bf 73} (2006) 105004,
    [arXiv:hep-ph/0602178].

\bibitem{method7}
    C.~Anastasiou,  R.~Britto, B.~Feng, Z.~Kunszt and P.~Mastrolia,
    Phys. \ Lett. \  B {\bf 645} (2007) 213,
    [arXiv:hep-ph/0612277].

\bibitem{Bern:rational}
    Z. Bern, L. Dixon and D. A. Kosower,
    \ Ann. \ Rev. \ Nucl. \ Part. \ Sci. \ {\bf 46} (1996) 109-148,
    [arXiv:hep-ph/9602280v1].


\bibitem{Papadopoulos:rational}
    G.~Ossola, C.G.~Papadopoulos and R.~Pittau,
    \ JHEP {\bf 05} (2008) 004,
    [arXiv:0802.1876[hep-ph]].

\bibitem{method7bis}
    R.~Britto and B.~Feng,
    \ Phys. \ Rev. \ D {\bf 75} (2007) 105006,
    [arXiv:hep-ph/0612089].

\bibitem{Papa:cut}
    G.~Ossola, C.G.~Papadopoulos and R.~Pittau,
    \ Nucl. \ Phys. \ B {\bf 763} (2007) 147-169,
    [arXiv:hep-ph/0609007].

\bibitem{Forde:2007mi}
    D.~Forde,
    \ Phys. \ Rev. \ D \ {\bf 75 } (2007) 125019,
    [arXiv:hep-ph/0704.1835].

\bibitem{Kilgore}
    W. B. Kilgore,
    [arXiv:0711.5015].


\bibitem{Catani1}
    S. Catani, M. H. Seymour and Z. Trocsanyi,
    Phys. \ Rev. \ D {\bf 55} (1997) 6819-6829,
    [arXiv:hep-ph/9610553].

\bibitem{SecondOrder}
    A.G. Morgan,
    Phys. \ Lett. \ B {\bf 351} (1995) 249-256,
    [arXiv:hep-ph/9502230].

\bibitem{Papa:6gamma}
    G.~Ossola, C.G.~Papadopoulos and R.~Pittau,
    JHEP \ {\bf 07} (2007) 085,
    [arXiv:0704.1271 [hep-ph]].

\bibitem{nagy}
  Z.~Nagy and D.~E.~Soper,
  Phys.\ Rev. \ D {\bf 74} (2006) 093006,
  [arXiv:hep-ph/0610028].

\bibitem{arbre}
    Z. Bern, L. Dixon, D.C. Dunbar and D.A. Kosower,
    Phys. \ Lett. \ B \ {\bf 394} (1997) 105-115,
    [arXiv:hep-th/9611127 v2]


\bibitem{Bern:pentagon}
    Z. Bern, L. Dixon   nd D.A. Kosower,
    Phys. \ Lett. \ B \ {\bf 302} (1993) 299-308; Erratum-ibid. \ B \ {\bf 318} (1993) 649,
    [arXiv:hep-ph/9212308v1].


\bibitem{method1}
  T.~Binoth,  J.~P.~Guillet, G.~Heinrich, E.~Pilon and C.~Schubert,
  JHEP {\bf 0510} (2005) 015,
  [arXiv:hep-ph/0504267].

\bibitem{reduction}
    T.Binoth, J.Ph. Guillet and G. Heinrich,
    Nucl. \ Phys. \ B \ {\bf 572} (2000) 361-386,
    [arXiv:hep-ph/9911342].

\bibitem{bernicot}
    C.Bernicot and J.Ph. Guillet,
    JHEP {\bf 01} (2008) 059,
    [arXiv:0711.4713].

\end{thebibliography}
\end{document}